\documentclass[useAMS,usenatbib]{mn2e}
\usepackage{graphicx}

\usepackage{graphicx}
\usepackage{amsmath}

\def\url{Url:}

\title[Optical-Near-IR Transmission Spectrum of WASP-19b]{An HST Optical to Near-IR Transmission Spectrum of the Hot Jupiter WASP-19b: Detection of Atmospheric Water and Likely Absence of TiO}
\author[C. M. Huitson et al.]{C. M. Huitson$^{1}$\thanks{E-mail: chuitson@astro.ex.ac.uk (CMH)}, D. K. Sing$^{1}$, F. Pont$^{1}$, J. J. Fortney$^{2}$, A. S. Burrows$^{3}$, P. A. Wilson$^{1}$, 
\newauthor G. E. Ballester$^{4}$, N. Nikolov$^{1}$, N. P. Gibson$^{5,6}$, D. Deming$^{7}$,
S. Aigrain$^{5}$, T. M. Evans$^{5}$, 
\newauthor G. W. Henry$^{8}$, A. Lecavelier des Etangs$^{9}$, A. P. Showman$^{4}$,  A. Vidal-Madjar$^{9}$, 
\newauthor K. Zahnle$^{10}$ \\ 
$^{1}$Astrophysics Group, School of Physics, University of Exeter, Stocker Road, Exeter, EX4 4QL, UK \\
$^{2}$Department of Astronomy and Astrophysics, University of California, Santa Cruz, CA 95064, USA \\
$^{3}$Department of Astrophysical Sciences, Peyton Hall, Princeton University, Princeton, NJ 08544, USA \\
$^{4}$Lunar and Planetary Laboratory, University of Arizona, Tucson, Arizona 85721, USA \\
$^{5}$Department of Physics, University of Oxford, Denys Wilkinson Building, Keble Road, Oxford OX1 3RH, UK \\
$^{6}$European Southern Observatory, Karl-Schwarzschild-Str. 2, 85748 Garching bei M\"{u}nchen, Germany \\
$^{7}$Department of Astronomy, University of Maryland, College Park, MD 20742 USA \\
$^{8}$Tennessee State University, 3500 John A. Merritt Blvd., P.O. Box 9501, Nashville, TN 37209, USA \\
$^{9}$CNRS, Institut d'Astrophysique de Paris, UMR 7095, 98bis boulevard Arago, 75014 Paris, France \\
$^{10}$NASA Ames Research Center, Moffett Field, CA 94035, USA }
\begin{document}

\date{Accepted 2013 July 5. Received 2013 May 24; in original form 2013 February 28}

\pagerange{\pageref{firstpage}--\pageref{lastpage}} \pubyear{2013}

\maketitle

\label{firstpage}

\begin{abstract}

We measure the transmission spectrum of WASP-19b from three transits using low-resolution optical spectroscopy from the Hubble Space Telescope (HST) Space Telescope Imaging Spectrograph (STIS). The STIS spectra cover a wavelength range of 0.29-1.03~$\umu$m, with resolving power $R=500$. The optical data are combined with archival near-infrared data from the HST Wide Field Camera 3 (WFC3) G141 grism, covering the wavelength range from 1.087 to 1.687~$\umu$m, with resolving power $R=130$. We reach S/N levels between 3,000 and 11,000 in 0.1~$\umu$m bins when measuring the transmission spectra from 0.53-1.687~$\umu$m.

WASP-19 is known to be a very active star, with the optical stellar flux varying by a few per~cent over time. We correct the transit light curves for the effects of stellar activity using ground-based activity monitoring with the Cerro Tololo Inter-American Observatory (CTIO). While we were not able to construct a transmission spectrum using the blue optical data because of the presence of large occulted star spots, we were able to use the spot crossings to help constrain the mean stellar spot temperature. 

To search for predicted features in the hot-Jupiter atmosphere, in addition to the transmission spectrum we also define spectral indices for differential radius ($\Delta R_P/R_\star$) measurements to specifically search for the presence of TiO and alkali line features. Our measurements rule out TiO features predicted for a planet of WASP-19b's equilibrium temperature (2050~K) in the transmission spectrum at the 2.7-2.9~$\sigma$ confidence level, depending on atmospheric model formalism. The WFC3 transmission spectrum shows strong absorption features due to the presence of H$_2$O, which is detected at the 4~$\sigma$ confidence level between 1.1 and 1.4~$\umu$m.

The transmission spectra results indicate that WASP-19b is a planet with no or low levels of TiO and without a high C/O ratio. The lack of observable TiO features are possibly due to rainout, breakdown from stellar activity or the presence of other absorbers in the optical.

\end{abstract}


\section{Introduction}
\label{intro}

Transiting planets allow us to study exoplanet atmospheres through transmission spectroscopy, as the light from the host star is filtered through the planetary atmosphere. For hot Jupiters, theory predicts the presence of strong, pressure broadened alkali lines at temperatures below 1500~K (\citealt{seagersasselov00,brown01,fortney10,burrows10}). Observations so far have had mixed results, with HD~209458b showing some of these features (\citealt{charbonneau02,narita05,sing08,snellen08}) while the spectrum of HD~189733b is instead dominated by a slope of increasing planetary radius with decreasing wavelength, attributed to a scattering signature similar to that of Rayleigh scattering (\citealt{pont08,lecavelier08,sing11}), where all but the narrow cores of the lines are obscured (\citealt{redfield08,jensen11,huitson12}). The discrepancy between the two well-studied hot Jupiters indicates that clouds or hazes may be very important in planetary atmospheres. Both the Na~I and K~I features have only been conclusively observed together in one exoplanetary atmosphere, the hot Jupiter XO-2b \citep{sing11b,sing12}. The XO-2b observations suggest that only the narrow line cores of the Na~I doublet are visible, but it is not known whether the wings are hidden by clouds or haze or whether the flat spectrum observed in the line wings is due to condensation of Na out of the gas phase on the planet's night side. Although the global average theoretical terminator temperature-pressure ($T$-$P$) profile does not cross the condensation curve of Na into Na$_2$S, it is possible for temperatures to be low enough for this condensation to occur on the planet's night side, if atmospheric horizontal wind speeds are slow enough (e.g. \citealt{iro05,showman09}). Alternatively, ansite (NaAlSi$_3$O$_8$) could trap Na at even at hotter temperatures, and could explain the lack of Na features \citep{burrows00,gibson13}.

At hotter temperatures, large TiO bands could obscure the alkali lines, depending on atmospheric dynamics. TiO is potentially important because it could contribute towards strong thermal inversions (stratospheres) and the splitting of planetary atmospheres into two distinct classes \citep{hubeny03,burrows07,burrows08,fortney08}. \citet{spiegel09} show that the fraction of TiO in the upper atmosphere is very dependent on macroscopic mixing, and find that even for WASP-12b, (a hotter planet than WASP-19b) with no cold trap on the day side, an effective eddy diffusion coefficient of $K_{zz}=1.6 \times 10^7$~cm$^2$~s$^{-1}$ or higher is required to lift enough TiO into the upper atmosphere to cause a strong inversion. 
Additionally, condensation may occur on the night side. \citet{parmentier12} use a 3D general circulation model (GCM) to model the night side cold trap and find effective $K_{zz}$ values that best match their 3D model results. Comparing their effective $K_{zz}$ values to those required to maintain enough TiO in the upper atmosphere to cause a strong thermal inversion, they found that the night side cold trap will be able to deplete TiO efficiently if the particles can condense into sizes above a few microns. Alternatively, an atmosphere with a high C/O ratio will reduce the amount of TiO in the atmosphere \citep{madhusudhan11,madhusudhan12}. 

TiO has not been conclusively observed yet, even in HD~209458b which has been observed to have a stratosphere \citep{desert08,knutson08}. At an orbital distance of 0.0166~AU ($P=0.789$~days), WASP-19b is a very close, highly irradiated planet, and with a temperature of $\sim 2050$~K should be a good candidate to have TiO in the upper atmosphere. It is also a good candidate for transmission spectroscopy, having a low surface gravity, with $R_P=1.386$~$\mathrm{R_J}$ and $M_P=1.186$~$\mathrm{M_J}$ \citep{hebb11,hellier11,2011AJ....142..115D}.

However, the star WASP-19 is very active, with chromospheric Ca II H \& K line emission ratios of $\log (R'_\mathrm{{HK}}) = -4.660$, compared to -4.501 for the active star HD~189733b and -4.970 for the inactive star HD~209458b (see \citealt{noyes84}, \citealt{knutson10} and references therein). 
Stellar activity also may have a significant effect on the atmosphere. \citet{knutson10} showed a correlation between stellar activity and the lack of strong planetary thermal inversions, as measured by the 3.6~$\umu$m and 4.6~$\umu$m \textit{Spitzer} occultation depths. They suggested that the correlation might be caused by stellar activity breaking down the molecules responsible for strong thermal inversions. WASP-19b is a good test for this hypothesis. Secondary eclipse data suggest that WASP-19b does not have a strong thermal inversion, implying that TiO should be absent from the upper atmosphere if it is responsible for strong inversions \citep{anderson11b,madhusudhan12}. Transmission spectroscopy across the optical regime can detect TiO, Na~I and K~I features. Spectroscopic observations blueward of 4000 \AA\ can also detect Rayleigh-like signatures and potential features from sulphur-containing molecules, which can cause millibar-level inversions (e.g. \citealt{zahnle09}).

Observations in the near-infrared also constrain the planetary type. There are many molecules predicted to be observable in the near-infrared, including H$_2$O, CO, CO$_2$ and CH$_4$, depending on planetary temperatures and chemical composition \citep{brown01,sharpburrows07,fortney10,burrows10}. Observed molecules can constrain the C/O ratio, with H$_2$O and CO$_2$ becoming more abundant when C/O $< 1$ and CH$_4$, HCN and C$_2$H$_2$ enhanced if C/O $> 1$ \citep{madhusudhan12,moses13}. 



Water has been detected in the day-side spectrum of the extrasolar planet HD~189733b \citep{grillmair08}, but spectroscopic observations in transmission have proved inconclusive, for this and other planets. Water features were observed in HD~189733b and XO-1b using the Hubble Space Telescope (HST) Near Infrared Camera and Multi-Object Spectrometer (NICMOS) \citep{swain08,tinetti10}. Followup observations using the same instrument found results to be consistent with no water, or a high altitude haze obscuring water features \citep{sing09}. 
However, the most recent reductions (\citealt{gibson11,gibson12,crouzet12}) show that the results for both planets are very dependent on the de-correlation methods used to deal with systematic trends. Therefore the NICMOS observations do not have the precision to either confirm nor deny the presence of molecular near-infrared features, either in transmission or emission, when uncertainties due to de-correlation methods are taken into account. 

The HST Wide Field Camera 3 (WFC3) has improved on the precision available using NICMOS, and new observations using WFC3 by \citet{gibson12b} suggest that in HD~189733b, the haze does extend into the near-infrared, covering the predicted solar-abundance molecular features. The observations are not precise enough to rule out weaker features, but solar abundance features can be ruled out. By putting all the existing optical and infrared observations together, \citet{pont13} also find results consistent with either haze or weaker than predicted molecular features redward of 1.1~$\umu$m. 
Followup observations of XO-1b using WFC3 by \citet{deming13} show the presence of near-IR water features, but at 0.84 times the amplitude predicted by solar-abundance models and significantly different from the NICMOS results, again showing that spectroscopic data from NICMOS do not have the required precision to detect features in transmission reliably. The new results suggest that hazes or clouds contribute a significant opacity source to the upper atmospheric transmission spectrum of XO-1b.

The picture of HD~209458b is more complicated, since there is still a discrepancy between the observed transmission spectra in the near-IR. \citet{barman07} used the STIS data of \citet{knutson07} to show evidence of water features in the near-IR consistent with their chemical equilibrium models. However, \citet{deming13} find muted water features in HD~209458b with WFC3, 0.57 times the expected amplitude from solar abundance models. Both instruments have proven more reliable for transmission spectroscopy than NICMOS, suggesting the possible presence of variable cloud or haze cover in the upper atmosphere of HD~209458b.
It is unclear whether clouds or hazes should be dominant in the upper atmosphere of WASP-19b because, although photochemistry may have a role in producing high-altitude hazes, the temperature of the planet is too hot for many condensates to form, such as MgSiO$_3$, Mg$_2$SiO$_4$ and iron-based condensates \citep{fortney05}.

The questions of the prevalence of TiO features, alkali features, sulphur bearing molecular features, near-IR molecular features and hazes, clouds and dust have prompted us to begin an optical spectroscopic survey of eight hot Jupiters across different regimes of planetary temperature, with temperatures ranging from 1080-2800~K (GO-12473, P.I. D Sing). The aim of the programme is to classify the observed planets into clear/hazy atmospheres, and atmospheres with and without TiO, and other molecular absorption. In bright targets, we can also identify alkali line absorption. Where possible, we combine the optical observations with near-IR WFC3 observations. Having a large number of observed planets will allow us to measure correlations between atmospheric properties and other parameters such as stellar activity and stellar type. Here, we present some of the first results from this programme; the transmission spectrum of WASP-19b in the optical and near-infrared.

\section{Observations}

Three transits of WASP-19b were observed using the HST Space Telescope Imaging Spectrograph (STIS) for programme GO-12473 (P.I.,~D.~Sing), with two transits observed with the G430L grating (2900-5700~\AA) and one transit observed with the G750L grating (5300-10300~\AA). Details are shown in Table \ref{table_stis_visits}. 

\begin{table}
\centering
  \begin{tabular}{c | c | c }
\hline
Visit &  Instrument Setup & Date of Observation (JD) \\
\hline
3 & STIS G430L & 2456047.584236	- 2456047.826377 \\
4 & STIS G430L & 2456051.517847 - 2456051.769039\\
18 & STIS G750L & 2456057.041354 - 2456057.297512 \\
WFC3 & WFC3 G141 & 2455743.874347 - 2455744.105609 \\
\hline
\end{tabular}
\caption{Table showing dates of HST STIS visits (top three rows) and the archive WFC3 visit. This STIS visits are referred to by their visit numbers in following sections. As there is only one WFC3 visit, we do not refer to it by visit number.}
\label{table_stis_visits}
\end{table} 

Each transit contains 38 spectra, with the first orbit containing 8 spectra and the three subsequent orbits containing 10 spectra, giving 10 in transit and 28 out of transit exposures for each observation. Both gratings have a resolving power of $R=500$, resulting in a resolution of $\sim 8.6$~\AA\ at 4300~\AA\ and a resolution of 15.5~\AA\ at 7750~\AA\ (the centre of the G750L band). The scale is $ \sim 2.75$~\AA\ per pixel for G430L and $\sim 4.88$~\AA\ per pixel for G750L. The $1024 \times 128$ subarray was used to reduce overheads. Exposure times were 293~seconds with 21~second overheads. 

We also analyse near-infrared archival HST WASP-19b data. One transit of WASP-19b was observed using the HST WFC3 in spectroscopic mode with the G141 grism (programme GO-12181, P.I. D.~Deming). The dates of the WFC3 observation are also given in Table~\ref{table_stis_visits}. The observations used the $128 \times 128$ pixel subarray to reduce overheads. The dataset consists of 274 spectra, covering the wavelength range 1.087-1.687~$\umu$m with a scale of 0.00465~$\umu$m per pixel and a resolving power of $R=130$. This gives a spectral resolution $\sim 0.01$~$\umu$m at 1.4~$\umu$m. Each exposure contains a zero read, then 4 non-destructive reads, with the first after 0.1~seconds and the next three every 7 seconds after that. Each exposure has a 19~second overhead. There are 70 exposures taken during the transit event. There is also one useable orbit before transit, which contains 70 exposures, and one after transit, which contains 70 exposures. 

As with many past transit studies with HST, for both the STIS and WFC3 observations, we did not use the first orbit, since the HST thermally settles into its new pointing position during this time, and the systematics are considerably worse during this orbit. The data were bias-subtracted, dark-subtracted and flat-fielded using the \textsc{calstis} and \textsc{calwf3} pipelines. 

For WFC3, Both the \_ima and the \_flt files are corrected by the pipeline for non-linearity in the pixel response by fitting a 3rd order polynomial to the differences in counts per second in the non-destructive reads for each image. This is a potential problem for exposures with a small number of reads, since the fitting algorithm will be sensitive to noise. However, the values are fitted per CCD quadrant rather than per pixel, so any variation in the correction should be common-mode rather than wavelength-dependent. There is a possibility of there then being an offset in flux between the two halves of the spectrum if the spectrum is spread over two quadrants (as is the case for WASP-19b). However, the first and second reads are assumed to be in the linear regime, and hence are not corrected. We found that extracting the spectrum using the second read (with an exposure time of $\sim 7.3$~seconds and maximum count level of 10,560~DN (data number), well within the linear regime) is very similar to the spectrum extracted using the final read, which contains a linearity correction. This suggests that the pipeline up-the-ramp fitting did not introduce any significant wavelength-dependent errors.

The data are not corrected with wavelength-dependent flat-fields in the pipelines. This can be a problem if the spectrum shifts in the dispersion direction during the observation, meaning that a given pixel will not always be receiving the same wavelength in each exposure and as such its response may vary from exposure to exposure. However, by cross-correlating with the spectrum from the first exposure, we found that these shifts were sub-pixel over the whole transit for each observation. 

Cosmic rays were subtracted from the images using custom routines. Previous studies of the bright targets HD~209458b and HD~189733b with STIS used the whole timeseries per pixel to flag outliers which are significantly different in flux from the mean value for that pixel after correcting for the transit (e.g. \citealt{sing08,sing11,huitson12}). WASP-19 is much fainter than HD~189733b (12.6 vs 7.7 V magnitude), meaning that, for our STIS data, much longer exposures were required to obtain enough counts for each measurement. Longer exposure times meant that the number of exposures per orbit in the WASP-19b data were nearly 3.5 times lower than for the HD~189733b observations. A small number of exposures per visit meant that the cosmic ray removal routines used previously were insufficient, since remaining outliers in the light curves dominated the transit fitting. A new routine was developed which uses the difference image of one exposure compared to the next and compares this with a batch of exposures to detect transient events, with the drop in flux due to the transit also taken into account. Using difference images has the advantage of subtracting almost all of the stellar spectrum in the images that are analysed for cosmic ray events. The affected pixel values are then interpolated from the profiles of columns either side of the bad pixel. 

\begin{figure}
\centering
\includegraphics[trim=0cm 2cm 4cm 0cm, clip=true,width=5.5cm,angle=270]{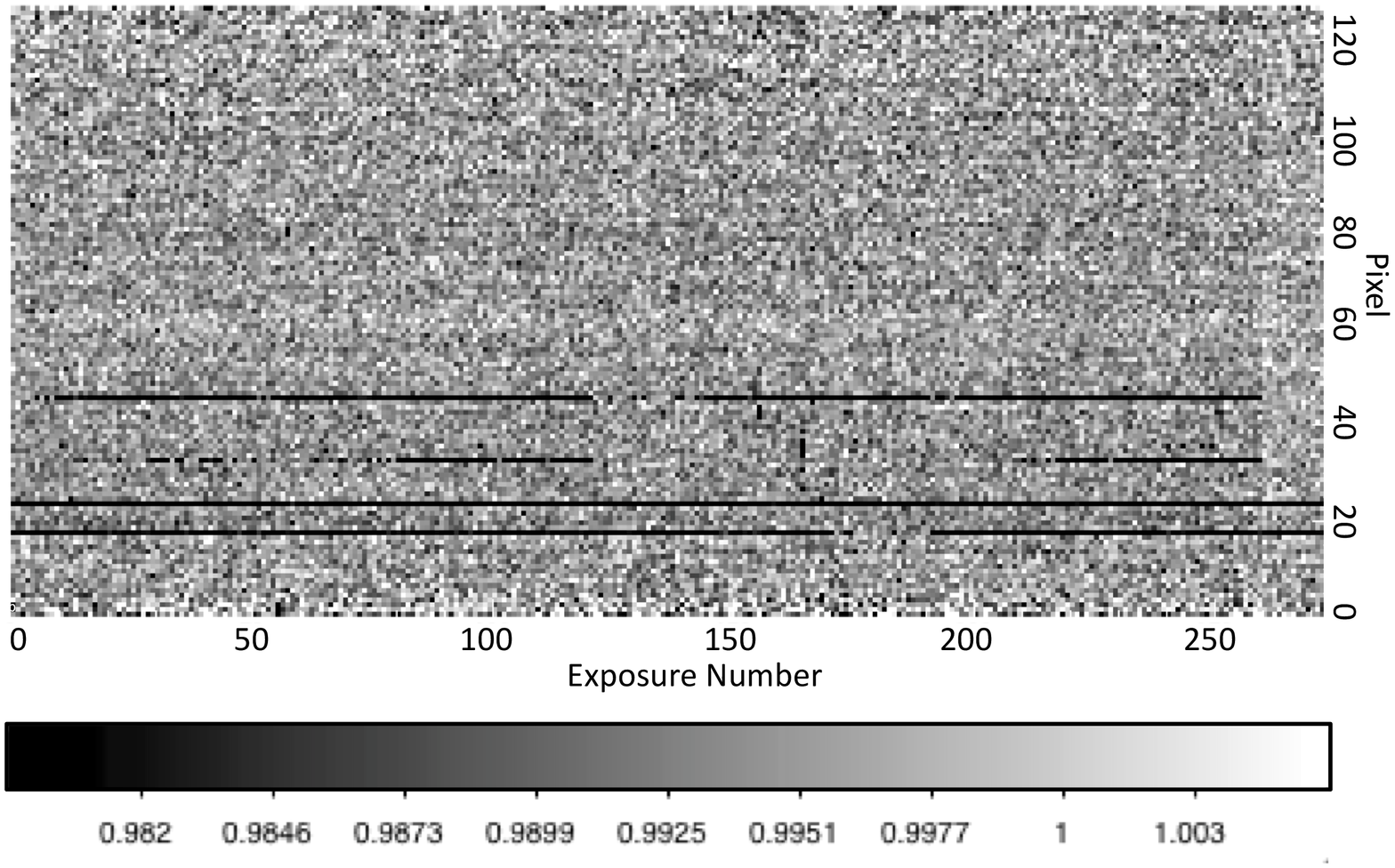}
\includegraphics[trim=0cm 2cm 5cm 0cm, clip=true,width=5.2cm,angle=270]{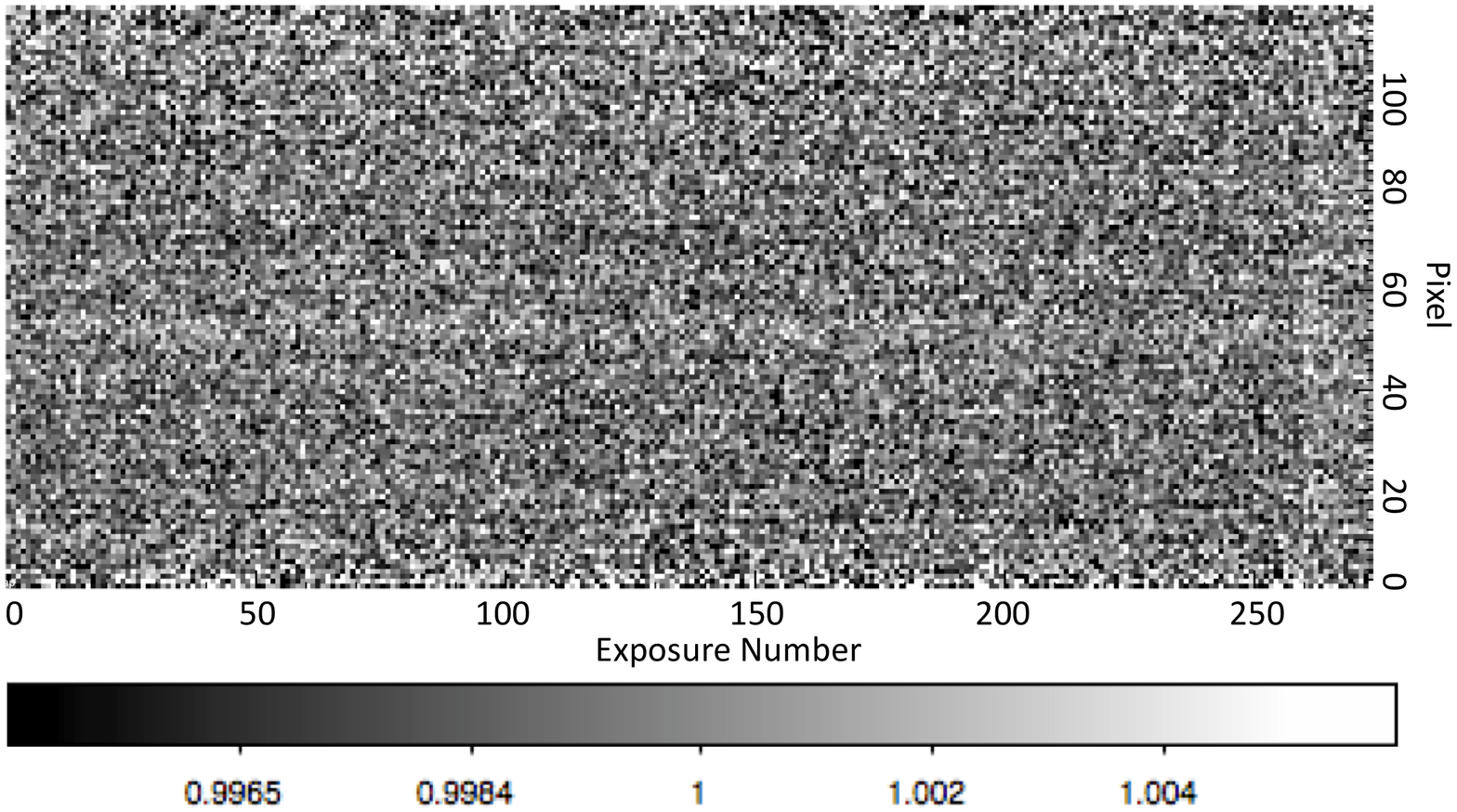}
\caption{Residual plot of each spectral element in the final read divided by the same spectral element in the previous read, in counts per second for each exposure, including all the pixels (top) and after removing the columns with saturated pixels (bottom). On the vertical axis, increasing pixel numbers correspond to increasing wavelength, with the shortest wavelength being 1.087~$\umu$m and the longest wavelength being 1.687~$\umu$m. The pixel scale is 46.5~\AA\ per pixel. The saturated regions are clearly seen as dark streaks corresponding to a low value of counts/second in the final read, where the pipeline has not corrected for the effects of nonlinearity. 
The increase in flux with increasing exposure number in the final few exposures is due to the ramp effect. Although visible by eye, the flux jump is beneath the white noise level when examined quantitatively. When using larger wavelength bins, the effect becomes more significant compared to photon noise, but is effectively removed by our de-trending procedures (section \ref{detrending_wfc3}), and in our spectral light curves other trends become more important. Furthermore, since the flux increase is time-dependent but not wavelength-dependent, relative transit depths as a function of wavelength (and hence spectral features) will not be affected.}
\label{turin_plot_linearity}
\end{figure}

Pixels flagged as bad by the HST pipeline were also interpolated over using the same procedure as for removing cosmic rays for STIS. For WFC3, we also interpolated over pixels which had been marked by the pipeline as ``unstable detector response" and ``unstable zero value". We discarded columns containing pixels flagged as ``saturated". Only the saturated pixels affected the spectrum by greater than 1~$\sigma$. The reason for this is that the pipeline does not correct pixels that it measures to deviate from linear response by more than 5~per~cent (the threshold for the ``saturation" flag), and so these pixels then become very different in flux from the rest of the pixel values. Since a given pixel reaching this threshold does not occur for every exposure, this produces an effect that is both time dependent and wavelength dependent. We therefore neglected the columns containing saturated pixels altogether if they were saturated in even one exposure. Figure~\ref{turin_plot_linearity} shows the variation in electrons/second between the final two reads before and after removing saturated columns. Four columns show clear evidence of saturated pixels, which show up as dark horizontal bands. Variations in electrons/second between the two final reads were at the photon noise level once the saturated pixels were removed. Note that the difference in value between the pixels flagged as saturated in the final read and the previous read is not at the 5~per~cent level (the difference between the second and final read is $\sim 3$~per~cent), suggesting that the pipeline flagged these pixels due to having only a small number of non-destructive reads per exposure and the fits being noisy. Since photon noise requires that we use large spectral bins for our final WFC3 spectrum, removing a small number of saturated pixels does not significantly affect our measurements. However, if we observed a brighter star and wished to use smaller spectral bins and hence all of the pixels, it may be worthwhile to re-run the pipeline with the up-the-ramp correction omitted, as long as count levels are sufficiently low.

After cosmic ray removal, the spectra were extracted using the \textsc{iraf apall} task. Transit light curves were produced by summing the photon flux over the full spectrum in each exposure. The spectral extraction was performed with and without background subtraction, using a range of background regions and different polynomial orders of background subtraction. The extraction was performed using a large range of aperture sizes. The best aperture size and background subtraction were selected from measuring white noise and the correlated noise in the out of transit exposures using the binning technique described in \citet{pont06} and \citet{winn08}. For WFC3, the levels of red noise are measured using $\sigma_N^2 = \sigma^2_w/N \left ( \frac{M}{M-1} \right ) + \sigma^2_r$, where $\sigma_N^2$ is the variance of binned fluxes in $M$ time bins of $N$ points, $\sigma_w$ is the white noise component (uncorrelated), and $\sigma_r$ characterises the red noise. For the WFC3 transit, the optimum aperture was determined for each non-destructive read. The optimum aperture size changes with different reads, as the source gets brighter on the CCD, the significance of the wings of the point spread function (PSF) increases and a wider extraction aperture is needed for later reads. For the low-cadence STIS observations, we used the out of transit standard deviation as a measure of noise.

For STIS, we used the \_flt files and found the optimum apertures for spectral extraction to be 13 pixels wide. For WFC3, we extracted both the \_flt and  the \_ima files, but used the \_ima files for the analysis since the out-of-transit standard deviation was found to be lower than for the \_flt files. The difference between the \_ima and the \_flt files is that the \_flt files combine all the reads in an exposure to give a mean value of electrons/second in each exposure. It is likely that the increased noise is due to the high photon noise in the first and second read being included in the average. 

\begin{figure}
\centering
\includegraphics[width=8cm]{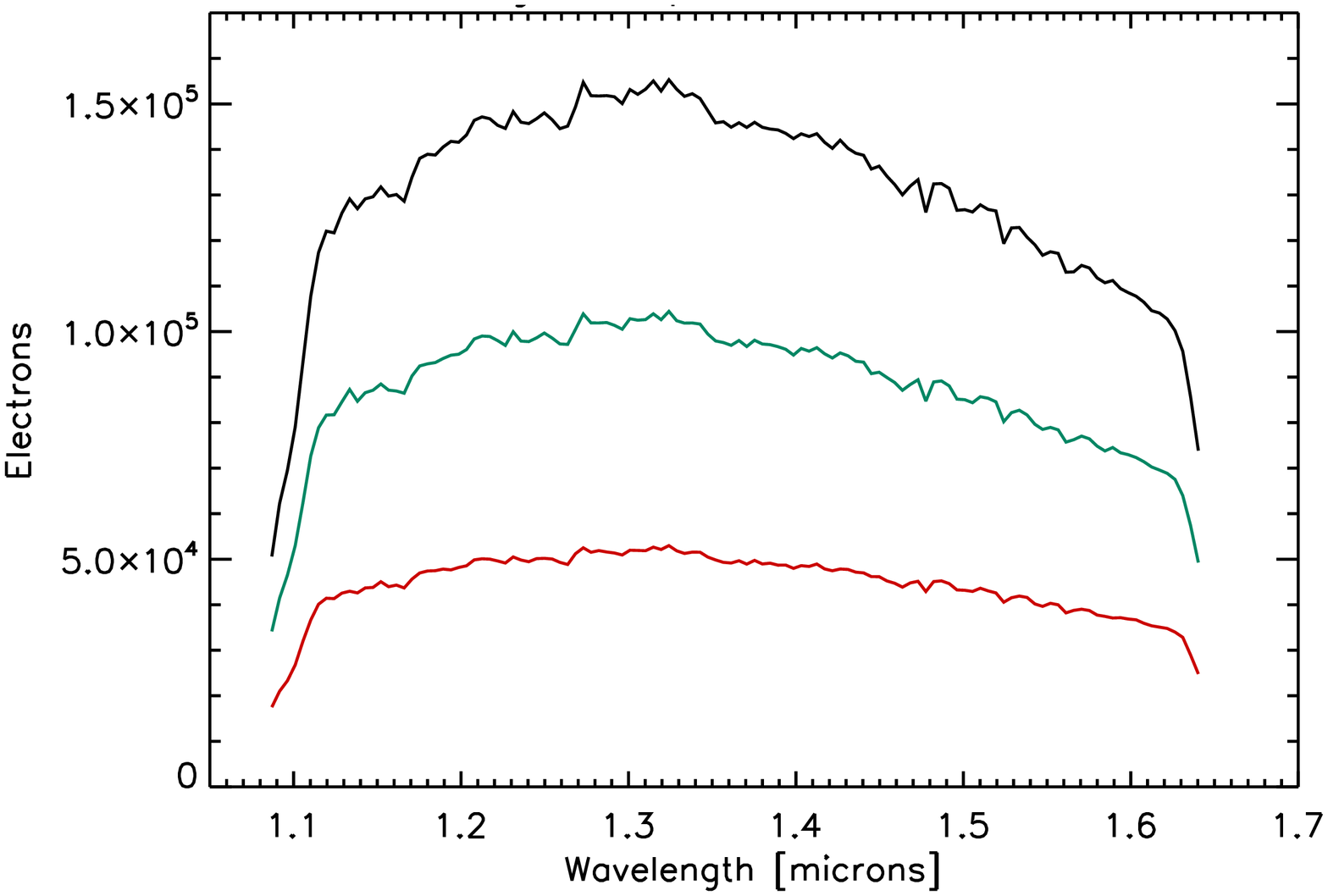}
\includegraphics[width=8cm]{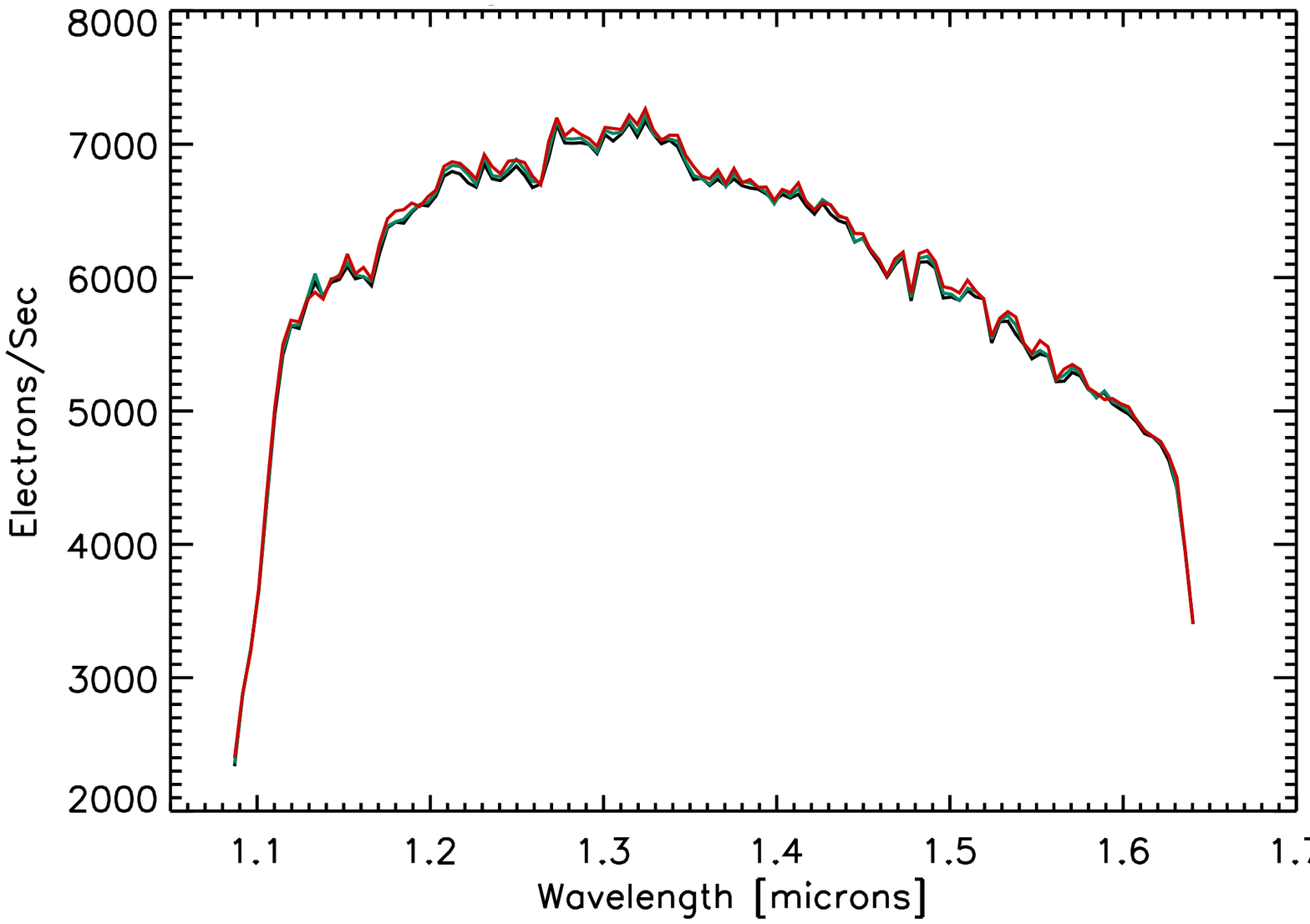}
\caption{The brightest spectrum extracted using each of the 3 last reads in counts (top) and counts per second (bottom). The different reads are shown in different colours. These are the spectra summed over the whole 13 pixel wide aperture. Photon noise error bars are not plotted for clarity, but the difference in counts per second between any two reads does not exceed the photon noise level. A colour version is available in the online journal.}
\label{fig_different_reads}
\end{figure}

According to the instrument manual\footnote[1]{http://www.stsci.edu/hst/wfc3/documents/handbooks/}, $> 5$ per cent nonlinearity (the threshold for the ``saturation'' flag) is observed with counts above 31,000~DN (or 78,000 electrons with a gain of 2.5), and we had saturation flags in 6 pixels of the spectrum. Also, nonlinearity begins at count levels of around 18,000~DN, well below this threshold, as shown in Instrument Science Report WFC3 2008-39 (Hilbert, 2009)\footnote[2]{www.stsci.edu/hst/wfc3/documents/ISRs/WFC3-2008-39.pdf}. Figure \ref{fig_different_reads} shows the brightest extracted spectrum in counts for each non-destructive read, except the first, and the same in counts per second, both after the linearity correction and after removing columns with pixels flagged as saturated. The variations in counts per second between one read and another are at the photon noise level. For the final read of each exposure, the best spectral extraction aperture was found to be 26 pixels wide. 


We used the extracted \_x1d spectra from the \textsc{calstis} pipeline for the STIS wavelength solution. The wavelength solution for the WFC3 spectrum has to be calibrated from the direct image taken at the beginning of the observation. This means measuring the $x$ and $y$ position of the target, and then locating what wavelength this corresponds to. This information of pixel location vs. $\lambda_o$ was taken from the Instrument Science Report WFC3-2009-17 (Kuntschner et al. 2009)\footnote[3]{www.stsci.edu/hst/wfc3/documents/ISRs/WFC3-2009-17.pdf}, which notes the different starting wavelengths for different pixel locations. 


The direct image provides the location of the zeroth order, and the observer can measure the number of pixels between the source in the direct image and the start of the spectrum on the CCD in the spectral images to get the starting wavelength for the spectrum. The reference pixel from the direct image is at $\lambda_o = 0.9005$~$\umu$m, and the start of the spectrum is 40 pixels higher in the $x$ (dispersion) direction. Using the scale of 0.00465~$\umu$m per pixel, this means that the spectrum begins at 1.0865~$\umu$m. For the 128 subarray, the image is sometimes moved between the direct image and the spectrum to ensure that as much of the first order spectrum as possible lands on the subarray. We checked that this was not done using the reference pixel information in the header. 

\section{STIS Analysis}

\subsection{Unocculted Spot Correction}

\begin{figure}
\centering
\includegraphics[width=8cm]{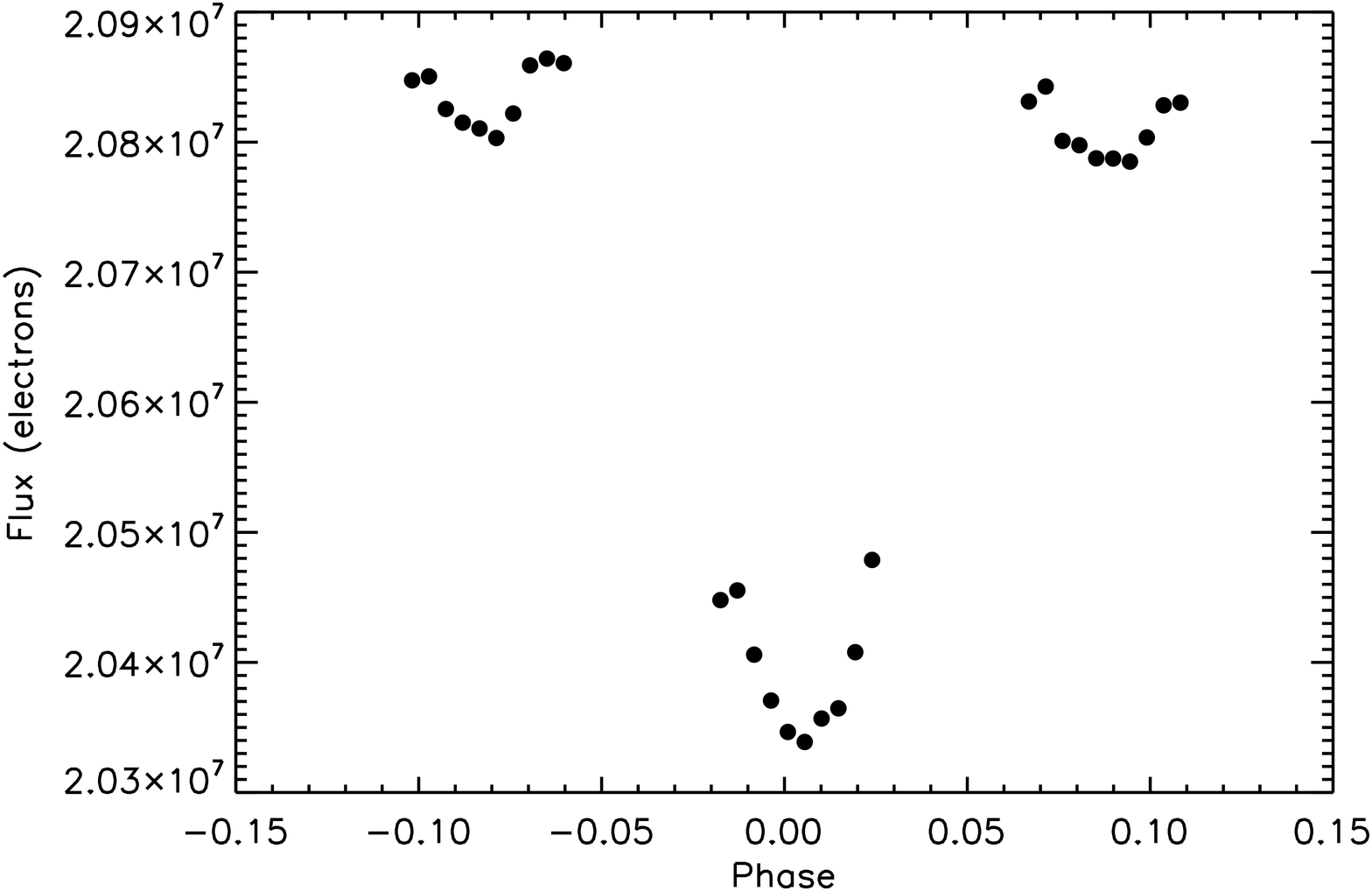}
\includegraphics[width=8cm]{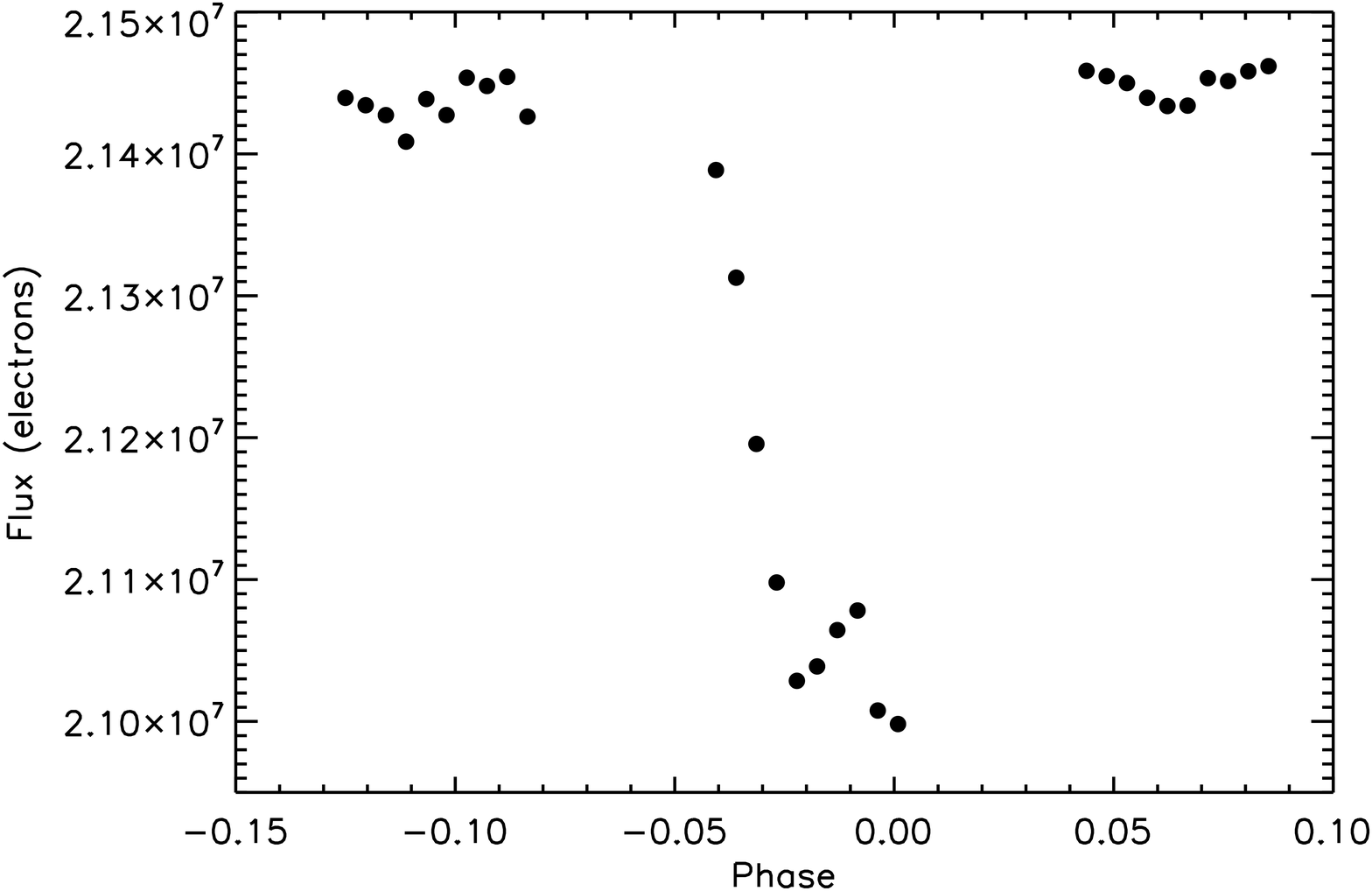}
\includegraphics[width=8cm]{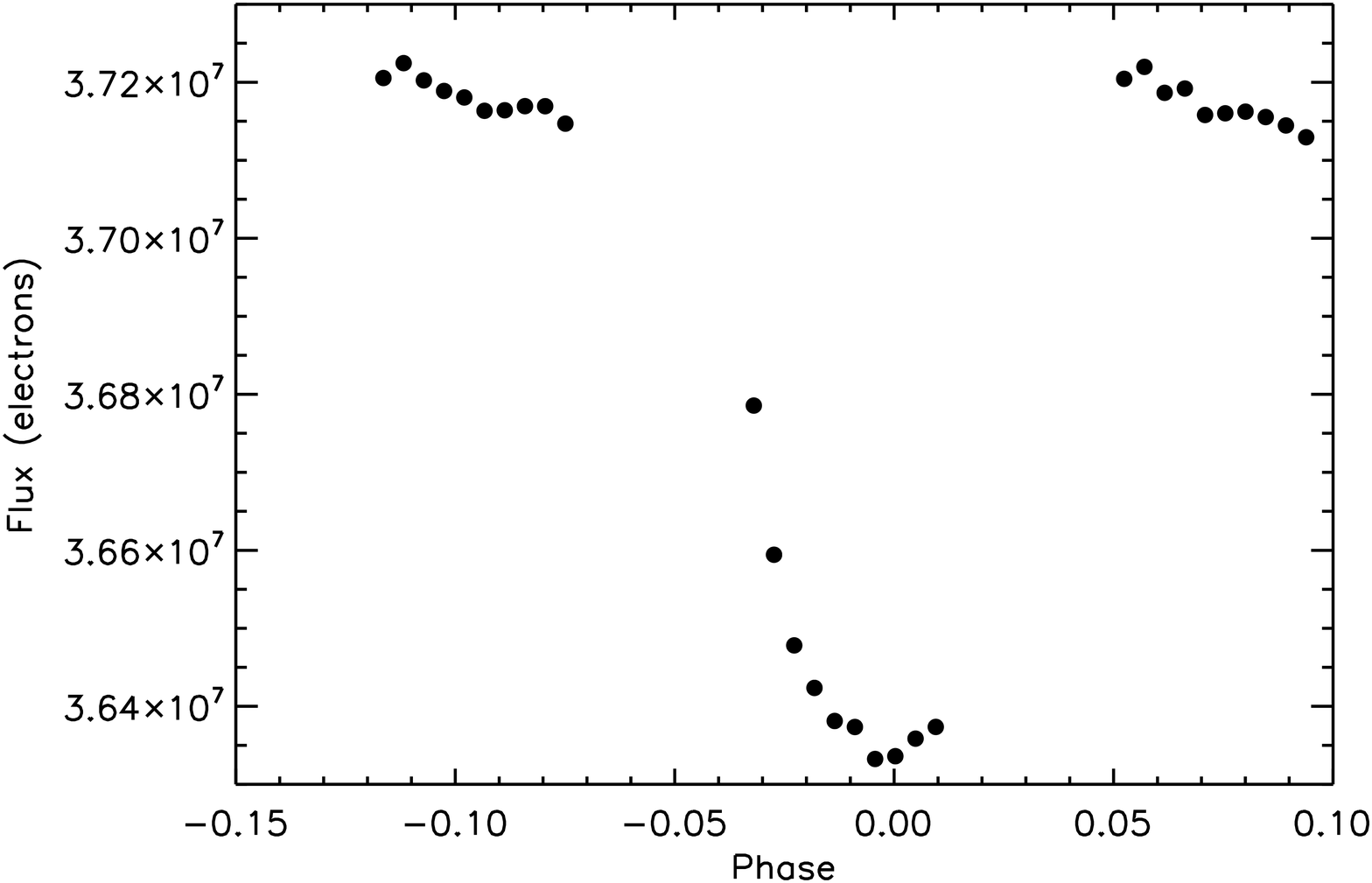}
\caption{STIS light curves before any correction for systematics, and before correcting for unocculted star spots. From top to bottom: visit 3, visit 4, visit 18.}
\label{fig_raw_stis_lc}
\end{figure}

Figure \ref{fig_raw_stis_lc} shows the STIS white light curves for each visit (the flux integrated over all wavelengths for each exposure). Systematic trends over an orbital timescale due to PSF variations caused by the thermal expansion and contraction of the HST optical assembly (``breathing" effect) can clearly be seen, which are dealt with in Section \ref{sec_detrending_stis}. The transits firstly need to be corrected for stellar activity as WASP-19 is known to be a very active star, and the stellar flux varies significantly over time due to this activity.

The flux variations over time occur because the number of star spots on the surface of the star changes with time, and also, as the star rotates, spots come into and out of view of the observer. Both of these effects cause the stellar flux to fluctuate quasi-periodically. Dark star spots on the stellar surface cause the overall flux from the star to be dimmer, and hence the observed planetary transit appears deeper if the planet is crossing a non-spotted stellar region, leading to an overestimation of the planetary radius. The effect on the measured planetary radii is wavelength dependent because the stellar spots have a different temperature and spectrum than the non-spotted stellar regions, and this may affect the transmission spectrum if the number of spots on the stellar surface is large. The effect of unocculted stellar spots needs to be accounted for before the planetary radii are fitted. Previous studies have done the correction after obtaining the spectrum, but doing the correction beforehand slightly changes the shape of the limb-darkened transit, since the limb darkening models assume a non-spotted star at the temperature of the non-spotted surface. We neglect the limb darkening of the spots, which should be negligible at our level of precision, and also do not consider the effects bright active regions on the star. The extensive data on the HD~189733b system shows no significant effects of bright active stellar regions \citep{pont13}.

\subsubsection{Variability Monitoring}

WASP-19 was monitored from the ground using the Cerro Tololo Inter-American Observatory (CTIO) 1.3~m telescope with the V filter (5400~\AA\ with FWHM $\sim 1200$~\AA) for 135 days, giving a light curve of flux over time with one exposure approximately every 4 days, although with increased cadence around the times of the STIS visits. We then obtained a second season of monitoring which covered a further 120 days, to better estimate the maximum stellar flux level.
All three of the HST STIS transit observations took place during the first season of this monitoring. Since there are CTIO photometric data points very close to our transit observations, we simply used these photometric data points themselves to give the flux dimming values for the times of our transit observations, using the rest of the light curve to normalise these points. 

The whole of the monitoring data was used to normalise the light curve after data points that were taken during transit events were removed. The flux values measured during the stellar monitoring need to be relative to the non-spotted reference level in order to be used to correct spectroscopic data. Assuming that there are always some spots on the stellar surface, the maximum flux observed during monitoring does not correspond to the non-spotted surface. \citet{aigrain12} found that the non-spotted flux can be estimated as $F_\star = \mathrm{max}(F) + k \sigma$,  where $F$ is the variability monitoring light curve, $k$ is a fitted value, and $\sigma$ is the scatter of the light curve. They also found that $k=1$ is a good value to use for active stars, based on data for HD~189733 and simulated test cases. We do not have enough data to further constrain the factor, so we fixed $k=1$. To determine percentage dimming, we normalised the photometric measurements to $F_{\mathrm{norm}}= F / F_\star$. It should be pointed out that the number of starspots on the stellar surface can change over long timescales. Since the amount of dimming is obtained by normalising the monitoring light curve using its maximum flux value and scatter, observing over multiple seasons increases the accuracy of the spot corrections. Upon obtaining the second $\sim 120$ day season of variability monitoring for WASP-19b, re-analysis of the CTIO data caused the flux levels measured for the first season to change by $\sim 0.6-0.9$~per~cent. Although this did not affect the relative radii determined (and hence spectral features) from the transits that occurred during the variability monitoring, their absolute level was affected by $\sim 0.0004$~$R_P/R_\star$ for wavelengths around the CTIO band. Although small, such an absolute correction could be larger if the stellar flux were observed for several years. Additionally, the effect will be larger in shorter wavelength bands. In comparing multiple datasets, variability monitoring is most useful if it is continuous throughout all transit observations, allowing the radii to be compared relative to one another on the same scale. Furthermore, the shift in normalised flux level obtained when using the second monitoring season illustrates that longer-term monitoring is useful to accurately determine the non-spotted stellar flux level, $F_\star$, for example as done for HD~189733b \citep{pont13}.

\begin{figure*}
\centering
\includegraphics[trim=0.5cm 11cm 0cm 10.5cm, clip=true,width=16.3cm]{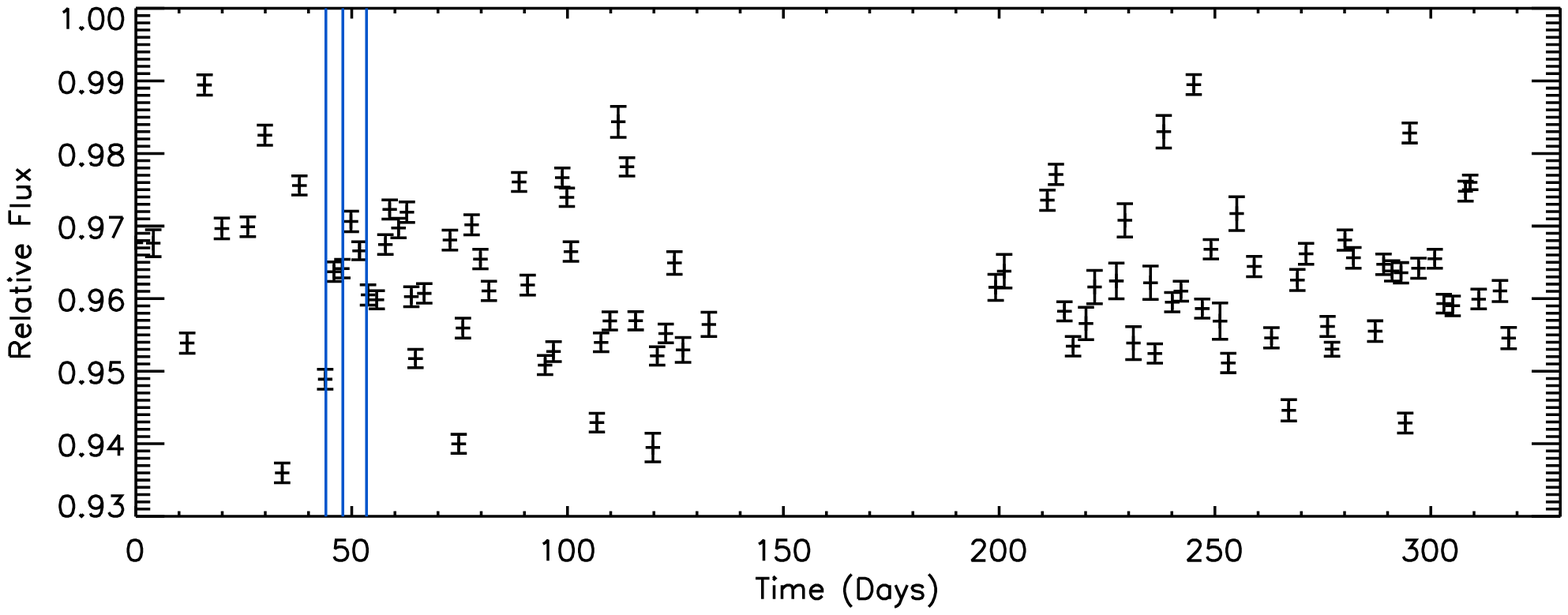}
\includegraphics[trim=2cm 13.2cm 2cm 4cm, clip=true,width=8cm]{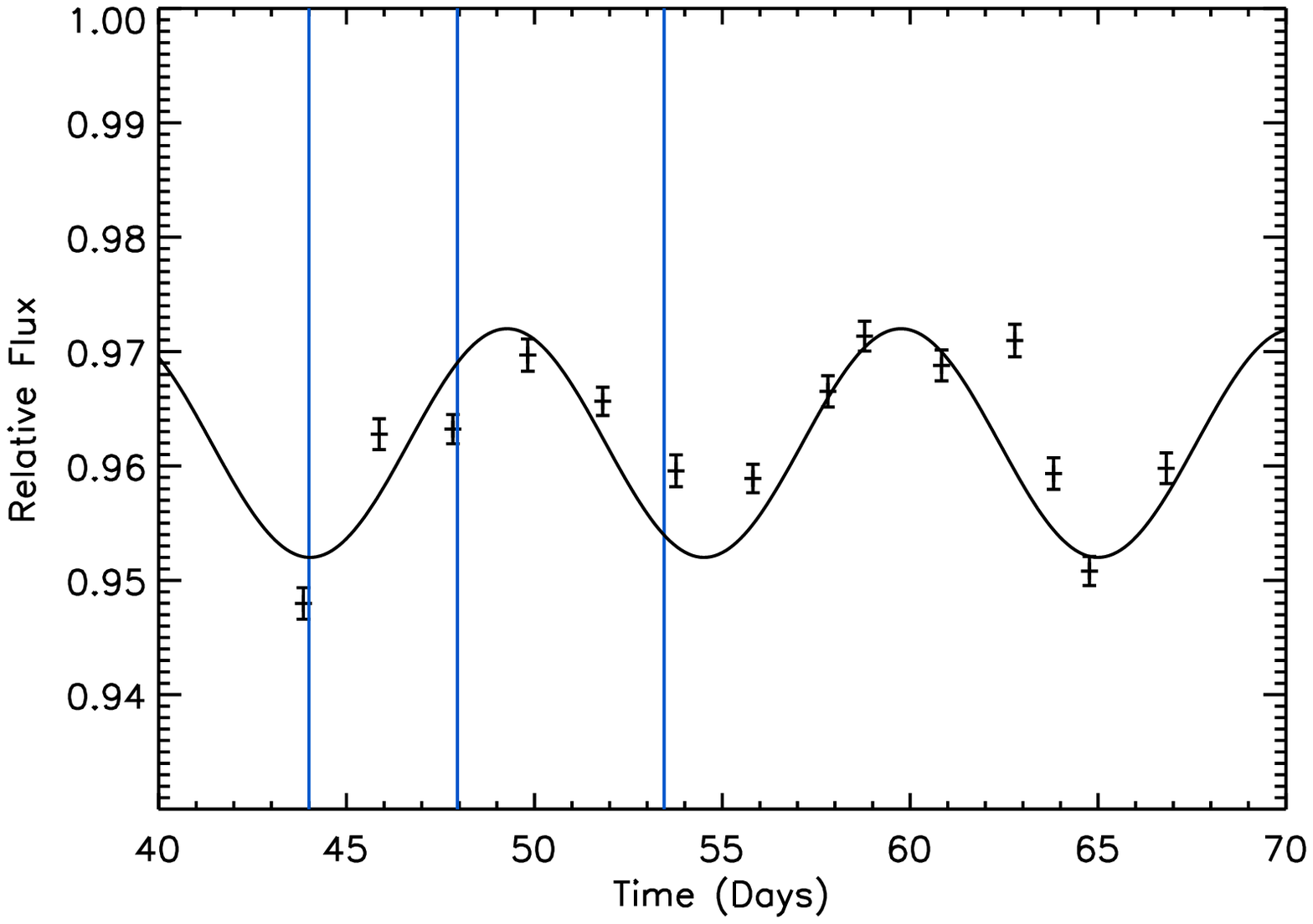}
\includegraphics[trim=0cm 0cm 0cm 0cm, clip=true,width=8cm]{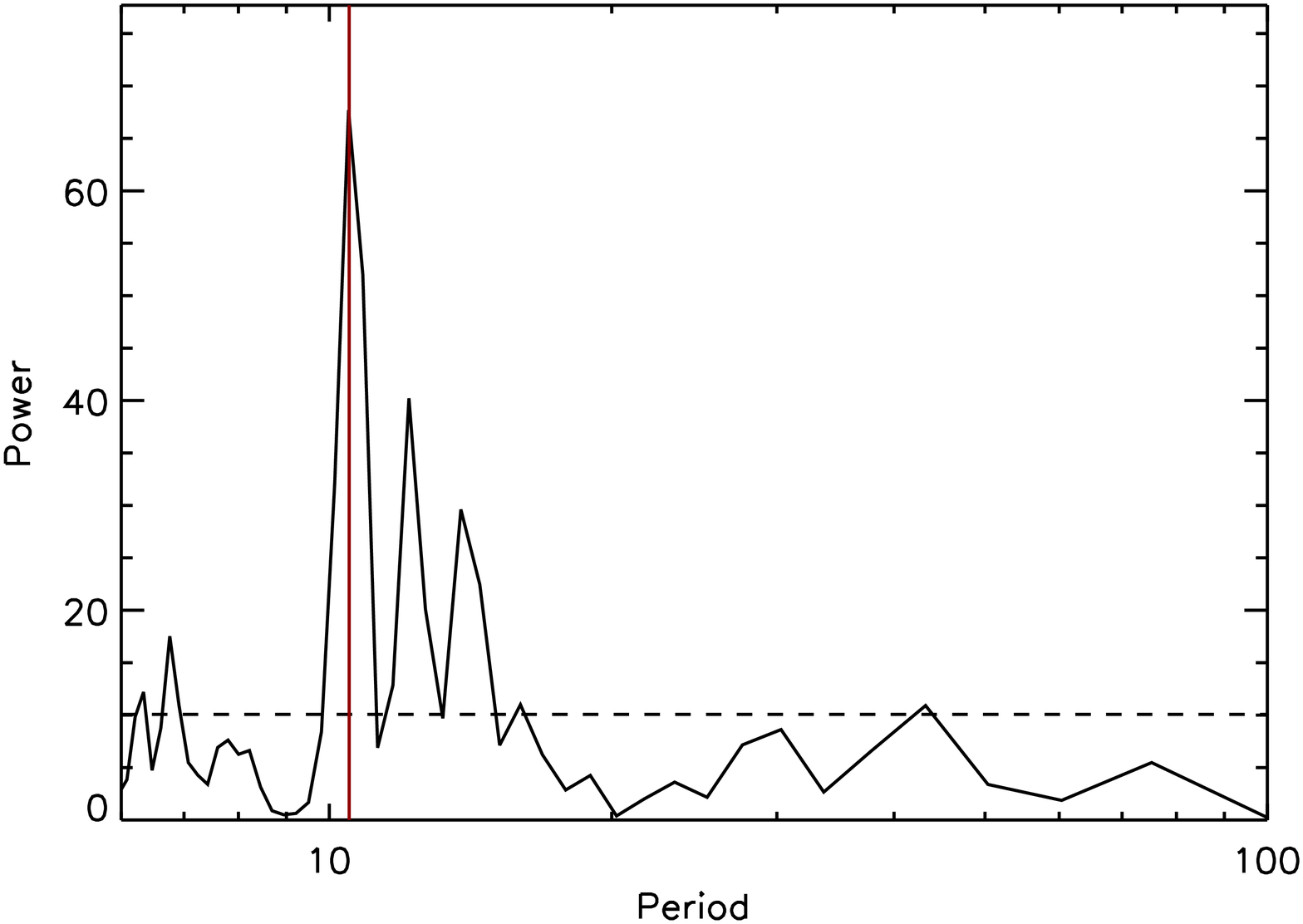}
\caption{\textit{Top:} CTIO variability monitoring data for both seasons, with all data points included, even those during transit events. The data are normalised to the assumed non-spotted stellar flux. Vertical (blue) lines show the starts and ends of the three STIS transits. The transits occur at days 44.0, 47.95 and 53.45, with a duration of $\sim 1.5$~hours. The Julian date of the first CTIO observation is 2456003.70253. \textit{Bottom Left:} Zoom-in on the variability monitoring data around the STIS transits, with a sine wave overlpotted with period of 10.5 days. The sine wave model is more simplistic than reality and it can be seen that the real photometry, and hence spot coverage, is more complex. Additionally, in the top plot, the amplitude clearly changes over periods longer than 10.5~days, most likely due to spot evolution. \textit{Bottom Right: }Lomb-Scargle periodogram showing dominant periodicities in the variability monitoring data. A vertical (red) line shows the stellar rotation period of 10.5 days determined in the literature (\citealt{hebb11,abe13}), which matches well with the strongest periodicity observed in the CTIO data (10.48 days). Two other strong periodicities are seen, less than 4 days from the main peak. There are not enough data to detect longer period modulations at higher than 3~$\sigma$ confidence level (the dotted line shows the 99.8~per~cent confidence level). The Lomb-Scargle periodogram made use of the IDL procedure \textsc{scargle.pro} written by J. Wilms.}
\label{fig_variability}
\end{figure*}

Table \ref{table_variability_monitoring} gives the dimming values corresponding to each of the STIS visits and Figure~\ref{fig_variability} shows the CTIO data along with a Lomb-Scargle periodogram showing a clear peak at 10.48 days, very close to the expected 10.5 day stellar rotation period (\citealt{hebb11,abe13}). 
The other significant peaks in the periodogram are likely aliases of the main peak, and are less than 2 days different from the main peak. With data sampled only every 4 days, we cannot draw any conclusions from these periodicities about possible spot configurations.

\begin{table}
\centering
  \begin{tabular}{c | c | c }
\hline
Visit &  $F_{\mathrm{norm}}$ & Error \\
\hline
3 & 0.949 & 0.0014 \\
4 & 0.964 & 0.0013 \\
18 & 0.961 & 0.0014 \\
\hline
\end{tabular}
\caption{Table showing values for the stellar fluxes as a fraction of the non-spotted flux, $F_\star$, for each STIS visit.}
\label{table_variability_monitoring}
\end{table} 

\subsubsection{Correcting the Light Curves for Unocculted Stellar Spots}

We define the flux dimming correction at the variability monitoring wavelength as $\Delta f_o = 1 - F_{\mathrm{norm}}$. So, for example, if the star is at 99~per cent brightness of the non-spotted flux, this is a 1~per cent dimming, or $\Delta f_o = 0.01$. The flux dimming at the CTIO wavelength then has to be extrapolated to other wavelengths, since the brightness contrast of spots compared to the non-spotted surface will be more severe in the blue and less severe in the red because the spots are cooler than the non-spotted surface. This wavelength-dependent correction factor is derived by \citet{sing11} as 

\begin{equation}
f(\lambda, T) = \left ( 1 - \frac{F_{\lambda, T_{\mathrm{spot}}}}{F_{\lambda, T_{\mathrm{star}}}} \right ) / 
\left ( 1 - \frac{F_{\lambda_o, T_{\mathrm{spot}}}}{F_{\lambda_o, T_{\mathrm{star}}}} \right ) ,
\label{spotscalewavelength}
\end{equation}
where $F_{\lambda, T_{\mathrm{spot}}}$ is the stellar model flux with a temperature $T_{\mathrm{spot}}$ and at the wavelength of the transit observations, $F_{\lambda, T_{\mathrm{star}}}$ is the stellar model flux with a temperature $T_{\mathrm{star}}$ and at the wavelength of the transit observations, $F_{\lambda_o, T_{\mathrm{spot}}}$ is the stellar model flux with a temperature $T_{\mathrm{spot}}$ and at the variability monitoring reference wavelength ($\lambda_o$), and $F_{\lambda_o, T_{\mathrm{star}}}$ is the stellar model flux with a temperature $T_{\mathrm{star}}$ and at the variability monitoring reference wavelength. $T_{\mathrm{spot}}$ and $T_{\mathrm{star}}$ are the temperatures of the spotted and non-spotted stellar surface respectively.

Therefore, the flux dimming correction at each wavelength becomes $\Delta f = \Delta f_o * f(\lambda, T)$, for a given $T$. Errors in $\Delta f_o$ do not significantly affect the wavelength-dependent spectrum, but rather the absolute baseline radius of the spectrum. For the wavelength-dependent correction, we set $T_{\mathrm{spot}}$ to be 5000~K, which seems to fit the occulted spot amplitude as a function of wavelength in the G430L data. The occulted spot temperatures were determined by splitting the G430L data into 5 bands and assuming that the occulted spot has a constant shape in each band and varies only in amplitude. We fit for this amplitude in each band along with $R_P/R_\star$. However, the spot amplitude is very degenerate with fitting for the planetary radius and HST systematic trends, and so we discuss differences in our transmission spectra that could result from assuming different temperature spots in Section~\ref{sec_discussion}. The fits to the G430L data are discussed in Section~\ref{sec_g430L_spec}. The factor $f(\lambda,T)$ varies from $\sim 1.2$ over the mean of the G430L band to $\sim 0.82$ over the mean of the G750L band for $T_{\mathrm{spot}} = 5000$~K. The flux dimming $\Delta f$ is then applied to each of the transits at each band, and then the radii are fitted. It is applied using the following derivation. 

The observed flux of the star (when not being crossed by the planet) is

\begin{equation}
F_{\mathrm{obs}} = (1 - \Delta f) * F_\star \Rightarrow F_\star = \frac{F_{\mathrm{obs}}}{(1 - \Delta f) } , 
\end{equation}
where $F_\star$ is the true brightness of the star if it were unspotted. We need to know $F_\star - F_{\mathrm{obs}}$, which is the flux dimming, so that we can add this value onto each exposure in the transit light curve. The flux dimming can be written as:

\begin{equation}
F_\star - F_{\mathrm{obs}} = F_\star - (1 - \Delta f) F_\star = \Delta f \times F_\star. 
\end{equation}
We do not know $F_\star$ explicitly, and so we write the dimming as a function of $F_{\mathrm{obs}}$ only:
 
 \begin{equation}
\Delta f \times F_\star =  \frac{\Delta f}{(1- \Delta f)} F_{\mathrm{obs}}
 \end{equation}
 Finally, the correction was added to the transit in the following way:
 
 \begin{equation}
 y_{\mathrm{corrected}} = y + \frac{\Delta f}{(1- \Delta f)}  \overline{y[\mathrm{OOT}]} ,
 \label{eqn_lc_spotcorr}
 \end{equation}
where $y_{\mathrm{corrected}}$ is the corrected light curve, $y$ is the original, uncorrected, light curve and $\overline{y[\mathrm{OOT}]}$ is the mean of the out of transit exposures and should be equivalent to $F_{\mathrm{obs}}$.
 
This model was tested on a non-limb-darkened transit. A transit was produced using the analytical models of Mandel \& Agol (2002) with all the limb darkening coefficients set to zero, and a fixed radius. The change in depth, $\Delta d / d$ should equal the flux dimming, $\Delta f$. We applied the unocculted spot correction to the model transit, and then used the Levenberg-Marquardt (L-M) least-squares technique, using the \textsc{idl mpfit} package (Markwardt 2009) to fit for the new transit radius. We tried several different values for $\Delta f$ and in each case, the new fitted depth, $(R_P/R_\star)^2$, was a factor $\Delta f$ smaller (e.g. $\Delta d / d$ was 1~per~cent smaller for an inputted $\Delta f$ of 1~per~cent).

\subsection{De-Trending the STIS White Light Curves}
\label{sec_detrending_stis}

\begin{figure}
\centering
\includegraphics[width=8cm,trim=-0.9cm -0.0cm 0.0cm 0.0cm]{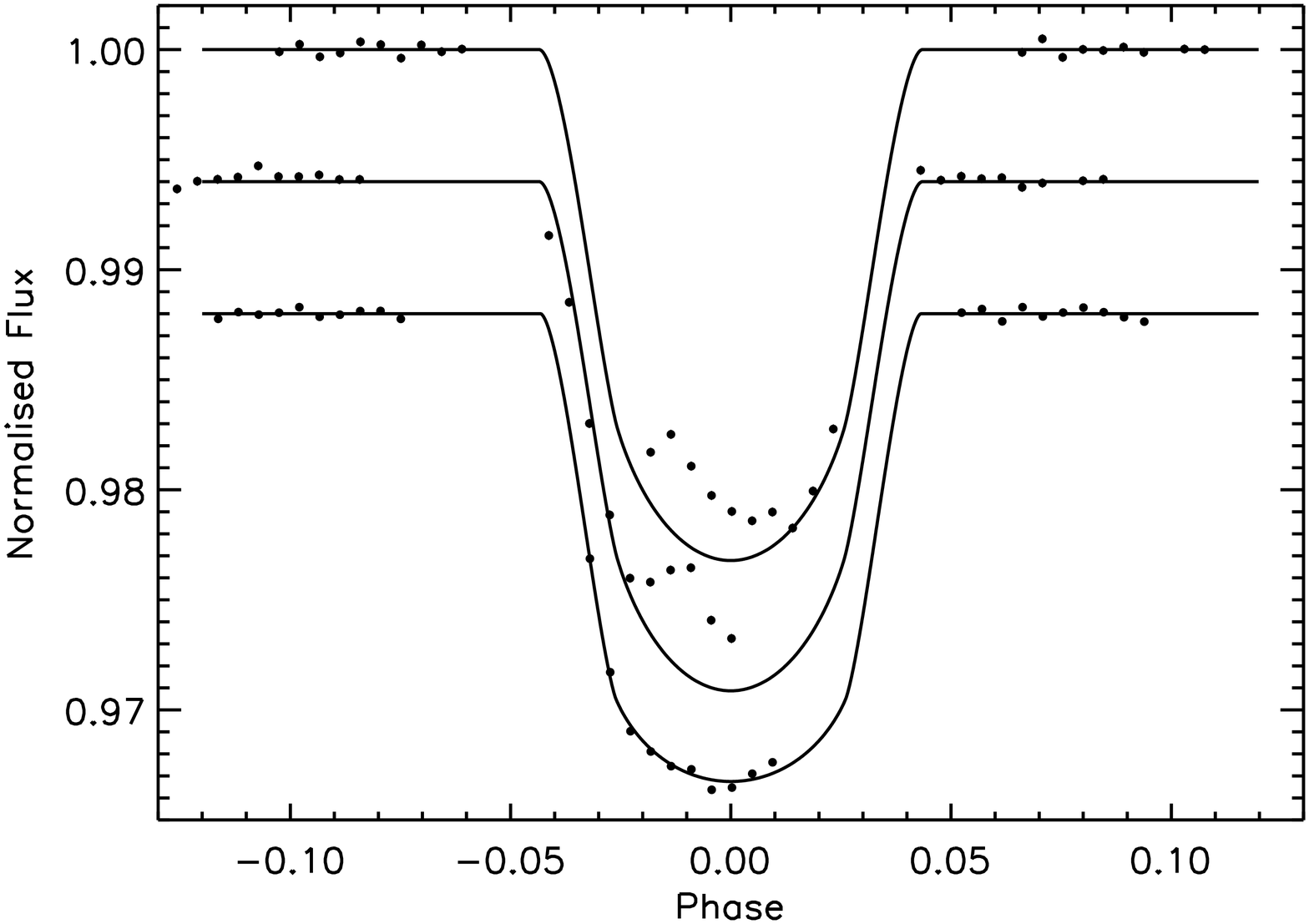}
\includegraphics[width=8cm]{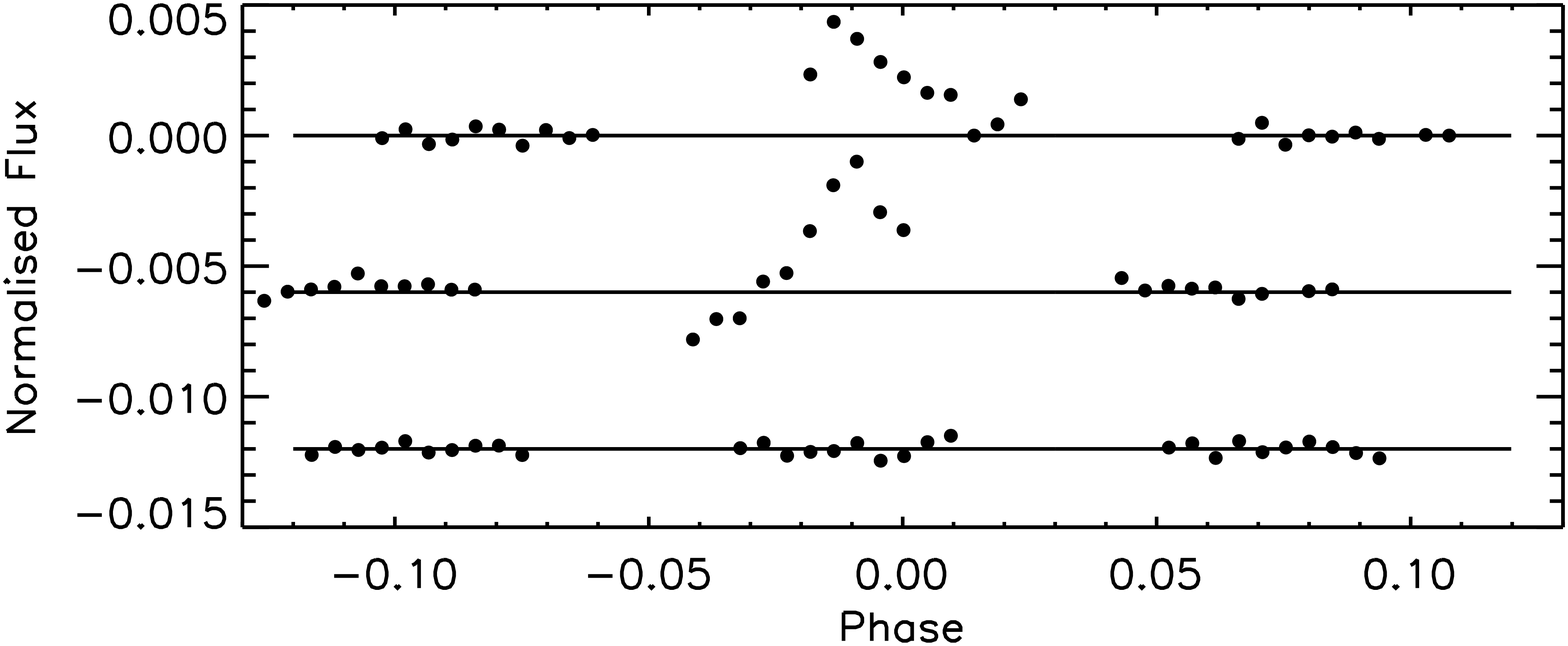}
\caption{STIS white light curves after de-trending, overplotted with the transit models of \citet{mandelagol02}. From top to bottom: visit 3, visit 4, visit 18. The two blue G430L light curves clearly show stellar spot crossings. Residuals are shown underneath in the same vertical order. The light curves and residuals have arbitrary flux offsets for clarity.}
\label{stis_white_detrended}
\end{figure}

The raw white light curves show the instrument systematics first described by \citet{brown01b}. The main systematic is the heating and cooling of the HST that occurs during an orbit, and results in telescope `breathing'. The resulting expansion and contraction of the telescope caused by changes in temperature result in changes in PSF and central position of the spectrum. The breathing effect was accounted for by fitting a 4$^{\mathrm{th}}$ order polynomial dependence of the fluxes on HST-phase simultaneously with transit depth. The fit also included a linear slope with time. We scheduled the first exposure of each orbit to be very short (1 second) and discarded this as well as the first orbit, due to problems with the first exposure of each orbit in the past (e.g. \citealt{sing11}). We used the L-M technique and the \textsc{idl mpfit} package, using the unbinned data to perform limb-darkened transit fits. We also found that fitting for linear dependences of the fluxes on $x$ and $y$ offsets of the spectra on the detector improved the fit, and improved the Bayesian Information Criterion (BIC), defined as $\chi^2+k \ln n$, where $k$ is the number of free parameters and $n$ is the number of data points \citep{schwarz78}. Higher order corrections with $x$ and $y$ offsets did not improve the BIC. Figure~\ref{stis_white_detrended} shows the de-trended light curves and the residuals overplotted with the transit models of \citet{mandelagol02}. We fixed the central transit time, orbital inclination and $a/R_\star$ to the parameters measured using the much higher cadence WFC3 data, which match well with literature values (see Section~\ref{wfc3_analysis}). The orbital parameters determined from the STIS data also match the literature values but have much larger error bars, most likely due to the lower cadence during the observations compared to the WFC3 dataset.

The two blue transits clearly show the occultation of starspots as `bumps' in the light curves where the flux in transit is higher than the non-spotted model predicts. Spots are darker than the surrounding stellar surface, and so the planet blocks less light when passing in front of a star spot than when passing in front of the non-spotted surface. The effect of occulted spots is therefore the opposite of unocculted spots. When performing the fits for the systematics, we did not include the points obviously crossing star spots. Star spot occultations can still clearly be seen in the residuals. 

For visit 4, the middle visit in Figure~\ref{stis_white_detrended}, the residuals drop below the predicted transit model during the start of ingress. Since the slope of data in ingress is different from the transit model, allowing the central transit time to be free did not allow this region to be fit better. Only a change in inclination or $a/R_\star$ can allow this region to be fit well, but such a fit deviates significantly from the transit parameters measured using the WFC3 data and literature values (see Section~\ref{wfc3_analysis}). We conclude that this shape is most likely due to residual systematic trends. Although there is a possibility of the planet crossing a bright region on the star, such crossings have not been observed with the more active system HD~189733b \citep{pont13} and we find in Section~\ref{sec_g430L_spec} that for 1000~\AA\ spectral bins, these deviations from the expected transit model are not correlated with wavelength as would be expected for a bright spot crossing.

We used the the Kurucz (1993) 1D ATLAS stellar atmospheric models\footnote[4]{See http://kurucz.harvard.edu/stars.html} and a 3-parameter limb darkening law of the form 

\begin{equation}
\frac{I(\mu)}{I(1)}=1-c_{2}(1-\mu)-c_{3}(1-\mu^{3/2})-c_{4}(1-\mu^2)
\end{equation}
as described in \citet{sing10}. Here, $\mu = \cos(\theta)$, and $\theta$ is the angle in radial direction from the disc centre. We used the closest match to the known information about WASP-19, with $T_{\mathrm{eff}}=5500$~K, $\log g$[cgs]$=4.5$ and [M/H] = 0. We then fixed the limb darkening coefficients to the model values when fitting for planetary radii. The white light $R_P/R_\star$ value for the G750L light curve is measured to be $0.1402 \pm 0.00053$. For this light curve, the residuals have a standard deviation of $2.4 \times 10^{-4}$ and the best fit gives $\chi^2_\nu = 2.2$ and BIC=74 assuming only photon noise uncertainties on the data points. Due to having no nonspotted points in the transits apart from ingress and egress, we do not attempt to determine absolute radii for the two blue transits.

We do not know the variation in stellar brightness due to activity between our observations and previous observations. However, Table~\ref{table_litradii} shows that our measured G750L white light radius is in agreement with the $R$ band literature value of \citet{hellier11}. Our measured radius ratio is larger than that measured by  \citet{2011AJ....142..115D} but smaller than those measured by \citet{2012arXiv1211.0864T} and that measured in \textit{z} band by \citet{hebb11}. The lightcurves of \citet{2012arXiv1211.0864T} were fitted with a transit model that includes occulted spots, but were not corrected for un-occulted spots due to there being no variability monitoring data available. All other literature radii are not corrected either for occulted or un-occulted stellar spots. 
The maximum dimming observed at our CTIO monitoring wavelength of 5400~\AA\ that does not coincide with a transit event is 5.7~per~cent. Table~\ref{table_litradii} gives the dimming at each wavelength used in the literature that a 5.7~per~cent dimming at 5400~\AA\ corresponds to, calculated using Equation~(\ref{spotscalewavelength}). The maximum dimming values for each wavelength quoted in Table~\ref{table_litradii} are the maximum overestimation of depth as a per~cent of the measured depth due to dimming from unocculted star spots. Apart from the measurements of \citet{2012arXiv1211.0864T}, unocculted star spots may reduce this effect. Using the values given in Table~\ref{table_litradii} means that the radii could change relative to one another by up to 0.0035~$R_P/R_\star$.

All the measured radii are consistent with one another within the limits of the unknowns on stellar variability and stated measurement uncertainties. It should also be pointed out that the differences in absolute radii between transits measured at different epochs could be even larger than the values quoted here due to long-term evolution of the number of star spots on the stellar surface. 
Alternatively, variations could be smaller, if observations are taken at similar periods in the variability cycle and the starspot configurations have not had long to evolve.
Comparing the sample of measurements shows that the variation between the different datasets taken over a period of 2 years is indeed smaller than the observed maximum variation observed during our variability monitoring seasons for the datasets that were not corrected for either occulted nor unocculted spots. The difference between the radii observed by \citet{2012arXiv1211.0864T} and \citet{2011AJ....142..115D} is larger than the difference expected due to unocculted spot variation, although the 1~$\sigma$ error bars of both observations overlap within the expected range. 

\begin{table}
\centering
  \begin{tabular}{c | c | c | c | c}
\hline
Band  & Radius & Reference & Maximum Flux \\
& & & Dimming (\%) \\
\hline
G750L			& $0.1402 \pm 0.00053$  & This work & 4.7 \\
z band			& $0.1425 \pm 0.0014$ & 1 & 4.0 \\   
Gunn r		& $0.1407 \pm 0.0043$	& 2 & 4.9 \\  
Cousins R	 	& $0.1342^{+0.0052}_{-0.0048}$ & 3 & 5.0 \\   
Gunn r	& 	$0.1435 \pm 0.0014$ & 4 & 4.9 \\   
Gunn r	& 	$0.1417 \pm 0.0013$ & 4 & 4.9 \\   
Gunn r	& 	$0.1430 \pm 0.0008$ & 4 & 4.9 \\   
\hline
\end{tabular}
\caption{Table showing a comparison of the STIS G750L white light radius with literature values. The maximum flux dimming values are the maximum flux dimming observed during the WASP-19b variability monitoring at 5400~\AA\ (5.7~per~cent), scaled for each wavelength using Equation~(\ref{spotscalewavelength}). Assuming that similar variations continue over long timescales, this constitutes the greatest dimming expected due to stellar spots. Literature references are: 1.\citet{hebb11}, 2.\citet{hellier11}, 3.\citet{2011AJ....142..115D}, 4.\citet{2012arXiv1211.0864T}. The three values from \citet{2012arXiv1211.0864T} are from three transits observed only 4 days apart.}
\label{table_litradii}
\end{table} 


The initial uncertainties on our fitted parameters are from a fit assuming photon noise uncertainties on the data points. We tried to scale the parameter uncertainties resulting from our fits with remaining red noise, by using the binning technique to determine $\sigma_r$ \citep{pont06} and then rescaling the fitted parameter uncertainties from the fits with $\beta$, where $\beta = \sigma_N / (\sigma_w/\sqrt{N})$ \citep{winn07}. Here, $\sigma_N$ is the standard deviation of fluxes in time bins of size $N$ and $\sigma_w$ is the white noise, which is measured from the standard deviation of the unbinned data compared to $\sigma_r$. As there are only 10 exposures per orbit for the STIS data, it is hard to estimate $\beta$ for large bin sizes, and we find $\beta=1$ for our light curve. For this reason, we were unable to re-scale our parameter uncertainties with red noise using this method.

An alternative way to estimate the level of red noise is by using the prayer bead method to quantify how any remaining systematics could affect the measured transit depth, by quantifying how flat the residuals are. We performed 30 fits, each time replacing the data with $y_i = m_i + r_{i+n}$, where $y_i$ is the new $i^{\mathrm{th}}$ exposure to be fitted, $m_i$ is the model value from the first fit for the $i^{\mathrm{th}}$ exposure, and $r_{i+n}$ is the residual for the $i+n^{\mathrm{th}}$ exposure. For each fit $n$ is increased by 1, until the residuals have completely cycled around. The variances from the prayer-bead method were below the white noise level, indicating low levels of red noise. 

Deviations from our model are therefore on short timescales and act like white noise. In order to take into account residual white noise above the photon noise level, we calculate $\beta_w$, which quantifies how the standard deviation of the unbinned data deviates from the formal photometric error (e.g. as used by \citealt{lendl12}). We then re-scaled the parameter uncertainties in the fits by $\beta_w$. In other fits in this paper, we do not always find $\beta=1$, so in this case, we re-scale with $\beta_w \times \beta$. Note that if either $\beta$ or $\beta_w$ is less than 1, we set it equal to 1 to avoid any shrinking of the error bars. The value of $\beta_w$ for the G750L white light curve was 1.68.

In case of unusual correlations between fitted parameters, we also tested rescaling our photometric error bars in the fitted data with $\beta_w \times \beta$ rather than rescaling the resulting parameter uncertainties from fits using photon noise error bars for the data. The results were equivalent, here and in all the other fits performed in this work, except in Section~\ref{sec_g430L_spec}.

It should be mentioned that such re-scaling of uncertainties is model-dependent as the residuals are used in the re-scaling. For the re-scaling purposes here, we assume that our model is correct (i.e. that it does not introduce trends) and therefore that we can re-scale our uncertainties to account for any remaining noise not taken into account in the model. We can be reasonably confident that our model describes the systematics well, since the dominant systematic trends of the STIS instrument are well known and understood \citep{sing08,sing11,huitson12}. Furthermore, in the cases where we are able to produce transmission spectra, we test that the choice of de-trending model does not significantly affect our transmission spectrum. We note also that re-scaling parameter uncertainties this way is only strictly valid if the uncertainties on each data point are similar \citep{bevingtonrobinson}. The assumption of similar uncertainties on each data point is valid for the case of space-based data taken during a short time period with the same instrument, which we have here.

\subsection{Optical Transmission Spectra}

\subsubsection{G750L}

\begin{figure}
\centering
\includegraphics[width=8cm]{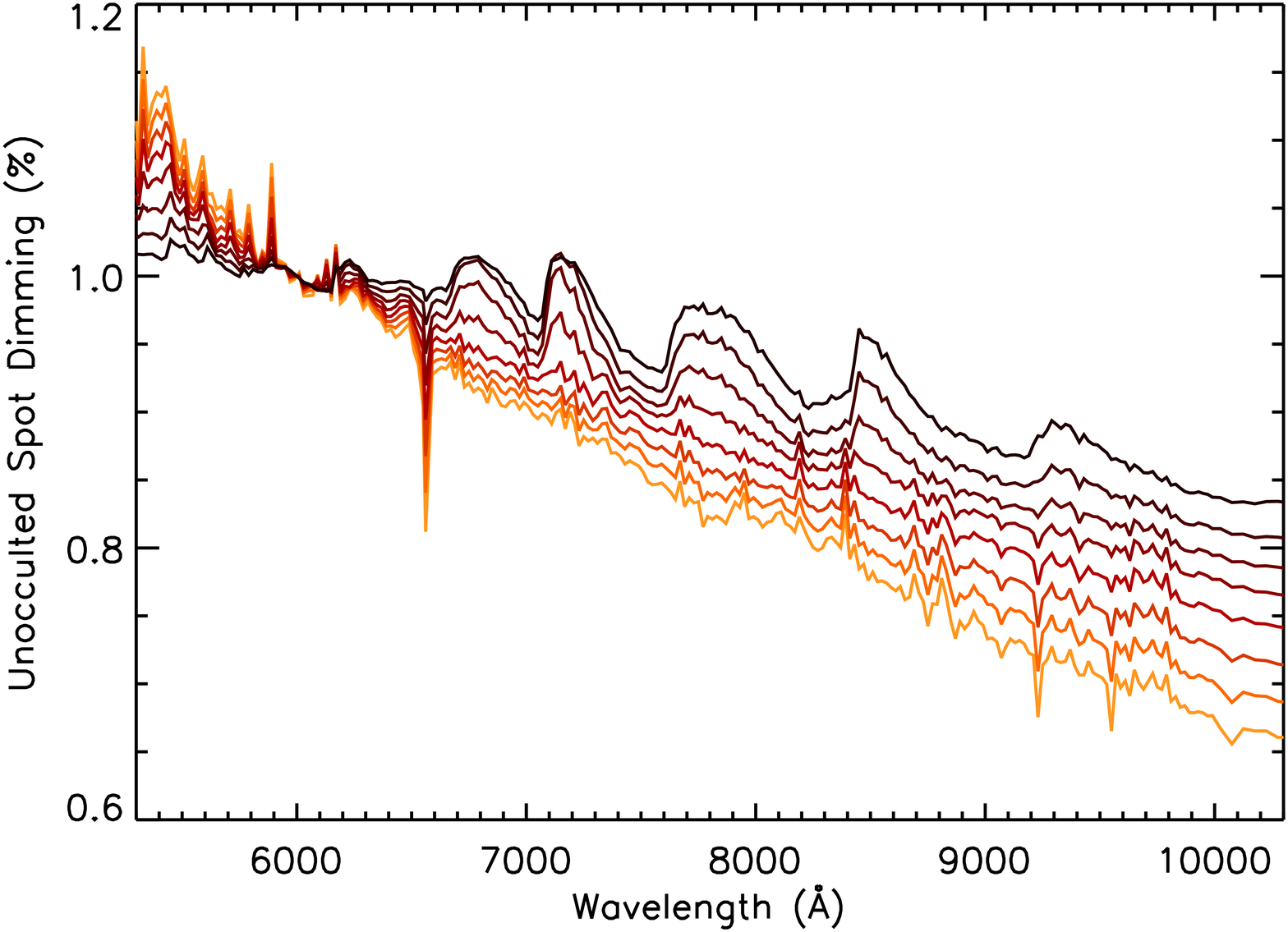}
\caption{Model wavelength dependent unocculted spot dimming over the STIS G750L wavelength range for spot temperatures of 5250-3500~K in increments of 250~K, assuming a 1~per~cent dimming at 6000~\AA. Hotter temperatures are shown in lighter reds, with decreasing temperatures shown with darker reds. The stellar spectrum models are from Kurucz (1993).}
\label{fig_spoctorr_stis_g750L}
\end{figure}

The spectra were split into 4 different bins $\sim 1000$~\AA\ wide, except the reddest bin, where the instrument response and stellar flux are lower, which was $\sim 2000$~\AA\ wide. The light curves were corrected for unocculted spots in each band using Equation~(\ref{eqn_lc_spotcorr}) and the $\Delta f_o$ values corresponding to the measured stellar fluxes given in Table~\ref{table_variability_monitoring}. The wavelength dependent unocculted spot correction is shown in Figure~\ref{fig_spoctorr_stis_g750L} for the STIS G750L wavelengths for different $T_{\mathrm{spot}}$ values ranging from 3500 to 5250~K, which is 250~K cooler than the non-spotted surface. We used the Kurucz (1993) model stellar spectra at different temperatures to work out the correction. The $\Delta f$ spot correction values used for each G750L band are given in Table~\ref{spotcorr_values_stis}, assuming a spot temperature of 5000~K, with a non-spotted stellar surface temperature of 5500~K. The limb darkening parameters were fixed to the model values, and are given in Table~\ref{ld_table_stis}.

\begin{table}
\centering
  \begin{tabular}{c | c }
\hline
Band (\AA)&  $\Delta f$ \\
\hline
5300 - 6300 & 0.037 \\
6300 - 7300 &  0.033 \\
7300 - 8290 & 0.030 \\
8290 - 10300 & 0.027 \\
\hline
\end{tabular}
\caption{Table of spot correction factors used for each STIS G750L spectral band, assuming $T_{\mathrm{spot}}=5000$~K. From Table~\ref{table_variability_monitoring}, the value for $\Delta f_o$ at the CTIO wavelength is 0.039.}
\label{spotcorr_values_stis}
\end{table} 

\begin{table}
\centering
  \begin{tabular}{c | c | c | c }
\hline
Band (\AA)&  $c_2$ & $c_3$ & $c_4$ \\
\hline
  5300 - 6300    &        1.2332    &  -0.6132    &  0.1399 \\
  6300 - 7300    &       1.3703     &  -0.9688     & 0.2774 \\
  7300 - 8290     &      1.3557      &  -1.1067      & 0.3560 \\
  8290 - 10300   &       1.3514    &  -1.2060      &  0.4061 \\
  \hline
\end{tabular}
\caption{Table of limb darkening coefficients from the Kurucz (1993) stellar atmosphere models used for each band in the STIS G750L transit light curve fits.}
\label{ld_table_stis}
\end{table} 

The transit radius was then measured for each bin. The inclination, $a/R_\star$ and central transit time were fixed to the values measured from the WFC3 white light curve. Using the same orbital parameters for each bin ensures that differences in the best fitting $R_P/R_\star$ between bins are due only to radius variations as a function of wavelength. We use the orbital parameters from the WFC3 white light curve since this represents the most constraining dataset that we have, and the derived values match well with those in the literature.

We fitted a 4$^{\mathrm{th}}$ order polynomial function of HST phase to all orbits, including the in-transit orbit, at the same time as fitting for the transit depth. In the bands with the highest signal-to-noise, linear trends with $x$ and $y$ position of the spectrum on the detector could also be fit. It was determined separately for each band whether $x$ and $y$ correlations should be fit or not, by calculating the BIC. Higher order corrections than linear with $x$ and $y$ were not justified in any of the bands. We also noticed a small quadratic trend with planetary phase. Fitting for this changed the measured planetary radii by much less than 1~$\sigma$ but increased the BIC, so the parameter was kept fixed to zero. In all cases, the systematic trends were fitted jointly with the planetary radius to enable covariances to be accounted for.

We do not see any distinctive occulted stellar spots. All the deviations from a flat line that we see in the residuals do not behave like spots as a function of wavelength (they are not of greater amplitude in bluer wavelengths). Without being sure that a correlated trend is due to a spot there is no justification to treat it like a spot, and therefore, we treated all remaining deviations in the residuals as noise (using the same method as for the white light curve, we used $\beta_w$ to rescale our  parameter uncertainties from the LM algorithm in each band, and the prayer-bead and binning techniques to test for red noise, finding $\beta=1$ for all light curves). We checked that the choice of fitted parameters describing the systematic trends had an insignificant effect on the measured radii. 

Additionally, in case small occulted starspots could affect the fitted parameters, we also tried using only the out-of-transit orbits to fit for the systematic trends. Once the fit to the out-of-transit exposures was performed, we fixed the parameters describing the systematic trends and then fitted the whole corrected light curve for only the planetary transit depth. We also tried using the divide-oot method, fitting for only a linear slope in phase and the planetary radius. The spectra from each method agree to within 1~$\sigma$. 

We also tried binning the spectrum into only two bands (5300-7300 \AA\ and 7300-10300 \AA), and the results from this agree very well with the results from using 4 bands. Fitting the white light radius as a whole is consistent as well. This indicates that using 4 bands has not introduced any new systematics from having less signal in each band. There is fringing in the reddest part of the G750L spectrum on the CCD. The bands were chosen deliberately to lie between fringes, but a test using bands deliberately in the middle of fringes produced the same spectrum. It is likely that the effect of fringing on our spectrum is negligible as we are using very large bins covering many fringe features.

\begin{figure}
\centering
\includegraphics[width=8.1cm]{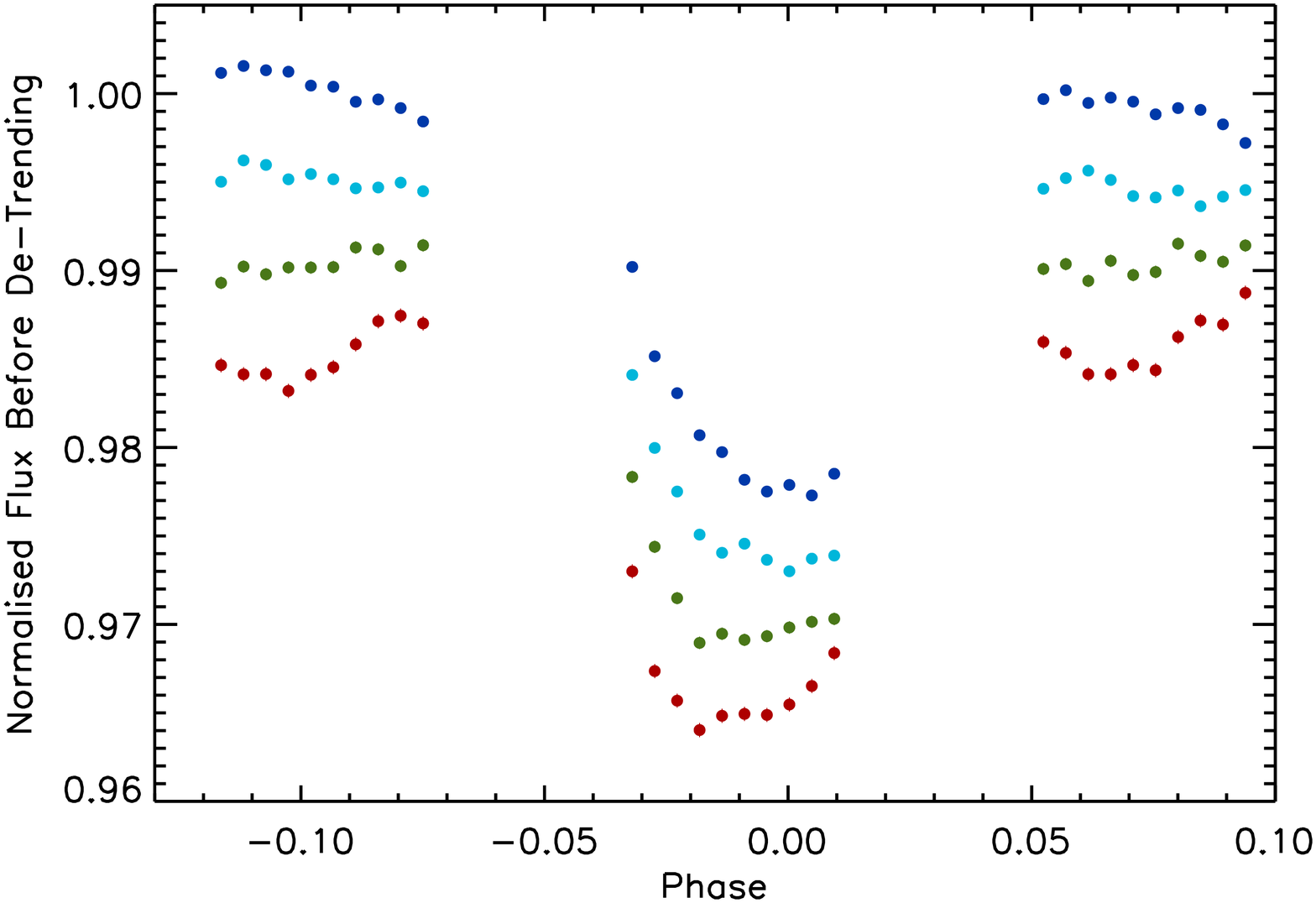}
\includegraphics[width=8.1cm]{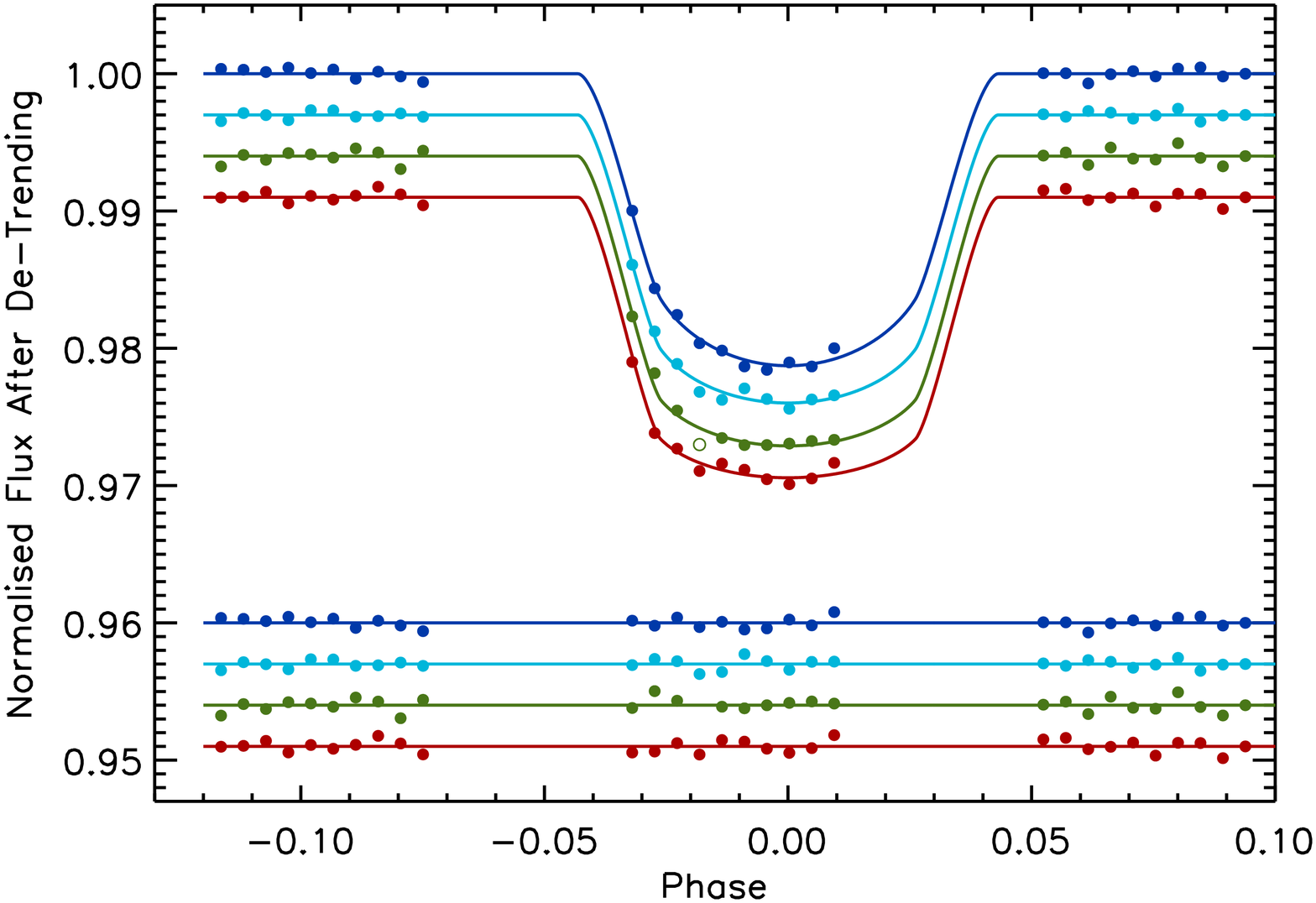}
\caption{\textit{Top: }Raw light curves for each of the G750L spectral bins, each normalised to their out-of-transit flux. The bluest bands are at the top and the reddest are at the bottom. Each light curve has an arbitrary flux offset for clarity. Wavelength-dependent trends as a function of HST orbital phase can clearly be seen. \textit{Bottom: } De-trended light curves for each of the G750L spectral bins with the bluest at the top and the reddest at the bottom overplotted with the analytical transit models of \citet{mandelagol02}. Underneath are the residuals with bluest at the top and reddest at the bottom with arbitrary flux offsets for clarity. One $> 3 \sigma$ outlier was clipped during the fit in the second reddest band (shown as an open circle), but the fit was performed again with this point not clipped and the measured $R_P/R_\star$ value was not significantly affected. For all plots, photon noise error bars are shown and are within the point symbols.}
\label{fig_g750l_spectral_lc}
\end{figure}

\begin{figure}
\centering
\includegraphics[width=8cm]{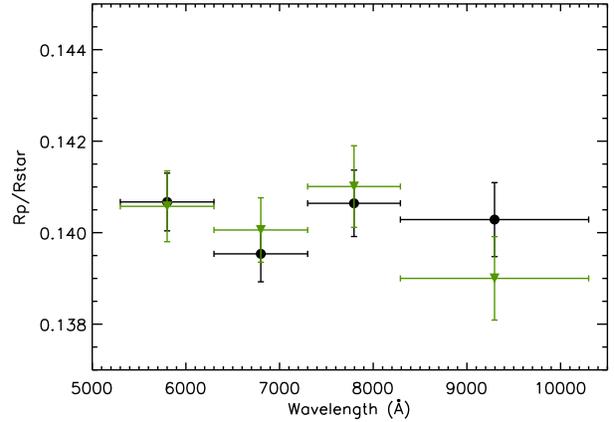}
\caption{STIS G750L transmission spectrum of WASP-19b. The fits were performed using photon noise uncertainties on the data points, and then the error bars from \textsc{mpfit} were re-scaled with $\beta_w$, since we do not detect any red noise. The black points show the spectrum extracted using 3.75~$\sigma$ clipping in the cosmic ray removal routine, and green points show the spectrum extracted using 5~$\sigma$ clipping of cosmic rays.}
\label{fig_g750l_spectrum}
\end{figure}

We tried both of our cosmic ray removal routines, and found that both methods produce similar spectra, although the spectrum produced with the ``difference" method is marginally flatter. The variance from the prayer bead method and also the white noise in the light curves are both lower for the ``difference" cosmic ray removal routine, and so this is the routine that we adopt. For the new routine, we used a conservative 5~$\sigma$ clipping in the cosmic ray removal and compared this with the results from using a 3.75~$\sigma$ clipping. In our experience with other similar datasets of the large HST STIS programme, 3.75~$\sigma$ is the level where the residuals strongly start to deviate significantly from Gaussian. For the WASP-19b dataset, the measured $R_P/R_\star$ value in the reddest bin changed the most, by $\sim 0.5$~$\sigma$, with the extracted light curve looking noticeably less noisy when using 3.75~$\sigma$ clipping. Figure \ref{fig_g750l_spectral_lc} shows the light curves determined from the 3.75~$\sigma$ cosmic ray clipping and Table \ref{table_g750l_spectrum} gives the radii for each band determined from the 3.75~$\sigma$ cosmic ray clipping and fitting the systematic trends and $R_P/R_\star$ simultaneously. Figure~\ref{fig_g750l_spectrum} shows the resulting spectrum for both clipping levels in the cosmic ray removal routine. It can be seen that the lower level clipping produces a marginally flatter spectrum than the higher level clipping, suggesting that un-removed cosmic rays can mimic spectral features in light curves with few exposures. The fits were performed using photon noise uncertainties on the data points, and then the error bars from \textsc{mpfit} were re-scaled with $\beta_w$, since we do not detect any red noise. We also tried subtracting the white light residuals from each spectral light curve, to remove common-mode trends, but found no significant change in noise level, indicating that the most significant trends are not common mode over our large wavelength range.

\begin{table}
\centering
  \begin{tabular}{c | c | c | c | c | c }
\hline
Band (\AA)&  $R_p/R_\star$ & $\chi^2$ & $\sigma_w$ & $\sigma_r$ & $\beta_w$ \\
\hline
5300 - 6300 & $ 0.1407 \pm 0.0007 $ & 25.3 & 0.00028 & 0 & 1.4 \\
6300 - 7300 & $ 0.1395 \pm 0.0006 $ & 26.1 & 0.00027 & 0 & 1.5 \\
7300 - 8290 & $ 0.1406 \pm 0.0008 $ & 24.6 & 0.00033 & 0 & 1.4 \\
8290 - 10300 & $ 0.1403 \pm 0.0008 $ & 21.6 & 0.00036 & 0 & 1.2 \\
\hline
\end{tabular}
\caption{Table of fitted planetary radii with respect to stellar radii for each STIS G750L band. Red and white noise values were determined using the binning technique. The $\chi^2$ values are based on the fits using only photon noise for the photometric uncertainties. Each fit has 18 degrees of freedom except the light curve where an outlier was clipped, which has 17 degrees of freedom.}
\label{table_g750l_spectrum}
\end{table} 

\subsubsection{G430L}
\label{sec_g430L_spec}

Due to the presence of severe occulted starspots, we were not able to obtain the transmission spectrum of WASP-19b in the G430L wavelength range. It was possible to obtain the differential transmission spectrum for HD~189733b by measuring the shape of the spot in the white light curve and then fitting jointly for a spot amplitude parameter and the planetary radius in each wavelength \citep{sing11}. However, fitting for the occulted spot amplitude was not possible for WASP-19b, as no exposures between 2$^{\mathrm{nd}}$ and 3$^{\mathrm{rd}}$ contact were ``spot-free". Furthermore, the necessary low cadence for the WASP-19b observations presents a problem when trying to disentangle the effects of stellar spots, differential planetary radii, systematic trends and any remaining low-level cosmic rays. We found that we could not construct a transmission spectrum because the measured planetary radii as a function of wavelength were very sensitive to the treatment of stellar spots, the de-trending model used, and to which exposures were used. Removing random single exposures from the transit changed the measured planetary radii by over $\pm 3$~scale heights, which is larger than any spectral feature we hope to detect in a given band. The parameter for the amplitude of the spot was very degenerate with the $R_P/R_\star$ parameter in our fits. If we fit the data using the MCMC package \textsc{exofast}, developed by \citet{eastman12}, it reports linear Pearson correlation coefficients of $\sim 0.5-0.6$ for some bands. The high correlation values are most likely due to not having any non-spotted exposures between 2$^{\mathrm{nd}}$ and 3$^{\mathrm{rd}}$ contact to anchor the transit model and so measure the departures from the non-spotted transit shape independently from the transit depth. To avoid unphysical results, we limited the radius variation across the wavelength range in the fits to 4 scale heights, which is the maximum predicted by current atmosphere models.

The correlation of the spot amplitude parameter with the HST-phase de-trending parameters was even higher than the correlation with fitted radius, with some linear Pearson correlation coefficients above $\pm 0.9$. To try to avoid the shape of the occulted spots affecting the de-trending for the structure of the systematics, we tried only fitting the two out-of-transit orbits for systematics and then fitted the transit and spot amplitude only, using the corrected light curve. However, it is still possible that the spot amplitude measurement could be affected by the instrumental systematics in each band. We rescaled our photometric uncertainties with $\beta$ and $\beta_w$ and re-fit for parameters rather than re-scaling the parameter uncertainties from fits assuming photon noise because we found small differences between the two methods in this case. 

\begin{figure}
\centering
\includegraphics[trim=0cm 0cm -0.6cm 0cm,width=8cm]{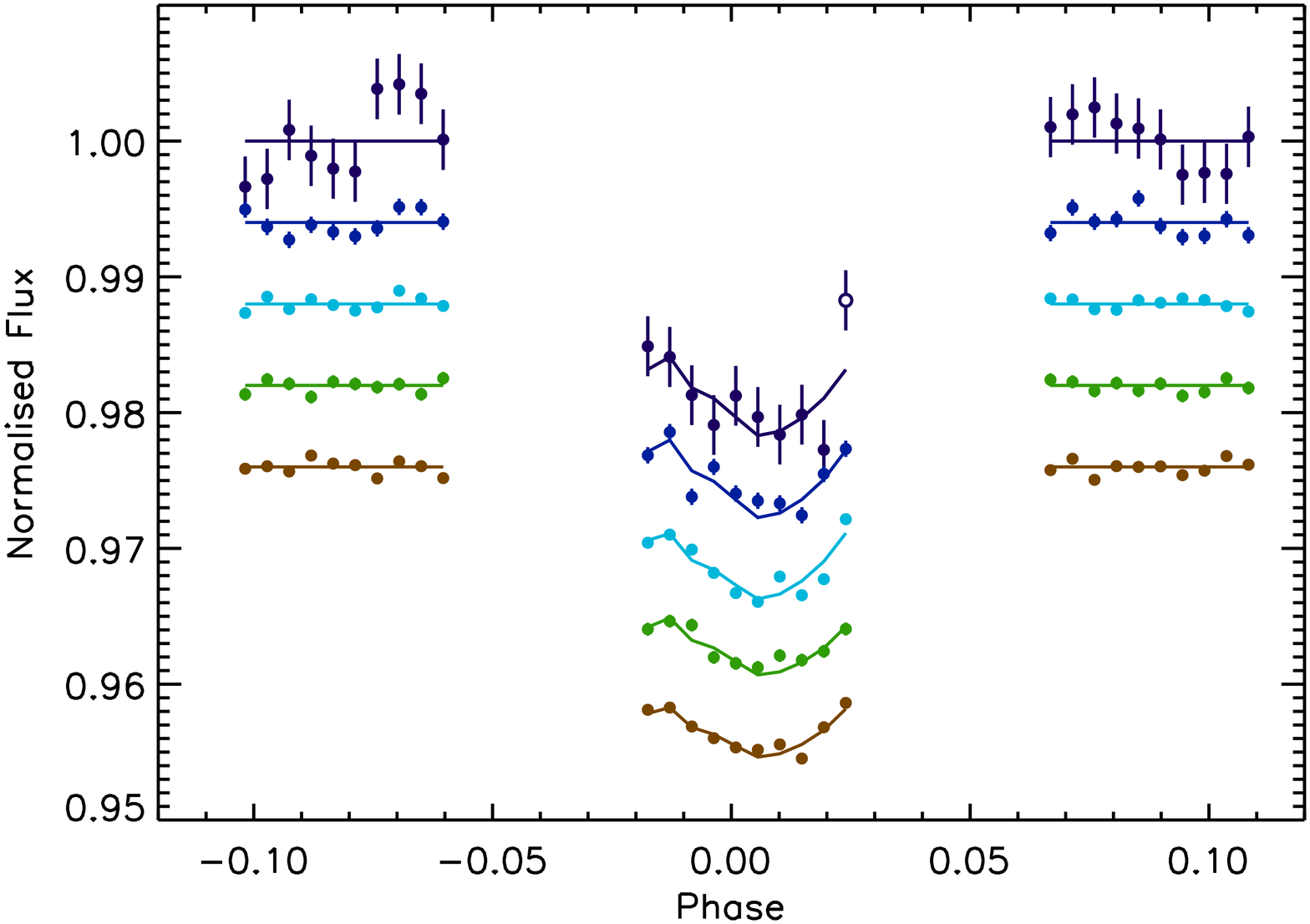}
\includegraphics[trim=0cm 0cm 0.5cm 0cm,width=8cm]{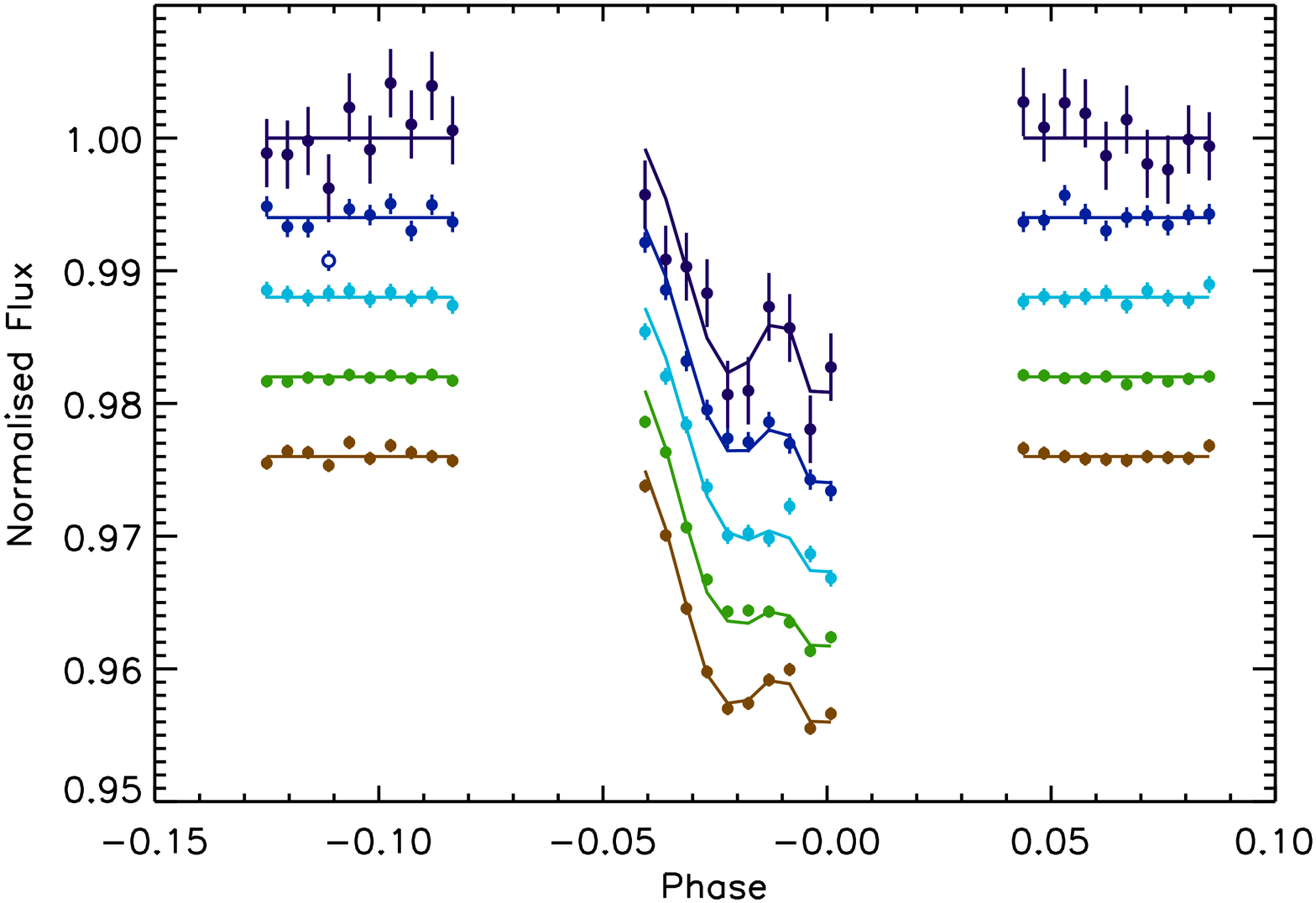}
\caption{STIS G430L light curves for each spectral bin, going from shorter wavelengths at the top of each plot to longer wavelengths at the bottom of each plot, after de-trending. The top plot shows visit~3 and the bottom plot shows visit~4. The transit models including the occulted spot are shown with lines. The last point in the transit was not fitted in the bluest band of visit 3, and one outlier in the exposure before transit was not fitted for the second bluest band of visit 4. The points not fitted are shown with open symbols. The error bars on the individual data points have been re-scaled with $\beta \times \beta_w$ to show the level of uncertainty in the data due to noise above the photon noise level.}
\label{fig_LCs_g430l}
\end{figure}

\begin{figure}
\centering
\includegraphics[width=8cm]{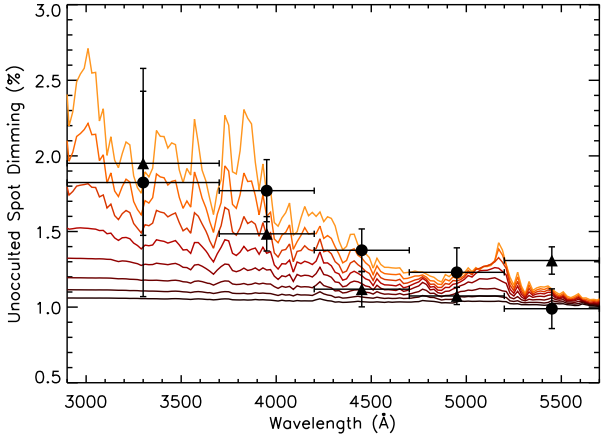}
\caption{Model wavelength dependent unocculted spot corrections over the STIS G430L wavelength range, with measured occulted spot magnitudes as a function of wavelength shown as circles (visit3) and triangles (visit4). Horizontal error bars show the data bin sizes. The model spot temperatures range from 5250-3500~K in increments of 250~K. Hotter temperature spot models are shown in lighter reds, with decreasing temperatures shown with darker reds. The stellar spectrum models are from Kurucz (1993).}
\label{fig_spotspec_g430l}
\end{figure}

Additionally, fitting the spot amplitude is not only very degenerate with other fitted parameters, but also very sensitive to noise. Since the white light spot includes noise, so does our model for the spot shape. We therefore smoothed our spot shape with a Gaussian function before fitting it for amplitude in each band. Using the prayer bead method gives 1~$\sigma$ uncertainties for the spot magnitudes that are $\sim 1/2$ of the 1~$\sigma$ uncertainties found using our rescaling method, suggesting that outliers and residual red noise do not substantially affect the fitted parameters. Figure~\ref{fig_LCs_g430l} shows fits to the spectral light curves using the spot shape from the data, with uncertainties scaled with $\beta \times \beta_w$. Figure~\ref{fig_spotspec_g430l} shows the resulting spot amplitude parameter for each wavelength, and Table~\ref{table_spotspec_g430l} gives the spot amplitude parameters for each wavelength band along with statistics of the fit. It is clear that, especially in the bluest band, not all trends have been removed.


\begin{table}
\centering
  \begin{tabular}{c | c | c | c | c }
\hline
Band (\AA) & spot amplitude & error & $\sigma_r$ & $\sigma_w$ \\
 & (per~cent) & (per~cent) & & \\
\hline
2900-3700 & 2.15 & 0.75 & 0.0015 & 0.0017 \\
3700-4200 & 2.10 & 0.21 & 0 & 0.00091 \\
4200-4700 & 1.70 & 0.14 & 0 & 0.00060 \\ 
4700-5200 & 1.56 & 0.16 & 0.00013 & 0.00050 \\
5200-5700 & 1.31 & 0.13 & 0 & 0.00050 \\
\hline
\end{tabular}
  \begin{tabular}{c | c | c | c | c }
\hline
Band (\AA) & spot amplitude & error & $\sigma_r$ & $\sigma_w$ \\
 & (per~cent) & (per~cent) & & \\
\hline
2900-3700 & 1.58 & 0.48 &  0.00034 & 0.0023 \\
3700-4200 & 1.11 & 0.11 &  0.00006 & 0.00075 \\
4200-4700 & 0.74 & 0.12 & 0 & 0.00063 \\
4700-5200 & 0.70 & 0.06 & 0.00009 & 0.00039 \\
5200-5700 & 0.93 & 0.09 &  0 & 0.00049 \\
\hline
\end{tabular}
\caption{Table showing fitted spot amplitude parameters for each band of G430L visits 3 and 4, where these are the relative amplitude values compared to the reference wavelength. Assuming a 1~per~cent flux dimming at 6000~\AA, the spot amplitude parameters and errors are in units of per~cent.}
\label{table_spotspec_g430l}
\end{table} 

We fitted the different spot spectra to the data using an amplitude offset constant across all wavelengths as a free parameter and using the weighted mean of the two datasets together. The fits suggest that higher spot temperatures in our range of 5250-3500~K give the best fits, with the best fitting temperature being 5000~K and with the fit using the lowest temperature of 3500~K worse by 2~$\sigma$. The fits using each visit individually give results within $\pm 250$~K of the value when using both visits. We find that the fitted spot magnitudes in each band are slightly different than when performing the fits using only photon noise uncertainties and re-scaling fitted parameters compared to re-scaling the photometric uncertainties, but our best fitting occulted spot temperature does not change. 
We adopt a 5000~K temperature for all spots on the stellar surface for the rest of the analysis. Our fits suggest that the spots are likely in the range from 4000-5250~K.
Due to the presence of obvious remaining systematic trends in the blue light curves, we still consider the possibility of different temperatures in Section~\ref{sec_discussion} and discuss the spectra resulting from assuming different temperatures for the unocculted spot correction.


We also find a suggestion of higher transit depths in the G430L wavelengths compared to the G750L prediction. If we fix the G430L radii to be at the measured G750L level, the fits are $\sim 2 \sigma$ worse than allowing the fit radius to be free, which gives a deeper transit. Deeper radii could indicate spectral features, however they could also be the result of unseen occulted stellar spots in the G750L data. Since we see spot crossings in the G430L transits, we can account for them and exclude the spot-crossing exposures in our fits. In the redder wavelengths, spots that are not easily visible by eye are not accounted for except as a source of correlated noise and if present, these can still make the measured planetary radii shallower. For this reason, and due to the dependence of the G430L radii on the de-trending methods used and occulted spot amplitude as a function of wavelength, we do not draw any conclusions from the measured deeper transit radii in the blue.

\section{WFC3 Analysis}
\label{wfc3_analysis}

\subsection{Spot Corrections}

There was no variability monitoring taking place during the WFC3 observation. However, we can estimate the maximum degree to which the measured $R_P/R_\star$ values can shift assuming that the flux of the star behaved in a similar way a year before the STIS observations. The maximum flux dimming in the CTIO band (which does not coincide with a transit) is $\Delta f_o = 0.06$. Averaged over the WFC3 band, the wavelength-dependent correction is 0.46, assuming a spot temperature of $T=5000$~K (with the non-spotted surface having $T_\star=5500$~K). The resulting maximum flux dimming, and change in depth would be $\Delta f =  0.06 \times 0.46 = 0.0276$, assuming that the star's inherent brightness variations have not changed over the time between the WFC3 and STIS observations. The dimming, $\Delta f$, is equal to the change in depth, $\Delta d / d$, or $\Delta{(R_P/R_\star)^2}/(R_P/R_\star)^2$ so the change in the planetary radius is 

\begin{equation}
\Delta (R_P/R_\star) \simeq \frac{1}{2} \Delta f (R_P/R_\star) ,
\end{equation}
to first order (as also used by \citealt{berta11b}, \citealt{sing11}, \citealt{pont13}).


Assuming that the stellar activity has remained similar over the course of a year, the unocculted spot correction is equal to $0.0138$~$(R_P/R_\star)$. The correction is likely an overestimate because it represents the maximum dimming observed during our monitoring. Also, any occulted starspots which are within the noise in the measurements have the same level of effect on the radius measurement but in the opposite direction (they will make the transit appear shallower) even though they are not seen. This is common in the near-IR, as the spot amplitudes are lower and photometric precision lower. Therefore, we do not correct for the effect of unocculted starspots. The unknown stellar activity level during the WFC3 observations means that we are not able to place the planetary radii on an absolute scale, but we can still draw conclusions from the relative transit depths to a level better than 1~per~cent. The effect of the wavelength-dependent correction on the planetary transmission spectrum is discussed in Section~\ref{significance_of_water}.

\subsection{De-Trending the WFC3 Light Curves}
\label{detrending_wfc3}

\begin{figure}
\centering
\includegraphics[width=8cm]{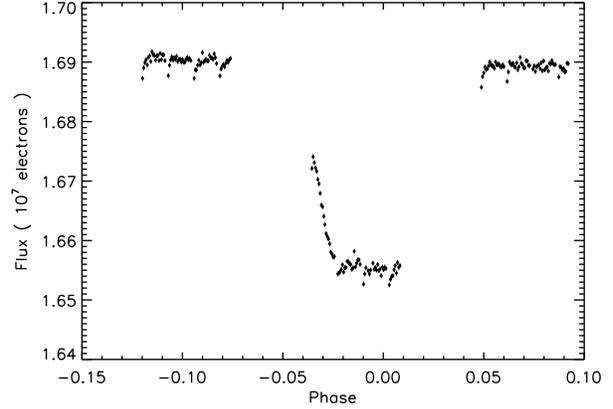}
\caption{WFC3 white light curve before removal of systematics. Ramp effects can be seen to occur on a timescale of 20 exposures.}
\label{fig_light_curves_not_fit}
\end{figure}

Figure \ref{fig_light_curves_not_fit} shows the white light curve for the WFC3 transit, which clearly shows an exponential-like ramp effect that occurs regularly every 20 exposures. \citet{berta11} suggest that the ramp trend could be due to charge trapping. Charge traps gradually fill up during one batch of 20 exposures and then are completely filled towards the end of the batch, causing the ramp to flatten off. After 20 exposures, there is a small delay before the start of the next batch, which could be due to a small instrument buffer dump, which would reset the charge traps. 

\begin{figure}
\centering
\includegraphics[width=8cm]{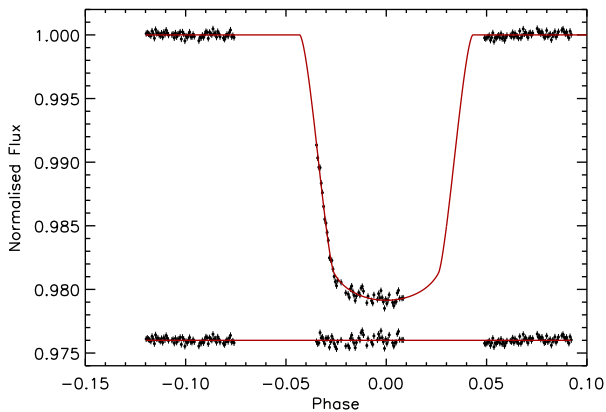}
\caption{WFC3 white light curve after using the divide-oot method. The transits are fitted with a linear slope in phase and the transit models of \citet{mandelagol02}, after performing the divide-oot method. Residuals are shown underneath with an arbitrary flux offset. The in-transit orbit has increased white noise from the divide-oot method compared to the two out-of-transit orbits which make the template, but white noise should not affect spectroscopic results. Photometric uncertainties compared to baseline flux are higher after performing divide-oot than for the raw light curve, because they include the uncertainty in each point of the template.}
\label{fig_white_light_divideoot}
\end{figure}

Firstly, we tried removing the systematics using the divide-oot method described by \citet{berta11}, where a template is made using the mean of the two out of transit orbits and then each orbit is divided by the template. The divide-oot method works very well if the systematics are repeatable over each orbit. Aside from the ramp-like effect, we also see the temperature-based variations as a function of HST phase, as also seen in the STIS data. HST phase trends should also be repeatable for each orbit and be removed using the divide-oot method. After performing the divide-oot method, we then fit for the transit using the transit models of \citet{mandelagol02} along with a linear slope as a function of planetary phase. Figure \ref{fig_white_light_divideoot} shows the corrected white light curve with the transit fit, and the corresponding residuals. Again, we use the \textsc{idl mpfit} package. Parameter uncertainties were re-scaled for remaining red noise in the residuals using the $\beta$ factor. No occulted spots are visible in the light curve and so we do not account for them in the fit. Any very small spots that are below the white noise level should not have a significant effect, and any remaining effects are treated in our procedures as a source of red noise. The standard deviation of the residuals is $2.8 \times 10^{-4}$, which is dominated by white noise. The fit assuming photon noise uncertainties on the data points gives $\chi^2_\nu=1.22$. Fitting also for any trends as a function of HST phase after performing divide-oot, did not change the fit, but significantly increased the BIC. The large increase in the BIC indicates that such extra parameters were not justified. 

The divide-oot method appears to remove the majority of the ramp effect, but we also tested whether we could remove the systematics by modelling the ramp with an exponential function of the form

\begin{equation}
F_0(a\phi_P + f(\mathrm{\phi_{HST}}) + 1)(1 - b e^{-(t-c)/\tau} + 1),
\end{equation}
where $a,b$ and $c$ are fitted constants, $\phi_P$ is the planetary phase, $f(\mathrm{\phi_{HST}})$ is a 4$^{\mathrm{th}}$ order polynomial of HST phase, $t$ is the time of an exposure within a batch between ramps and $\tau$ is the timescale of the ramp (20 exposures). The first term deals with linear variations over a whole visit, the second term deals with orbital variations, primarily due to thermal breathing, and the last term deals with the ramp effect.

We also fitted an exponential polynomial of the form 

\begin{multline}
F_0(a\phi_P + f(\mathrm{\phi_{HST}}) + 1) \times \\
(1 - b_1 e^{-(t-c_1)/\tau}  - b_2 e^{-2(t-c_2)/\tau} - b_3 e^{-3(t-c_3)/\tau} - b_4 e^{-4(t-c_4)/\tau}+ 1), \\
\end{multline}
The results from both ramp models are less satisfactory than using the divide-oot method, partly because each ramp in a given orbit does not have exactly the same shape. We also tried simply removing the first 3 exposures in each batch and fitting a 4$^{\mathrm{th}}$ order polynomial as a function of HST phase. This also was not as satisfactory as using the divide-oot method, although not using these exposures along with doing divide-oot was an improvement on including them. Table \ref{table_fitting_stats_wfc3} shows the results of the different fits.


\begin{table}
\centering
  \begin{tabular}{c | c | c | c | c | c | c}
\hline
Model &  Fit First & $\chi^2_\nu$ & BIC & $\sigma_r$  \\
& 3 points? & & & ($\times 10^{-5}$) \\
\hline
Divide-oot & no & 1.22 & 280 & 6 \\
Divide-oot & yes & 1.36 & 310 & 6 \\
Exponential & no & 1.79 & 350 & 6 \\
Exponential & yes & 1.85 & 409 & 11 \\
Exponential polynomial & no & 1.96 & 422 & 11 \\
Exponential polynomial & yes & 1.83 & 417 & 9 \\
HST phase fit only & no & 1.92 & 354 & 8 \\
\hline
\end{tabular}
\caption{Table of fitting statistics for different systematic removal methods for the WFC3 white light curve. The $\chi^2_\nu$ values are from using only photon noise as the uncertainties on the fitted data points.} 
\label{table_fitting_stats_wfc3}
\end{table} 

Given that the statistics of the divide-oot technique are better than the other models and the residuals appear flatter, we decided to use this method for analysing the planet's spectrum. However, we also produced the transmission spectrum using all of the other methods, and fitting only a linear slope in time, and found the same spectral features, simply with larger error bars. The fact that different de-trending methods agree on the relative transit depths confirms the assumption that the systematics are very repeatable from orbit to orbit and hence do not significantly affect the relative transit depths. 

\subsection{System Parameters}

The WFC3 data are of good enough quality that we can also constrain the system parameters. The system parameters resulting from the L-M fit to the white light timeseries are given in Table \ref{table_system_params}. We also measured the system parameters using \textsc{exofast}, and these results are also shown in Table~\ref{table_system_params}. Conversions between MJD and BJD$_{\mathrm{TDB}}$ made use of the online time utilities developed by J. Eastman \footnote[5]{http://astroutils.astronomy.ohio-state.edu/time/}. It is found that the two analysis methods produce similar results, although the value of $a/R_\star$ is higher when fitting with \textsc{exofast}. The discrepancy in the fitted $a/R_\star$ value is most likely due to the fact that \textsc{exofast} uses a quadratic limb darkening model free to vary within the stellar model grid but constrained by priors on the stellar properties. In addition, in \textsc{exofast} the stellar parameters themselves are free to vary and constrained by the \citet{torres10} mass radius relation. The resulting parameters match well with those determined by \citet{hellier11}, which are shown in the table also. The stellar radius and mass were given very tight priors based on the values quoted by \citet{hellier11}.

\begin{table}
\centering
  \begin{tabular}{c | c | c | c }
\hline
Parameter &  \textsc{mpfit} & \textsc{exofast} &Hellier et al. \\
& & & (2011) \\
\hline
$i$ (deg) & $79.8 \pm 0.5$ & 79.7 $\pm$ 0.6 & $79.4 \pm 0.4$ \\
$a/R_\star$  & $3.60 \pm 0.05$ & $ 3.872 \pm 0.16$ & $3.60 \pm 0.04$ \\
$T_0$  & 2455168.96898 & 2455168.96843 & 2455168.96879  \\
(BJD$_{\mathrm{TDB}}$) & $\pm$ 0.0006 & $\pm$ 0.0019 & $\pm $ 0.00009 \\
$b$ & $0.635 \pm 0.02$ & $0.695^{+0.028}_{-0.03}$ & $0.657 \pm 0.015$ \\
\hline
\end{tabular}
\caption{System parameters for the WASP-19b system determined from the WFC3 white light curve. The red noise in the residuals for \textsc{exofast} was higher than when using mpfit, which is why the error bars are larger. The larger correlated noise level is most likely because of having to use quadratic limb darkening, affecting the shape of the fit.}
\label{table_system_params}
\end{table} 

\subsection{Near-IR Transmission Spectrum}

\begin{figure}
\centering
\includegraphics[trim=-0.5cm 0cm 0.5cm 0cm,width=8cm]{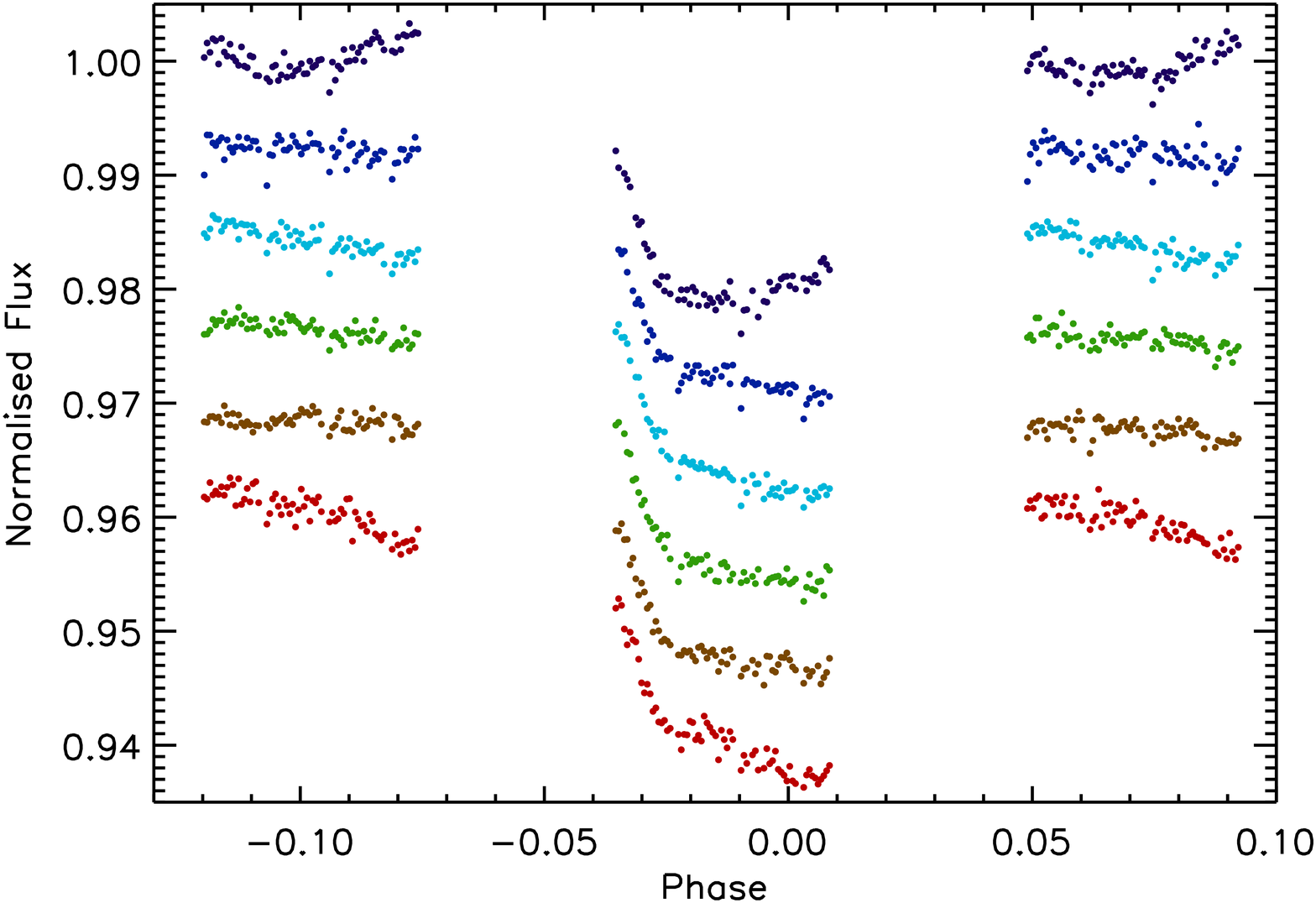}
\includegraphics[trim=-0.5cm 0cm 0.5cm 0cm,width=8cm]{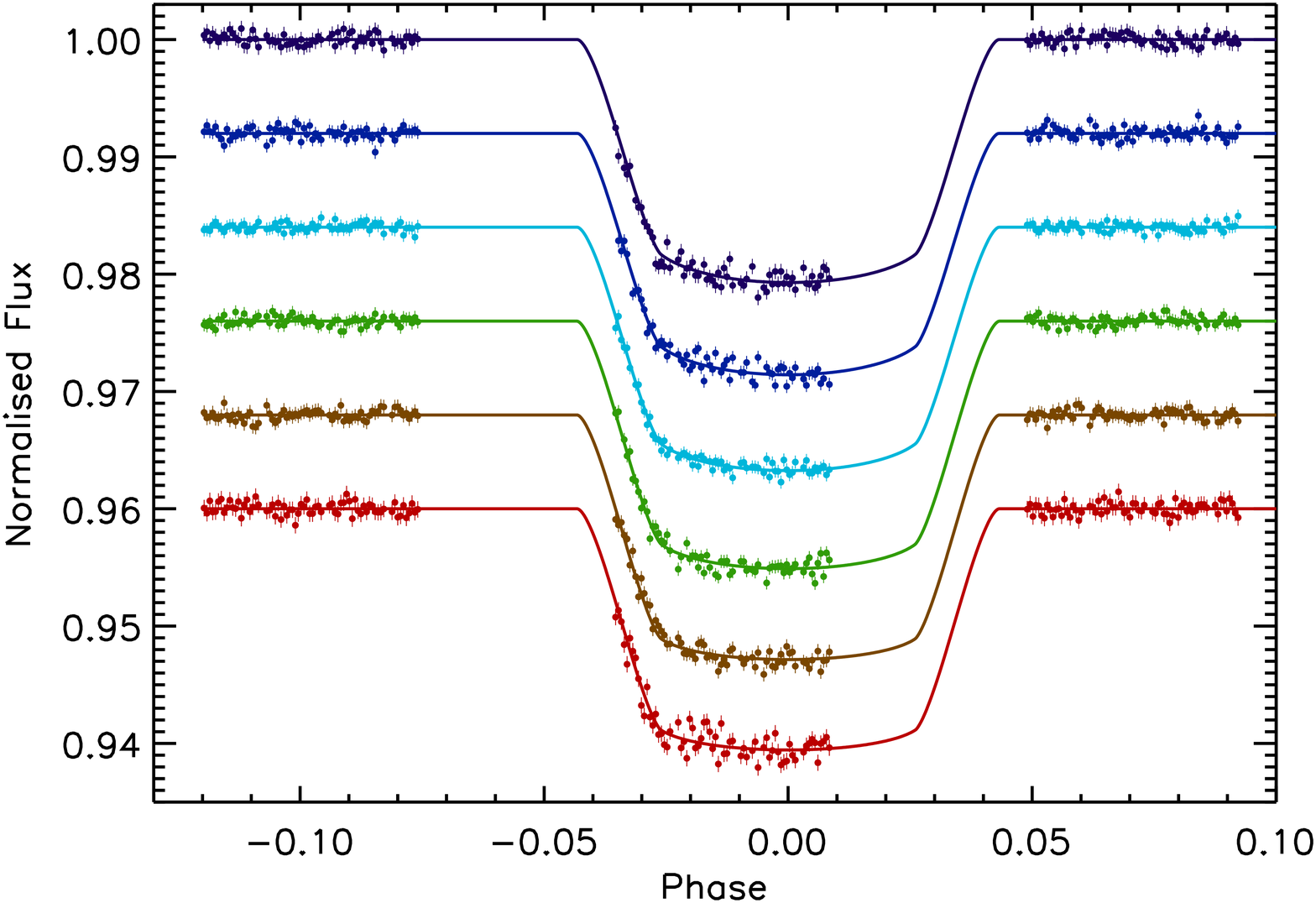}
\includegraphics[trim=0cm 0cm -1cm 0cm,width=8.5cm]{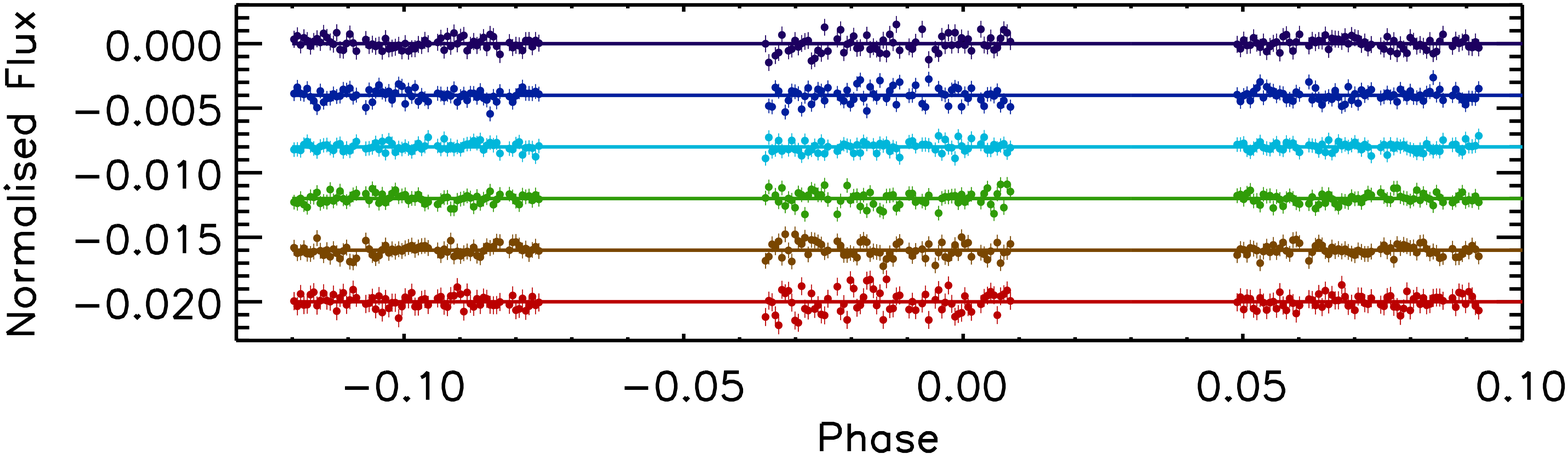}
\caption{\textit{Top: }Raw spectral light curves before performing divide-oot, with each light curve normalised to its out-of-transit flux level. The bluest bands are at the top and the reddest bands at the bottom, with arbitrary flux offsets for clarity. The ramp effect in the spectral bins is less significant compared to white noise scatter than it was for the white light curve. Error bars assume photon noise and are within the point symbols. \textit{Middle: }Corrected spectral light curves overplotted with transit models from \citet{mandelagol02}. The bluest bands are at the top and the reddest are at the bottom, with arbitrary flux offsets for clarity. \textit{Bottom: }Residuals with the bluest bands at the top and the reddest bands at the bottom, with arbitrary flux offsets for clarity. For the corrected data and the residuals, error bars assume photon noise and also include the error in the out-of-transit template used for the divide-oot procedure for each point.}
\label{fig_spectral_light_divideoot}
\end{figure}

The system parameters, \textit{i}, $\rho_\star$ and $T_0$ were fixed to the values found for the white light curve. The spectral timeseries was broken up into spectral bins $\sim 1000$ \AA\ wide. We then subtracted the white light residuals from the spectral light curve to remove common-mode trends. This reduced the red noise error bars on the measured radii but had no other effect on the measured spectrum. Once the white-light residuals were removed, the divide-oot method was performed on each light curve. Each light curve was then fitted for $R_P/R_\star$ and a linear trend over the whole visit. The raw and corrected light curves for each band are shown in Figure \ref{fig_spectral_light_divideoot} along with residuals. Again, we see no clear occulted starspots, so we cannot justify correcting the spectrum for occulted starspots, and any remaining low-level occulted spots are treated as red noise. The limb darkening values are fixed to the model outputs, given in Table~\ref{ld_table_wfc3}.

\begin{table}
\centering
  \begin{tabular}{c | c | c | c | c }
\hline
Waveband ($\umu$m) & $c_2$ & $c_3$ & $c_4$ \\
\hline
1.087 - 1.187 & 1.4098    &  -1.3968    &  0.48859 \\
1.187 - 1.287 & 1.4080   &   -1.3847    &  0.48165 \\
1.287 - 1.387 & 1.5450     &  -1.6282     & 0.58416 \\
1.387 - 1.487 & 1.7834    &  -2.0855    &  0.79436 \\
1.487 - 1.587 & 2.0182   &   -2.5803    &   1.0382 \\ 
1.587 - 1.687 & 1.9108    &  -2.4396    &  0.9865 \\
\hline
\end{tabular}
\caption{Limb darkening coefficients for the WFC3 bands, from the Kurucz (1993) stellar atmospheric models.}
\label{ld_table_wfc3}
\end{table} 

\begin{figure}
\centering
\includegraphics[width=8cm]{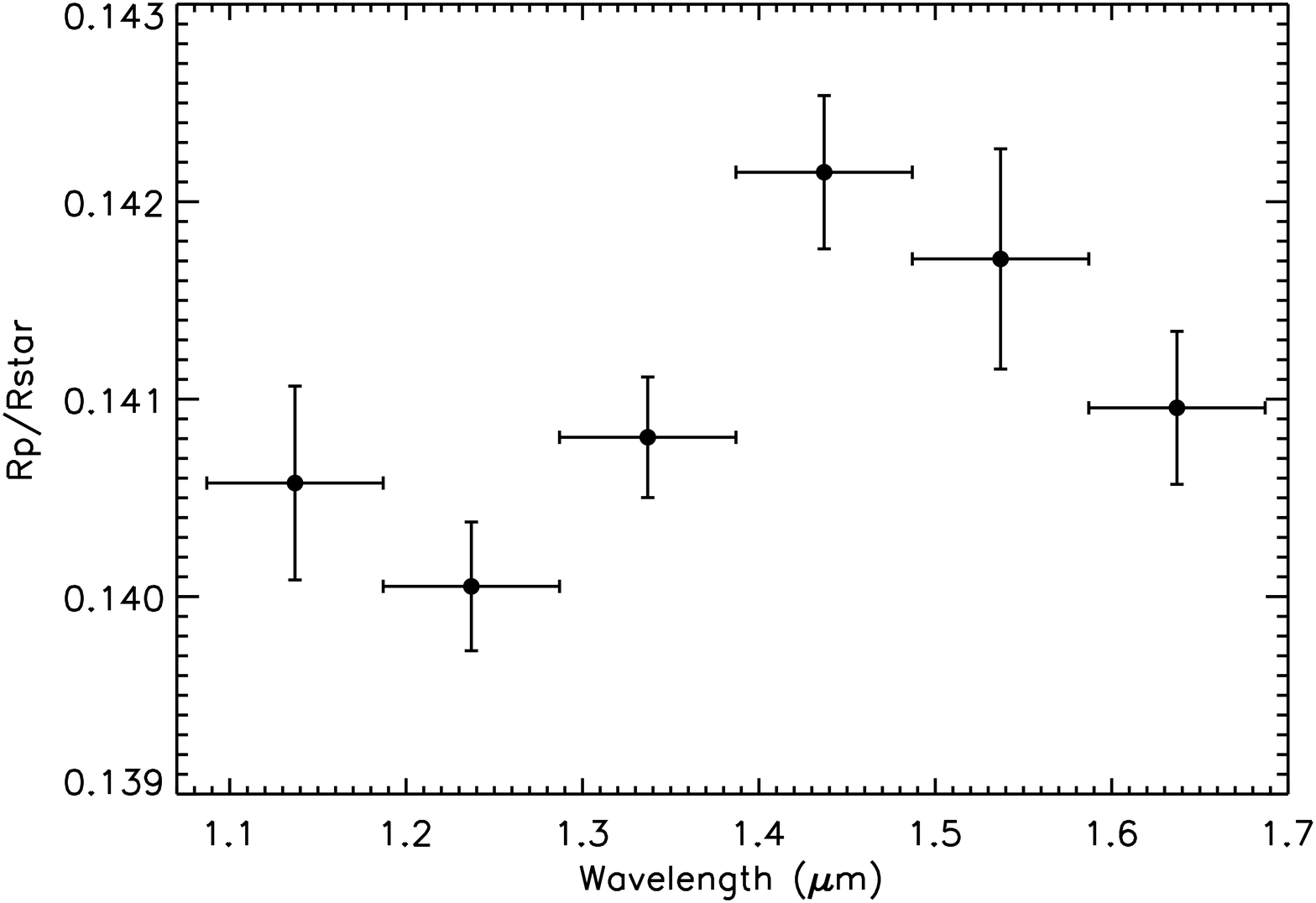}
\caption{Spectrum of WASP-19b from WFC3 G141. Errors are from a fit based on photon noise uncertainties on the individual exposures, with fitted parameter uncertainties scaled with $\beta$.}
\label{fig_wfc3_spectrum}
\end{figure}

The resulting spectrum is shown in Figure \ref{fig_wfc3_spectrum}. The fitted $R_P/R_\star$ values are given in Table \ref{table_spectrum}. We also tried bins of 500 and 250 \AA\ and also shifting these bins by half the bin size. The resulting spectra were very similar, but considerably noisier due to the dominance of photon noise in smaller bin sizes. The errors on the fitted $R_P/R_\star$ values are the parameter uncertainties from \textsc{mpfit} based on a fit using photon noise uncertainties for the data points, scaled to include remaining red noise by $\beta$. We find that the standard deviation of the white noise is at the photon noise level, so we do not also re-scale our parameter uncertainties with white noise. As a final check for any remaining nonlinearity, we also extracted the spectrum using the middle two reads, where the number of counts per exposure are within the linear regime, and the resulting spectra agree to well within 1~$\sigma$ with the spectrum extracted from the final read.

\begin{table}
\centering
  \begin{tabular}{c | c | c | c | c | c | c }
\hline
Waveband&  $R_P/R_\star$  & $\chi2$ & DOF & $\sigma_w$ & $\sigma_r$ & $\beta$ \\
 ($\umu$m) &    &  & & \multicolumn{2}{c}{($\times 10^{-4}$)} &   \\
\hline
1.087 - 1.187 & $ 0.1406 \pm 0.00049 $ & 153 & 201 & 5.4 & 1.3 & 1.41 \\
1.187 - 1.287 & $ 0.1401 \pm 0.00033 $ & 202 & 198 & 5.5 & 0.7 & 1.05 \\
1.287 - 1.387 & $ 0.1408 \pm 0.00031 $ &  93 &  201 & 3.7 & 0 & 1.00 \\
1.387 - 1.487 & $ 0.1421 \pm 0.00039 $ & 172 & 201 & 5.0 & 0.9 & 1.27 \\
1.487 - 1.587 & $ 0.1417 \pm 0.00056 $ & 136 & 201 & 4.7 & 1.5 & 1.68 \\
1.587 - 1.687 & $ 0.1410 \pm 0.00039 $ & 207 & 202 & 6.9 & 0.3 & 1.02 \\
\hline
\end{tabular}
\caption{$R_P/R_\star$ values for the WFC3 spectrum of WASP-19b. Uncertainties on fitted parameters are based on using photon noise for the photometric errors in the fits and then re-scaling the fitted parameter uncertainties with $\beta$. The red noise, $\sigma_r$, white noise for the unbinned data, $\sigma_w$, and the scaling factors, $\beta$ are also listed along with fitting statistics. Note that, for $\beta$ values less than 1, we set $\beta=1$ so that there is no scaling of parameter uncertainties.}
\label{table_spectrum}
\end{table} 



\section{Discussion}
\label{sec_discussion}

The transit depths of the spectra for the two different datasets (STIS and WFC3) are measured at different times and so will not be on the same absolute scale as one another because the stellar brightness will be different at the two epochs, by an unknown amount. Unocculted star spots could cause the infrared $R_P/R_\star$ values to be overestimated by up to 1.38~per~cent
relative to the optical data, assuming a spot temperature of 5000~K and non-spotted stellar surface temperature of 5500~K, 
and assuming that the unocculted spot correction is accurate for the STIS radii. We also have to assume no significant changes in stellar variability amplitude between the times of the two observations. Unseen occulted spots could reduce this effect.


For the STIS spectrum, there are uncertainties in the unocculted spot correction through uncertainties in $\Delta f_o$ and uncertainties in the spot temperature. There is also an uncertainty on the estimated value of the non-spotted stellar flux. The error in $\Delta f_o$ and the non-spotted stellar flux will alter the planet's effective baseline radius, and will not significantly change the shape of the transmission spectrum (relative $R_P/R_\star$ as a function of wavelength). From Table~\ref{table_variability_monitoring}, the errors in $\Delta f_o$ translate to errors of less than 1~per~cent on the absolute $R_P/R_\star$ values. Uncertainty in the spot temperature will change the shape of the spectrum by altering its slope as a function of wavelength, and due to features in the stellar spectra could affect the measured planetary atmospheric spectral features. We discuss the effects of uncertainty in assumed unocculted stellar spot temperature on the STIS results in Section~\ref{sec_optical_features}.

\begin{figure}
\centering
\includegraphics[width=8cm]{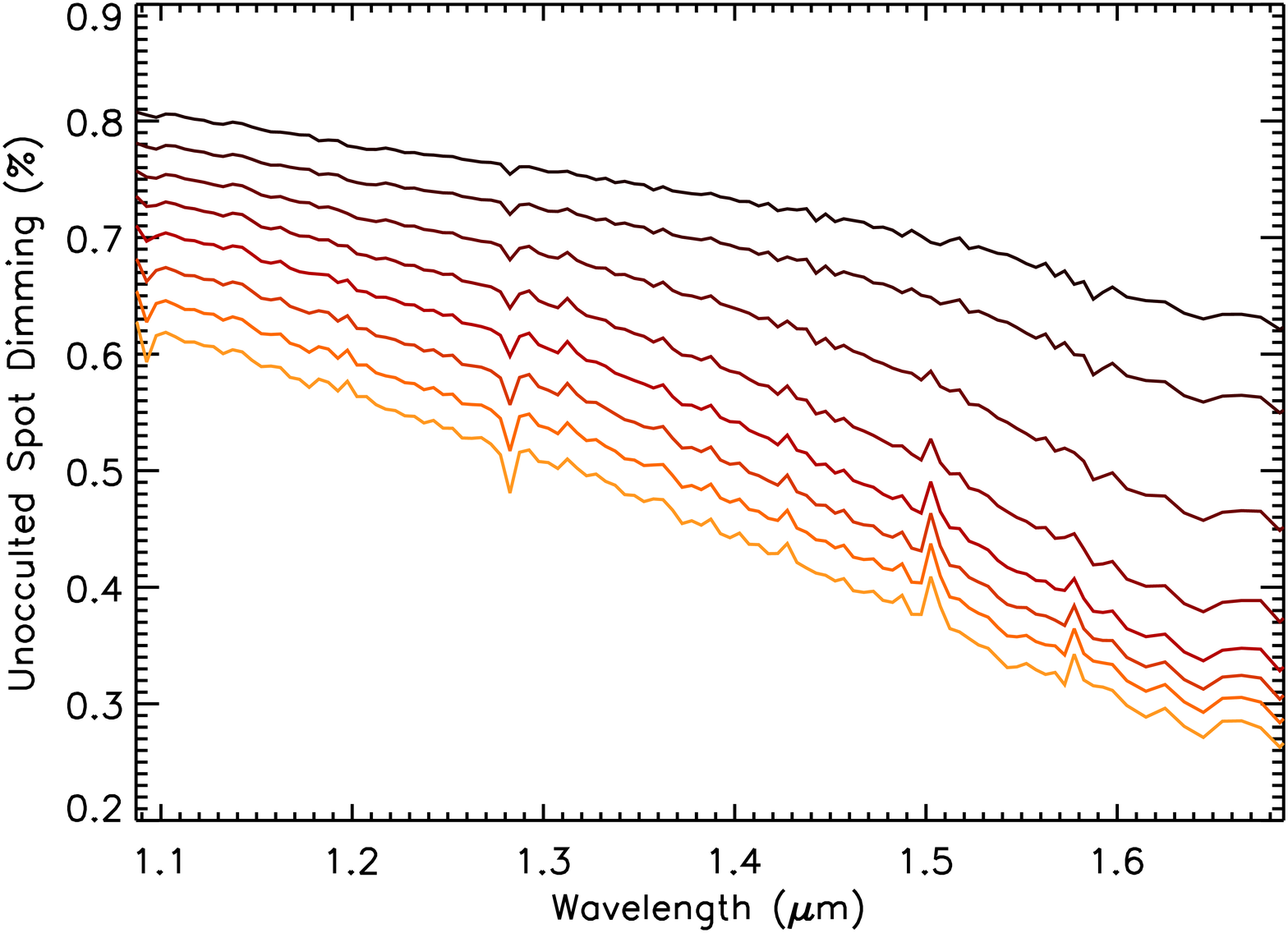}
\caption{Model wavelength dependent unocculted spot dimming over the G141 wavelength range for spot temperatures of 5250-3500~K in increments of 250~K, normalised to 1~per~cent at 6000~\AA. Hotter temperatures are shown in lighter reds, with decreasing temperatures shown with darker reds. The stellar spectrum models are from Kurucz (1993).}
\label{fig_spoctorr_g141}
\end{figure}

For the WFC3 spectrum, the unocculted spot corrections are less severe, and to some extent may be cancelled out by small occulted spots below our noise level. We are not able to correct for this effect, but expect that differential water features between the stellar spots and the non-spotted surface will be washed out to insignificant levels in our large wavelength bins, meaning that the water detection will not be significantly affected. Figure~\ref{fig_spoctorr_g141} shows the unocculted spot dimming over the G141 wavelength range for different temperature spots using the models of Kurucz (1993). These grid models are at low resolution and do not display all the lines present in the stellar spectra, and we discuss the potential of contamination of measured water features in transmission by unocculted stellar spots in Section~\ref{significance_of_water}.

As the stellar activity level during the WFC3 transit is not known, the two transits cannot be joined together to make a single spectrum without an unknown correction in the relative baseline planetary radii of up to 1.38~per~cent. We now draw conclusions from the individual datasets alone, and then use these conclusions together to try to understand the planetary atmosphere.

\subsection{Optical to Near-Infrared Spectrum of WASP-19b}

For both datasets, we compare to two different sets of models; one set based on the formalism of \citet{fortney10,fortney08} and the other set based on the formalism of \citet{burrows10} and \citet{howe12}. Both model sets are calculated specifically for the WASP-19b system. In order to compare these different models to the data, we used the pre-calculated models, allowing them to shift up or down to match the data, with the base radius as the only free parameter. It is important to use more than one set of models since different model sets have been calculated using different methods and not all models in the literature agree (see \citealt{shabram11}).

The models from \citet{fortney10,fortney08} are specifically generated for the WASP-19 system, and include a self-consistent treatment of radiative transfer and chemical equilibrium of neutral and ionic species. Chemical mixing ratios and opacities assume solar metallicity and local chemical equilibrium accounting for condensation and thermal ionisation but no photoionisation \citep{lodders99,loddersfegley02,lodders02,visscher06, lodders09,freedman08}. The transmission spectrum calculations are performed using 1D $T$-$P$ profiles for the day side, which should dominate the spectrum at the terminator, with $\sim 20$~per~cent of incoming energy lost to the un-modelled night side. The surface gravity used was $g=15.9$~ms$^{-2}$. 

The models from \citet{burrows10} and \citet{howe12} are generated specifically for the WASP-19 system and use a 1D dayside $T$-$P$ profile with stellar irradiation, in radiative, chemical, and hydrostatic equilibrium. Chemical mixing ratios and opacities assume solar metallicity and local chemical equilibrium accounting for condensation and thermal ionisation but no photoionisation, using the opacity database from \citet{sharpburrows07} and the equilibrium chemical abundances from \citet{burrowssharp99} and \citet{burrows01}. 

\begin{figure}
\centering
\includegraphics[width=8.8cm]{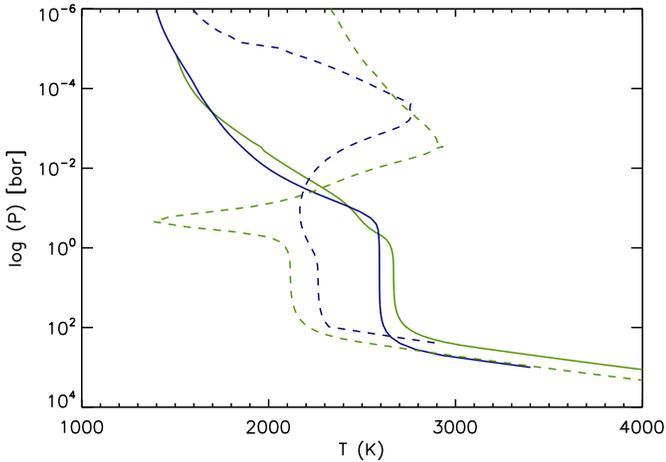}
\caption{Model $T$-$P$ profiles used in the calculation of the model transmission spectra, for atmospheres containing no TiO and VO (solid lines) and atmospheres with a solar abundance of TiO and VO (dotted lines). Blue lines show the models of \citet{fortney10,fortney08} and green lines show the models of \citet{burrows10,howe12}.}
\label{model_tp}
\end{figure}

For both formalisms, the models including TiO and VO opacities have a strong thermal inversion in the $T$-$P$ profile, while the models lacking these opacities do not have this thermal inversion. The $T$-$P$ profiles used to calculate the model transmission spectra re shown in Figure~\ref{model_tp}.

\subsection{Looking for Optical Features}
\label{sec_optical_features}

\begin{figure*}
\centering
\includegraphics[width=8.8cm]{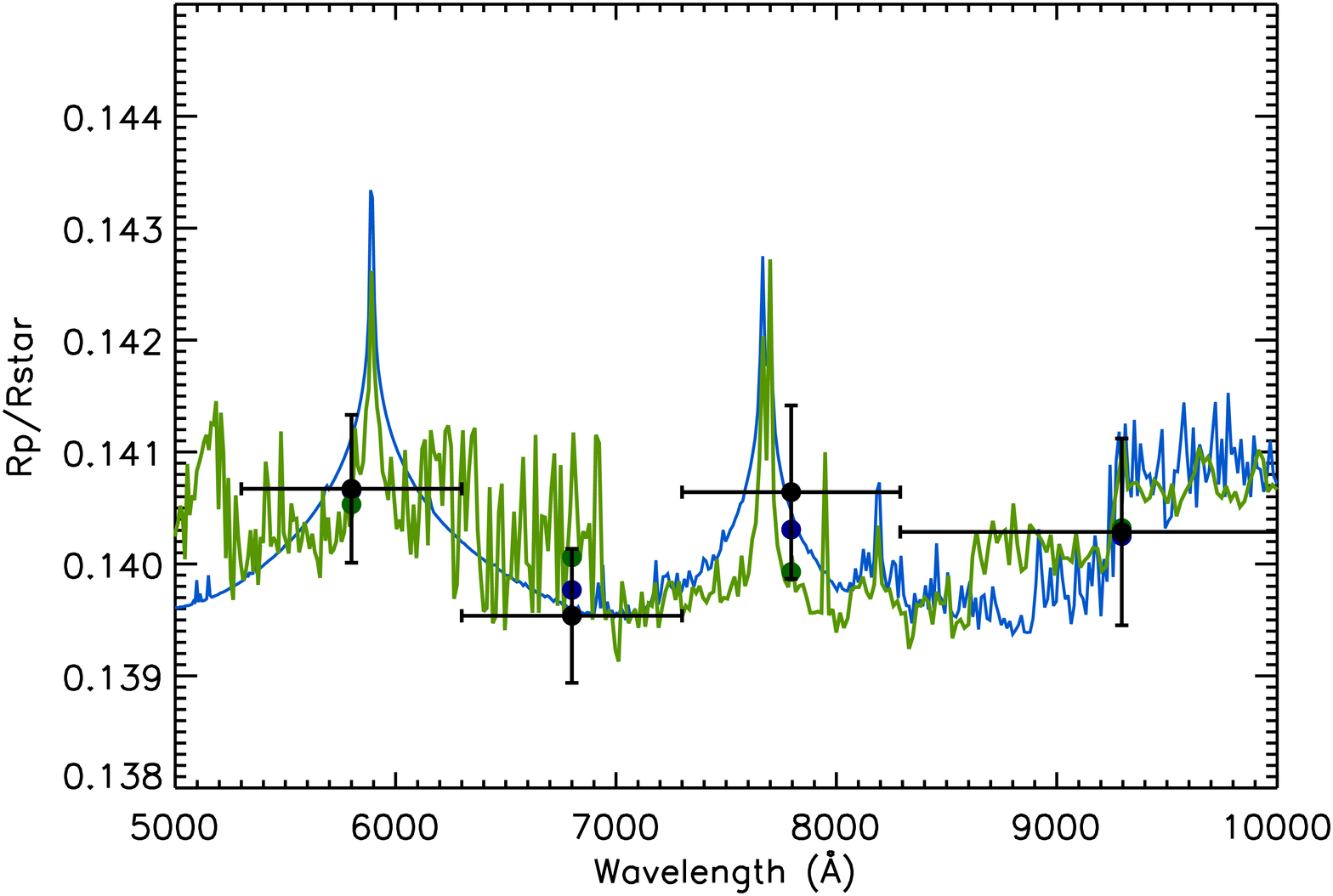}
\includegraphics[width=8.8cm]{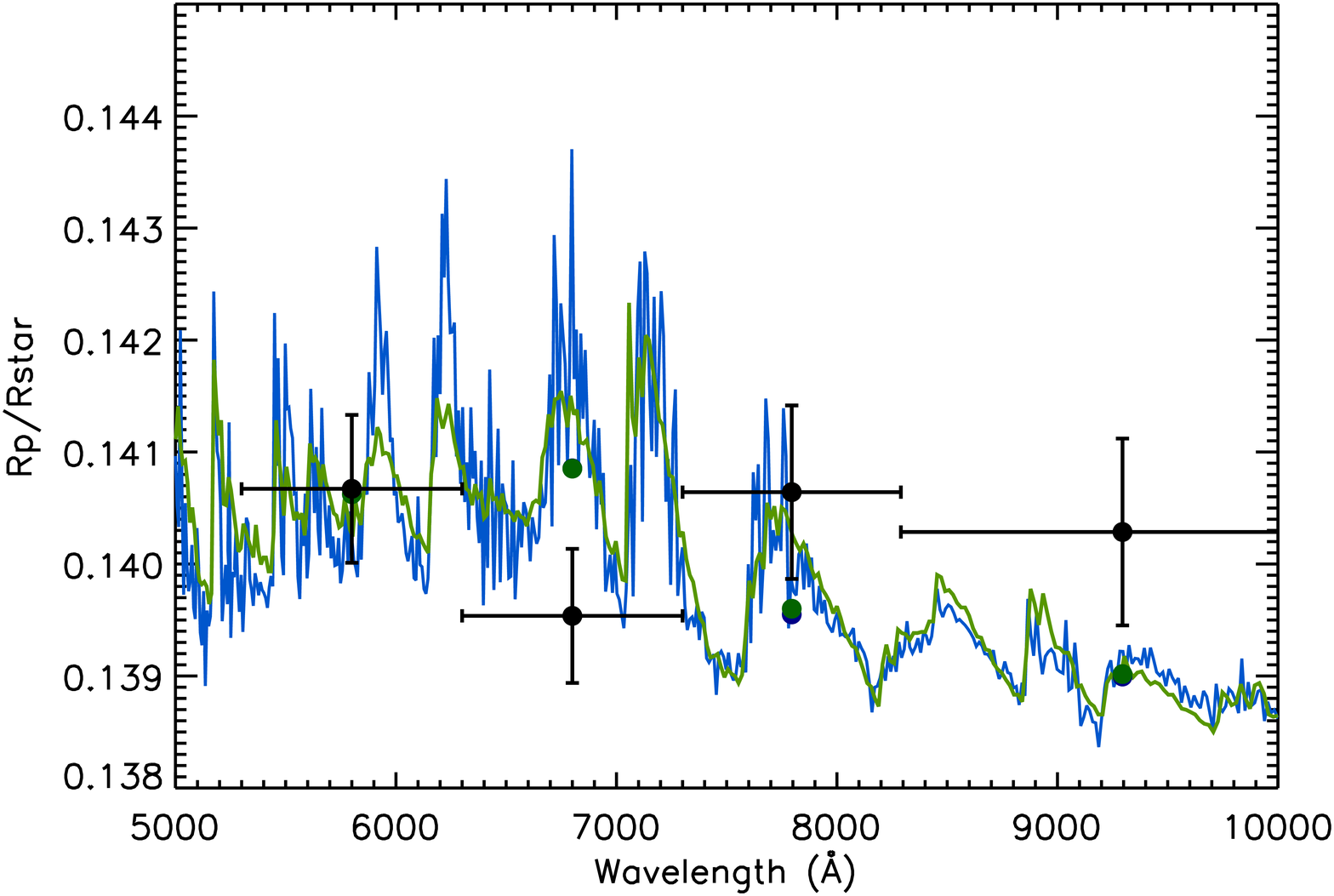}
\caption{The STIS G750L data, with pre-calculated models, fitted to the G750L data baseline level with baseline $R_P/R_\star$ as a free parameter. The left panel shows models with no TiO and the right panel shows models with a solar abundance of TiO. The data points are shown in black with horizontal error bars indicating the bin sizes, the unbinned models are shown with lines and binned model values for our data bins are shown as circles (blue for the \citet{fortney10,fortney08} models and green for the \citet{burrows10, howe12} models).}
\label{stis_modelfits}
\end{figure*}

The STIS radii are consistent with a flat spectrum ($\chi^2=2$ for 3~DOF when fitting a flat line to the spectrum), but the predicted alkali line features in the G750L wavelength range are smaller than the observational error bars. Using clear atmosphere models fitted with an offset in baseline $R_p/R_\star$ gives $\chi^2=1-1.7$ depending on the model used, and using models with solar abundance TiO opacities obscuring the alkali lines, again with a fitted $R_P/R_\star$ offset, gives $\chi^2 = 9.1-9.4$, both for 3~DOF, suggesting a lack of observed TiO features at the 2.7-2.9~$\sigma$ level. The data and fitted models are shown in Figure~\ref{stis_modelfits}. The only free parameter in the spectral model fits was an absolute shift in $R_P/R_\star$, with the shape of the spectrum as a function of wavelength not varying. 

We also produced the spectrum with slightly different binning but similar bin sizes, so that bins were specifically centred around the sodium and potassium lines, and also the region between the lines. It was thought that the differences between the TiO and no-TiO models would be greatest using this binning. However, the difference between the TiO and non TiO models did not increase any further than when using the original binning, due to some of the features in the TiO model being in similar places to the alkali lines. Combined with the bin sizes needing to be decreased, and hence the points having larger photon noise, the method using specific bins centred on alkali lines was no more constraining than the original binning. The original binning is therefore shown in Figure~\ref{stis_modelfits}. 

Additionally, we investigated decreasing the bin sizes for our transmission spectrum. If we use bin sizes of $\sim 400$~\AA\ with binning designed such that two bands are centred on the Na~I and K~I doublets, there is no significant difference in fits between a flat line and a clear atmosphere case. The atmospheres without TiO are favoured over the atmospheric models including TiO opacities at the 2~sigma level. The lowering of the significance level is most likely due larger photon noise uncertainties. We also calculated the spectrum using 200~\AA\ binning (for all but the 4 reddest bins, where the STIS response curve is so low, we kept the bin sizes at 500~\AA). The spectrum calculated using 200~\AA\ wide bins suggests that the flat spectrum is favoured over either the clear or equilibrium model cases. However, the fit of a flat line to the spectrum is not a good fit, with a $\chi^2_\nu$ of 1.7. Possibly other absorbers are present in the optical, but any conclusions about other absorbers not in the model is too tentative at this stage to warrant further discussion. When calculating the spectrum with 100~\AA\ bin sizes, the errors on radii measured in the bins are so large due to photon noise that there is no significant difference between the fits of the clear atmosphere models, equilibrium models and a flat spectrum. 

Our measured spectrum of the change in $R_P/R_\star$ as a function of wavelength suggests an atmosphere without the predicted TiO features. Tests assuming spot temperatures of 3500~K and 5250~K instead of 5000~K to fit the spectral bands give very similar results, suggesting that our choice of spot temperature does not significantly affect the spectral shape and measured features. For further confirmation, we now look for specific features using differential light curves, where the differences could be more significant. Using differential light curves has the advantage that common-mode systematic trends naturally cancel out. For bands that are close together, this may provide an improvement over subtracting the white light residuals, since the systematics may be more similar in bands which are not very far apart in wavelength. 

We define two bands to look for the presence of TiO features. Firstly, we define a ``red edge" band to measure the very strong slope of decreasing planetary radius with increasing wavelength characteristic of TiO in red wavelengths, and refer to the differential measurement as [TiO]$_\mathrm{red}$. Secondly, we define a comb of ``in-TiO" bands and ``out-of-TiO" bands, to be sensitive to the smaller features within the large TiO visible feature, and refer to the differential measurement as [TiO]$_\mathrm{comb}$. We also define bands around the alkali line features, with an ``in" band centred around the doublet core and an ``out" band a combination of a band either side of this, with the differential measurements referred to as [Na] and [K] for the sodium doublet and the potassium doublet respectively compared to the surrounding continuum. We use a band either side to avoid our measurements being skewed by any slopes in the spectrum. We restrict the TiO comb filter to the flat region of the broad TiO feature, to ensure that it is a separate measurement from the red edge measurement, and is not sensitive to spectral slopes. We then apply these bands to the WASP-19b models and use the model values to compare with the differential measurements from the data. The wavelength ranges for each index and the predicted TiO and non-TiO model values for the WASP-19b models are given in Table~\ref{table_model_indices}. We give the model values in units of scale heights so that more than one planet can be directly compared using the index measurements.

\begin{table*}
\centering
  \begin{tabular}{c | c  c c c }
\hline
Band Name &  [TiO]$_\mathrm{red}$ & [TiO]$_\mathrm{comb}$ & [Na] & [K] \\
\hline
``in" bands (\AA) & 6616 - 7396 &  5443 -         5543  & 5597- 6190 & 7280-8080 \\
 &  & 5608 -         5700&   & \\
&  & 5839 -         6002 &    \\
&    &  6150 -         6348 \\
&    &   6641 -         6929 \\
\hline
``out" bands  (\AA) & 7396 - 8175 &   5346 - 5443    & 5300-5597 & 6880-7280 \\ 
& & 5543 -         5608   &     6190-6486 & 8080-8480 \\
 &   &   5700 -         5838    \\
 &   &      6002 -         6150 \\
 &   &       6348 -         6641 \\
 &   &       6929 -         7046 \\
\hline
WASP-19b & -0.71$^{[1]}$ & 0.09$^{[1]}$ & 1.15$^{[1]}$ & 0.60$^{[1]}$ \\
Model Value (no TiO) & -0.07$^{[2]}$ & 0.27$^{[2]}$ & 0.40$^{[2]}$ & 0.36$^{[2]}$ \\
\hline
WASP-19b & 1.75$^{[1]}$ & 1.39$^{[1]}$ & -0.10$^{[1]}$ & -0.60$^{[1]}$  \\
Model Value (TiO) & 1.71$^{[2]}$ & 0.82$^{[2]}$ & 0.10$^{[2]}$ & -0.57$^{[2]}$ \\
\hline
\end{tabular}
\caption{Table showing the wavelength ranges for differential bands and predicted differential model values. The model values are in scale heights, where we take the scale height to be $\sim 490$~km for WASP-19b. The models are made specifically for WASP-19b, but using units of scale heights has the advantage that the model values will be similar between different planets of the same atmospheric type. Model values with [1] next to them are based on the formalism of \citet{fortney10,fortney08} and model values with [2] next to them are based on the formalism of \citet{burrows10,howe12}.}
\label{table_model_indices}
\end{table*} 


The models containing TiO have very similar values in our index measurements aside from the [TiO]$_\mathrm{comb}$ index value. The two models have different resolution, with the models from \citet{fortney10,fortney08} being binned in our plots by a factor of 5, and it is reasonable to suppose that the contrast between small ``in" and ``out" TiO bands will decrease in a lower resolution model. Additionally, there are differences between the clear atmosphere models. In our defined wide bands, the [Na] and [K] signals are smaller in the models based on \citet{burrows10,howe12} than the models based on \citet{fortney10,fortney08}. The reason for lower [Na] and [K] signals in the \citet{burrows10, howe12} models is that these models contain opacities that are not included in the models of \citet{fortney10,fortney08}, which obscure the broad Na~I and K~I line wings. The reason for the [TiO]$_\mathrm{red}$ signal being less negative is the same, because the bands used for the [TiO]$_\mathrm{red}$ measurement intersect with the potassium feature, which will cause a negative differential absorption depth measurement in a clear atmosphere and a positive measurement in an atmosphere containing TiO opacities.

\begin{figure}
\centering
\includegraphics[width=8cm]{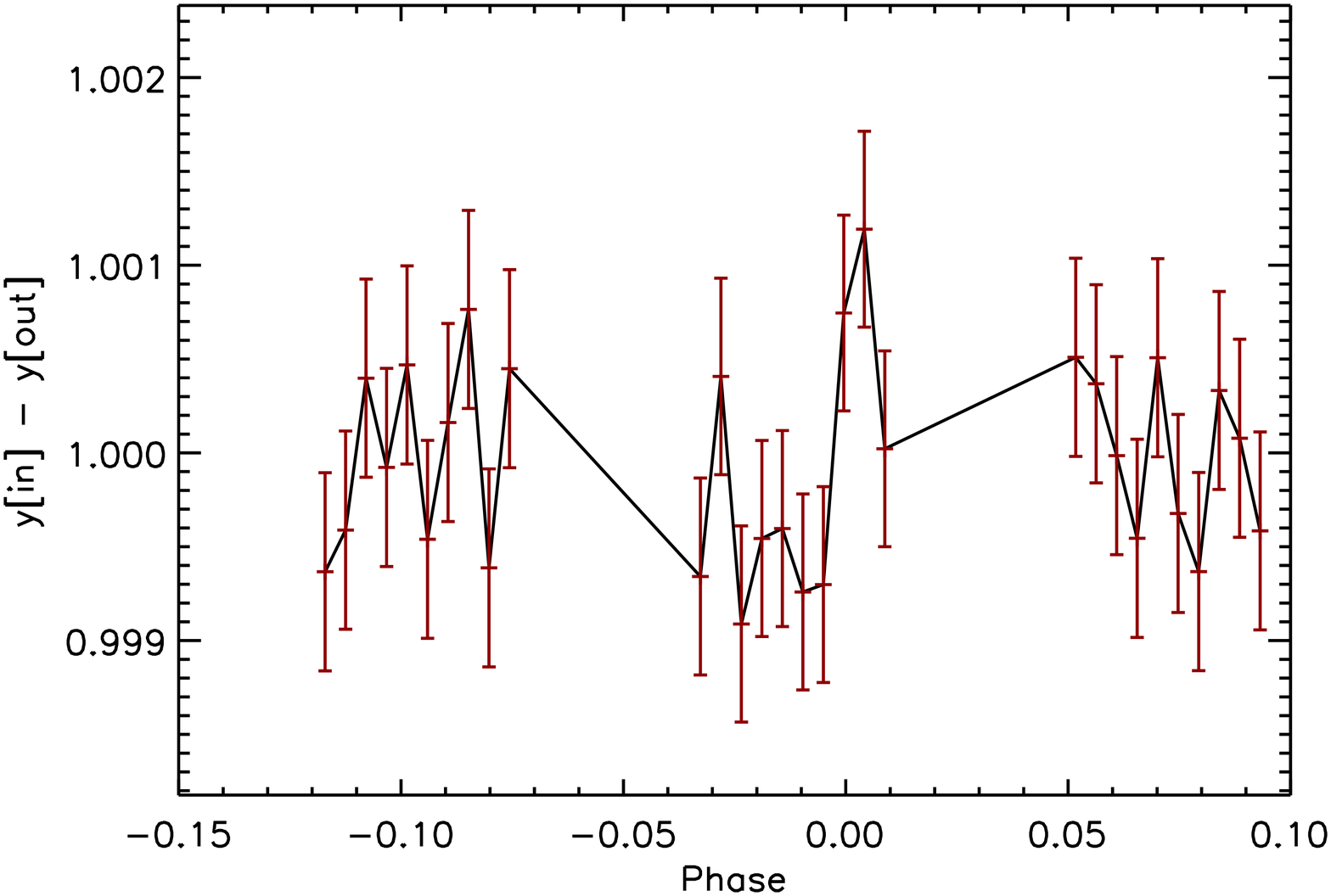}
\includegraphics[width=8cm]{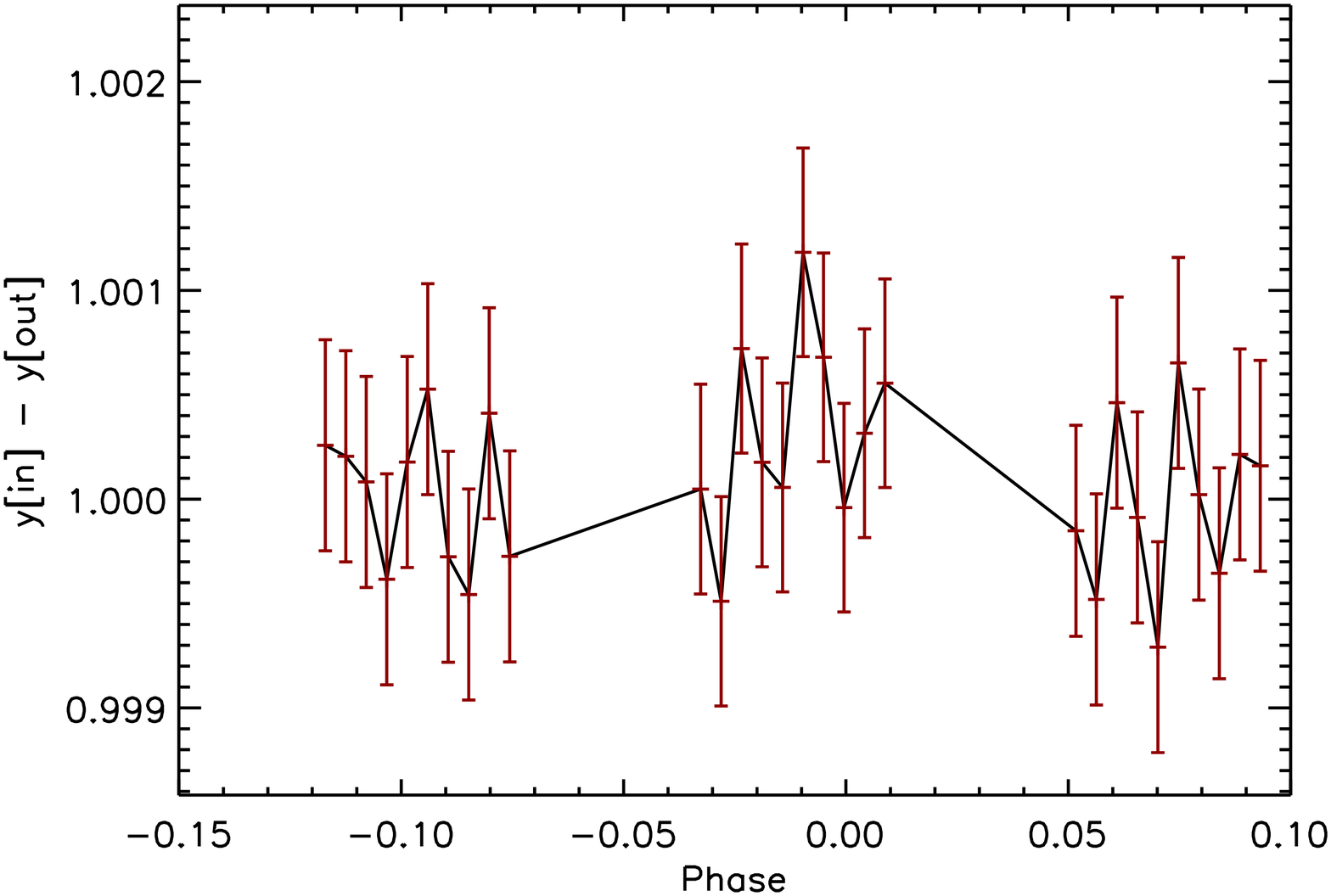}
\includegraphics[width=8cm]{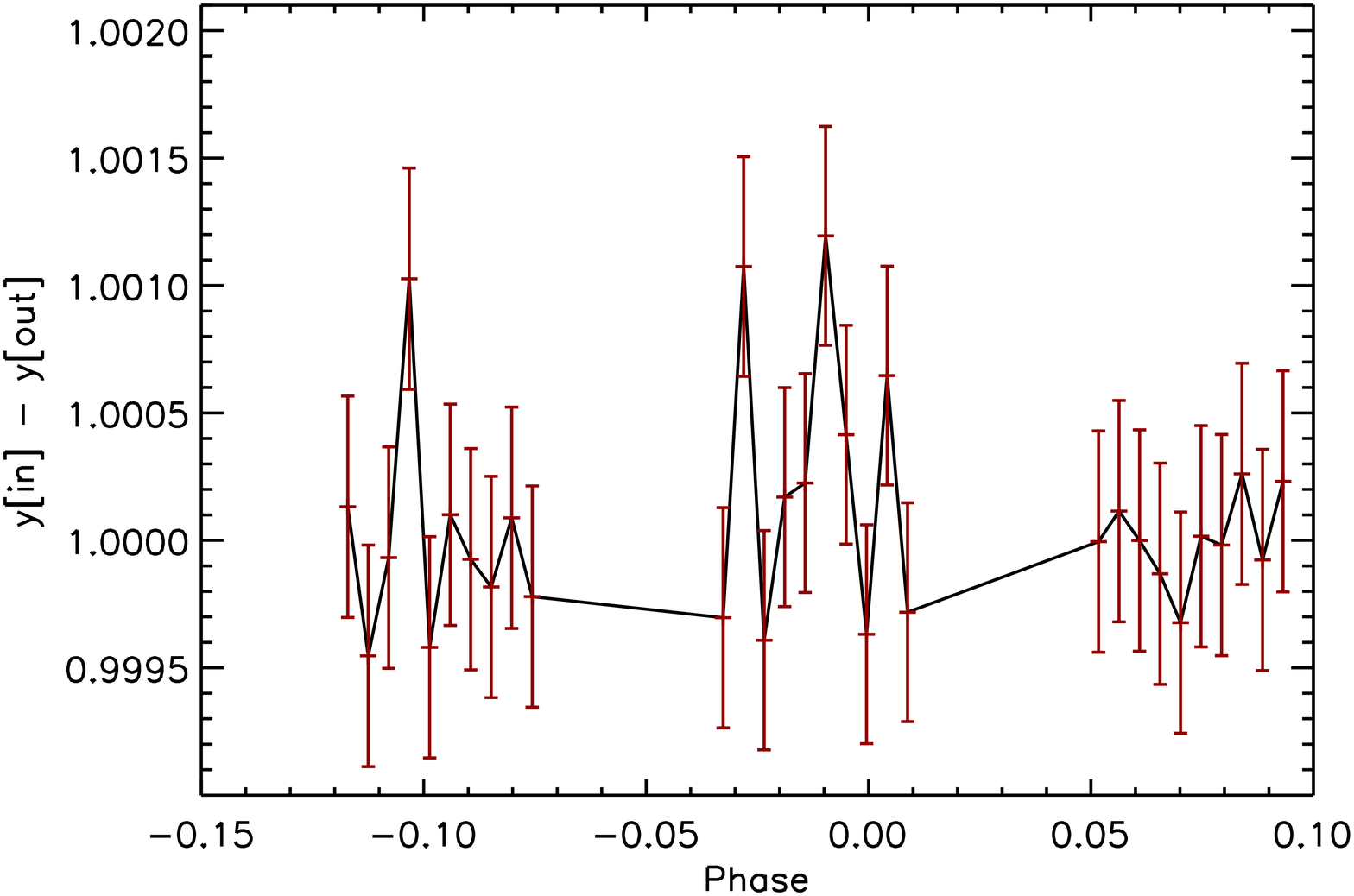}
\caption{Differential light curves. \textit{Top: }K minus continuum ([K]), \textit{middle: }TiO ``red edge" ([TiO]$_\mathrm{red}$), \textit{bottom: }TiO ``comb" ([TiO]$_\mathrm{comb}$). Each differential light curve is the ``in" - ``out" light curve + 1. Uncertainties on the data points are from photon noise.}
\label{fig_differentialLC_stis}
\end{figure}

We measured the differential transit depths for each ``in" and ``out" band combination by first normalising the light curve in each wavelength to the out-of-transit flux, and then subtracting one light curve from the other. We corrected for differential limb darkening by using the Kurucz (1993) stellar models to predict the limb darkening coefficients for each wavelength, and found the differential limb darkening coefficients. These were then subtracted from the differential light curves. We then measured the differential absorption depth in transit compared to out of transit, as previously done for alkali line features (e.g. \citealt{charbonneau02}). The differential light curves are shown in Figure~\ref{fig_differentialLC_stis} and Table~\ref{table_index_measurements} lists the results. We used the out-of-transit regions of the light curves to de-trend with remaining wavelength-dependent systematics, based on HST phase and the position of the spectrum on the CCD. We found that de-trending using only a linear slope with time produced differential absorption depths that agree to within $\sim 6$~per~cent (less than $1/20 \sigma$ for the broad features). Additionally, we tried fitting for the differential transit (one limb-darkened transit minus the other) simultaneously with the systematics and found similar results (both with a full model and fitting only a linear slope with phase) with differences of less than $1/2 \sigma$. The uncertainties on each mean ``in" minus ``out" value is the standard deviation of the differential light curve. For bands other than sodium, we found that fitting different models for the systematics gives similar results. Therefore, we can assume that white noise mostly dominates in the differential light curves. 

We found a more significant dependence on fitted parameters for the sodium feature, however, which indicates that wavelength-dependent systematics dominate in this band, and so we are not able to use this measurement. The measured [Na] value changed by $\pm 2$~scale~heights depending on the model chosen. 
It is well known that STIS exhibits wavelength-dependent breathing trends, which are worse at the spectral edges (e.g. \citealt{sing11}). While it was hoped that the differential measurements would alleviate this problem to some extent by using bands relatively close together, it is not sufficient for the [Na] measurement. Wavelength-dependent trends clearly remain and the method of their removal significantly affects the measured differential [Na] measurement.
For the sodium feature, the blue ``out" of Na~I band lies very near the spectral edge of the STIS response curve. Therefore, we also tried to calculate the differential absorption in the [Na] index using smaller ``out" bands. The results were very similar to using the full ``out" bands. We tried using only the red ``out" band also, but found that this still gave significantly different results depending on the de-trending model used. When constructing the transmission spectrum using 400~\AA\ bins, the prayer bead analysis showed significantly higher variance in the band centred on the sodium feature and the bands either side of it compared to other bands with similar signal levels. The problem may therefore be in the ``in" band itself. We tried one further method, which was to make the ``in" and ``out" bands even smaller, so that the dominant systematics would be similar in each band. However, in order to do this, we required bands of $\sim 100$~\AA\ width, which meant that high photon noise uncertainties prevented us from making a useful measurement. 

We also defined some bands in the blue G430L wavelengths to see if we could distinguish spectral slopes characteristic of hazes from spectral slopes induced by the occulted spots. However, we found that the results we obtained varied significantly (by $\pm 3$~scale~heights) for a given index depending on how we chose to treat the systematics. We therefore do not quote values from these data.


\begin{table}
\centering
  \begin{tabular}{c | c | c | c}
\hline
& & \multicolumn{2}{c}{model $\Delta z/H$ value} \\
Index & $\Delta z/H$ & \multicolumn{2}{c}{TiO / no TiO}\\ 
& & [1] & [2] \\
\hline
[K] &  $0.71	\pm 1.32$ & -0.6 / 0.6 & -0.57 / 0.36 \\ 
$[\mathrm{TiO}]_\mathrm{red}$ & $-1.73 \pm 1.59$ & 1.75 / -0.71 & 1.71 / -0.07\\ 
$[\mathrm{TiO}]_\mathrm{comb}$ & $-1.42 \pm 1.37$ & 1.39 / 0.09 & 0.82 / 0.27 \\ 

\hline
\end{tabular}
\caption{Table showing differential absorption depth values scaled with scale height, $H$ ($\sim 490$~km). Model values in column [1] are based on the formalism of \citet{fortney10,fortney08} and model values in column [2] are based on the formalism of \citet{burrows10,howe12}. }
\label{table_index_measurements}
\end{table} 

The [TiO]$_\mathrm{red}$ and [TiO]$_\mathrm{comb}$ measurements show a tentative lack of predicted TiO features, assuming solar TiO abundance. The [TiO]$_\mathrm{red}$ measurement is 1~$\sigma$ away from the model values for the clear atmosphere model and 2~$\sigma$ away from both model values assuming an atmosphere with solar TiO. The [TiO]$_\mathrm{comb}$ measurement is 2~$\sigma$ away from the TiO-containing model value from \citet{fortney10,fortney08} and only 1.6~$\sigma$ away from the lower resolution TiO-containing model value from \citet{burrows10,howe12}. The [TiO]$_\mathrm{comb}$ measurement is 1.1-1.2~$\sigma$ away from the clear atmosphere model values.

The [K] measurement is within 1~$\sigma$ of the clear and TiO model values, and as such the precision is not high enough to distinguish between the cases of whether the potassium feature is present or whether it is not. We tested whether the measured potassium feature becomes stronger when using smaller ``in" bands for the potassium feature. For example, the measurement when using a 50~\AA\ ``in" band around the K~I feature instead of the band defined in Table~\ref{table_model_indices} and using the same ``out" band as defined in Table~\ref{table_model_indices} gave a differential radius of $2.0 \pm 3.5$~scale~heights, while the model values are 3.0-3.4~scale~heights. The measurement when using a 100~\AA\ band for the ``in" potassium band is $1.9 \pm 2.8$~scale~heights, where the model values are 2.1-2.8~scale~heights. The differences between the model values as a function of ``in" bandwidth used to measure the K~I feature are within the uncertainties of the measurements and so we cannot confirm or reject the hypothesis that K~I is present in abundance in the upper atmosphere of WASP-19b. 

We found that the measurements of differential depth in the sodium feature were still dependent on the choice of de-trending model used and so we do not quote those values here. However, we find that the spectrum around the sodium region is better behaved than the spectrum in the `out' bands, with the measured $R_P/R_\star$ values changing less significantly depending on the de-trending model than the differential measurement. We find an $R_P/R_\star$ value of $0.1383 \pm 0.0032$ in a 30~\AA\ band centred around the Na~I feature, where photon noise dominates. In a wider, 100~\AA\ band centred around the Na~I feature, instrumental systematics become more important and we obtain an $R_P/R_\star$ value of $0.1359 \pm 0.0019$~(random)~$\pm 0.0038$~(syst). The difference between measured $R_P/R_\star$ values is less than the measurement uncertainties. The corresponding difference in model values between the $R_P/R_\star$ values binned over a 30~\AA\ band and the $R_P/R_\star$ values binned over a 100~\AA\ band, both centred on the Na~I feature is 0.0004 to 0.0008 depending on the model. We are therefore not able to confirm or rule out the presence of the predicted Na~I feature in the transmission spectrum of WASP-19b.


It is possible that features in the transmission spectrum can change depending on the assumed stellar spot temperature, and hence assumed features in the stellar spectrum, meaning that any of our band measurements could be affected by which temperature we use for the unocculted spot correction. We investigate possible spot temperatures in increments of 250~K from 5250~K to 3500~K, which is 2000~K cooler than the nonspotted stellar surface. We find that altering the spot temperature from the extremes of this range, from 3500 to 5250~K, affects the measured index measurements by $ \le 0.05$ scale heights.

\subsection{Significance of the Near-Infrared Water Features}
\label{significance_of_water}

\begin{figure*}
\centering
\includegraphics[width=13cm]{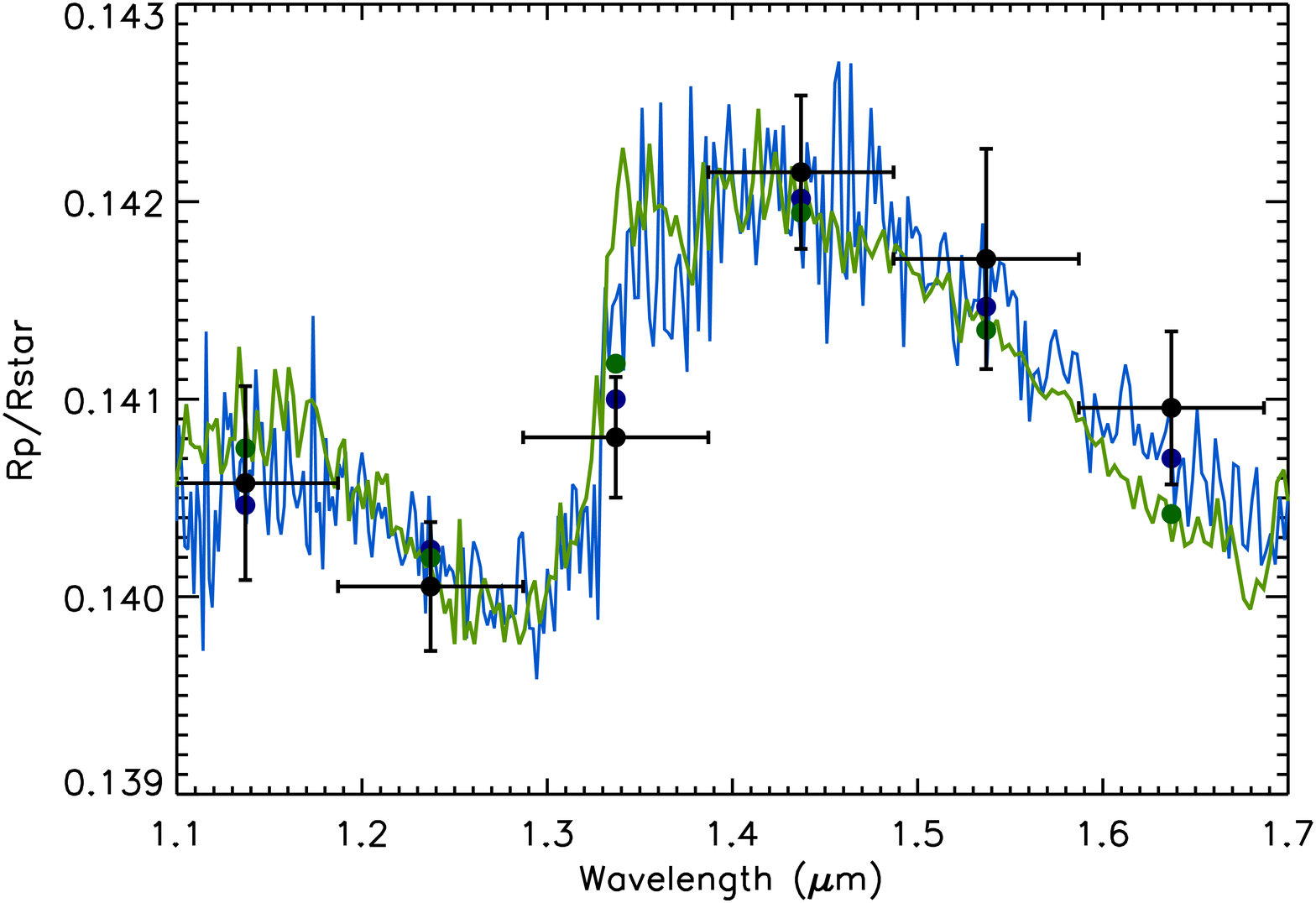}
\includegraphics[width=8cm]{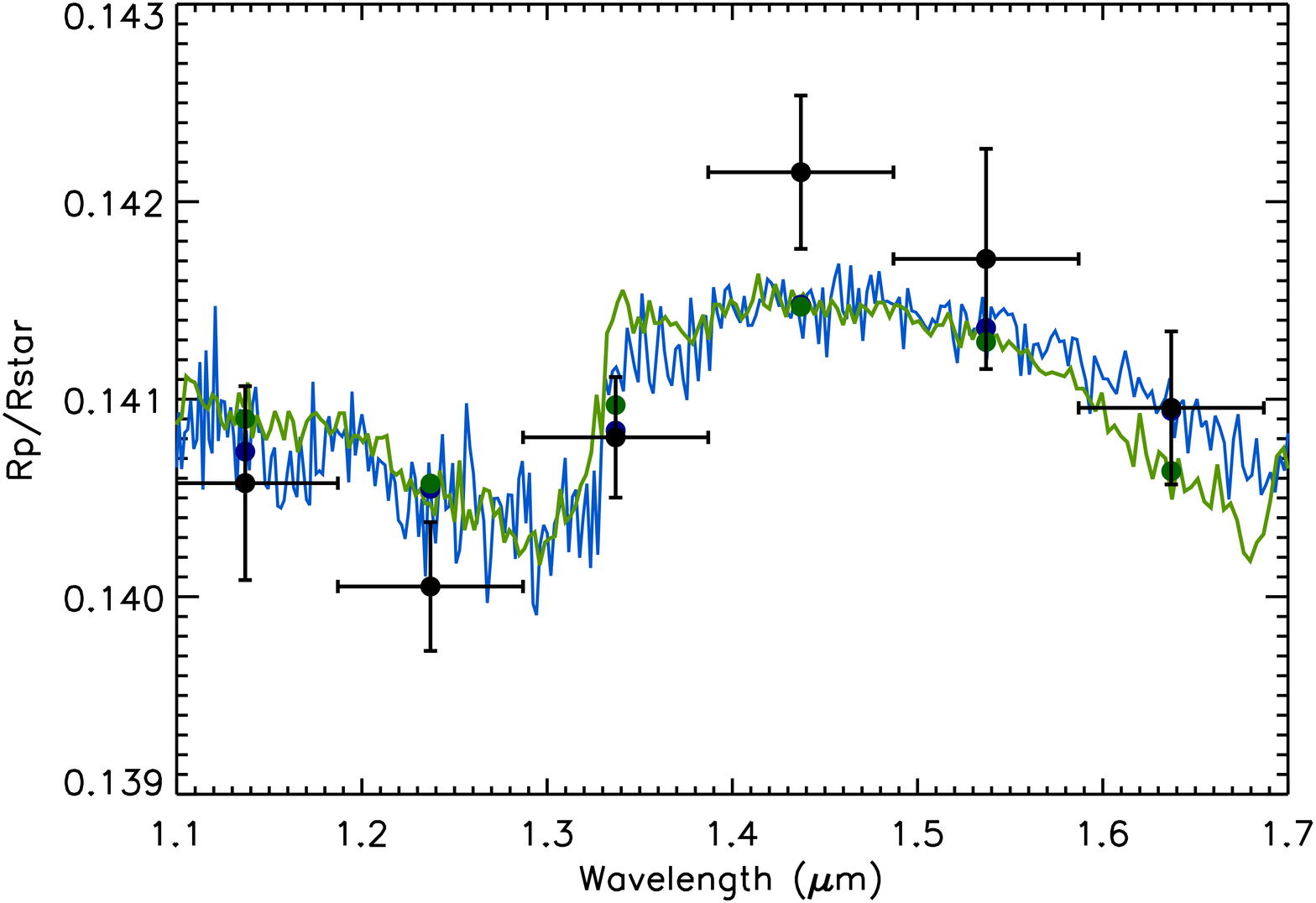}
\includegraphics[width=8cm]{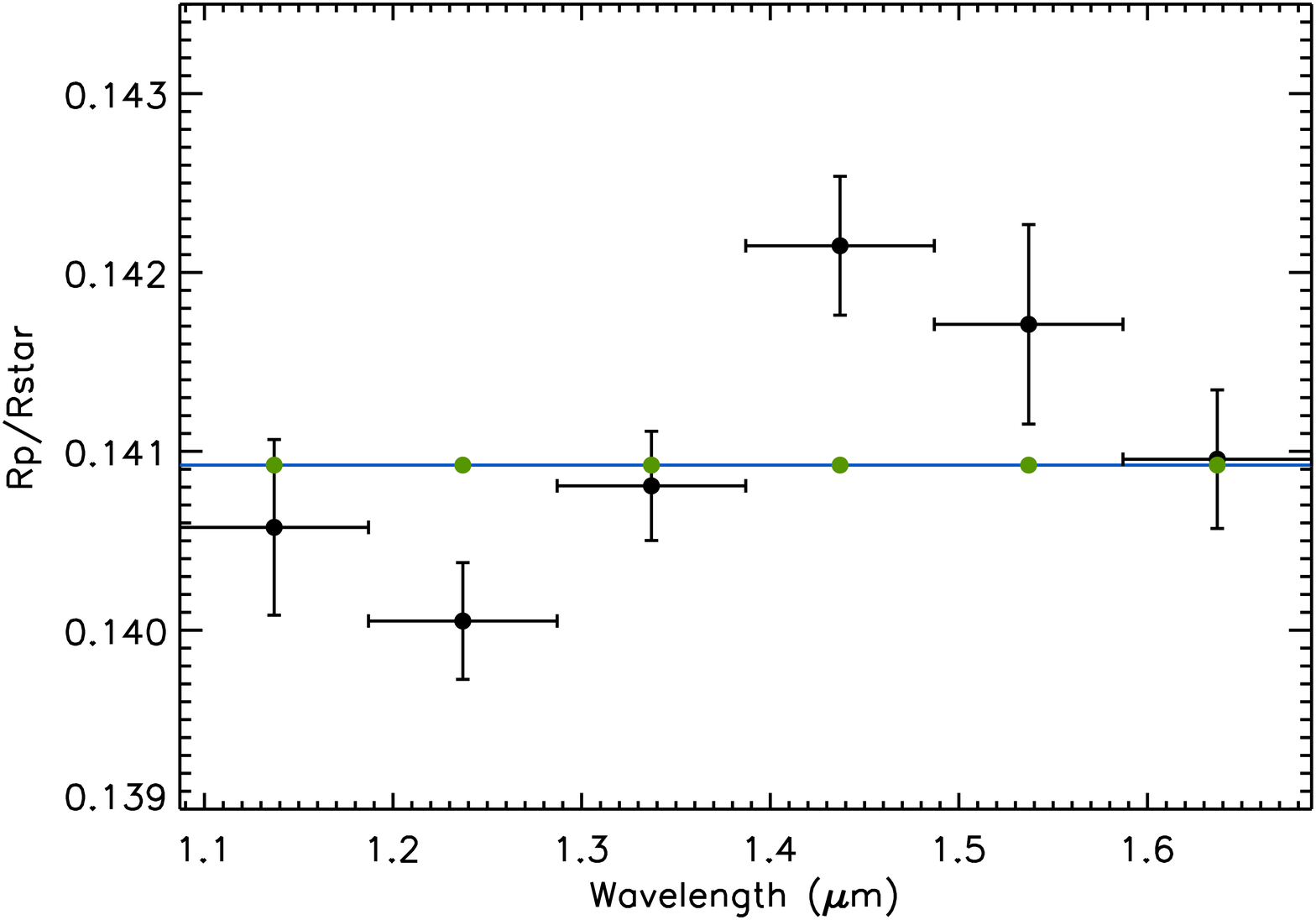}
\caption{\textit{Top: }WFC3 data and models including water opacities but no TiO opacities with non-inverted $T$-$P$ profiles, fitted to the WFC3 data level with baseline $R_P/R_\star$ as a free parameter. \textit{Bottom Left: }WFC3 data and models including water and TiO opacities and a strong temperature inversion, fitted to the WFC3 data level with baseline $R_P/R_\star$ as a free parameter. \textit{Bottom Right: }Flat line fit to the WFC3 data. The data points are shown in black with horizontal error bars indicating the bin sizes, the models are shown with lines and binned model values are shown as circles (blue for the models of \citet{fortney10,fortney08} and green for the models of \citet{burrows10,howe12} for the top and bottom left plots).}
\label{wfc3_modelfits}
\end{figure*}

Comparing the clear-atmosphere WASP-19b models to the water feature with fitting only an absolute offset in $R_P/R_\star$ as a free parameter gave $\chi^2=1.5-4.1$ for 5 DOF, which are our best fitting models. A flat line, on the other hand, gave $\chi^2=22.1$ for 5 DOF, indicating a 4~$\sigma$ confidence in the water dominated model over the null hypothesis. Using models that include opacities from TiO and VO gave $\chi^2=5.8-7.6$ for 5 DOF, indicating that the WFC3 data suggest a lack of observable TiO features. Comparing each model including TiO and VO opacities to the corresponding clear atmosphere model shows that the clear atmosphere case is a better fit at the $1.8-2$~$\sigma$ level. The data and models fitted to the baseline $R_P/R_\star$ of the data are shown in Figure~\ref{wfc3_modelfits}. 

We have not corrected the WFC3 spectrum for occulted or unocculted stellar spots. If present, unocculted spots would increase the measured $R_P/R_\star$ values compared to the non-spotted case, while occulted stellar spots would do the opposite. If the star is uniformly spotted, both the effects of occulted and unocculted spots should cancel out (e.g. \citealt{pont13}). The most significant effect on the infrared spectrum if they do not cancel out would be to introduce a slope in the transmission spectrum. The slope would be $\sim 0.00027$~$R_P/R_\star$ from the bluest to reddest bin, assuming $T_{\mathrm{spot}}=5000$~K and a dimming of 1\%. This slope should not highly affect the detection of water features, particularly not the feature at 1.4~$\umu$m, which is well encompassed by the G141 band, and so has a peak in $R_P/R_\star$ as a function of wavelength rather than a slope as a function of wavelength. Furthermore, we do not see a significant slope in the spectrum compared to the spectral features, suggesting that occulted and unocculted spots have indeed cancelled out to a large extent.

Another possible effect is due to the presence of water features in the stellar spots. If the water features in the spots are significantly different from the non-spotted stellar spectrum, unocculted spots will artificially increase the significance of water features in the planet's transmission spectrum (e.g. \citealt{deming13}). There are no significant water features seen in the Kurucz (1993) grid of stellar spectra for the G141 wavelength range. To test whether our assumed unocculted spot dimming models as a function of wavelength are reasonable, we also investigated the much higher resolution models of R.L. Kurucz for HD~209458 and HD~189733 where water features in the stellar spectra can clearly be seen. We used the stellar model of HD~209458 ($T=6100$~K) as the non-spotted surface and the stellar model of HD~189733 ($T=5050$~K) as the spot spectrum. We found that, although water features are clearly seen in stellar spectra at higher resolution, the differential effect compared to the non-spotted surface is negligible in the large wavelength bins used for the transmission spectrum.



\subsection{Interpretation}

Although we cannot yet put the spectra from the STIS G750L and WFC3 G141 instruments together on an absolute scale due to unknown differences in stellar brightness between the two observations, we can still place strong constraints on the atmospheric type. We detect water in the near infrared, and our observations in the optical suggest that we do not observe large TiO features. 

The C/O ratio of WASP-19b is currently unconstrained from secondary eclipse observations \citep{anderson11b,burton12,lendl12}, but models predict that the water abundance will drop dramatically compared to a solar-composition atmosphere if the C/O ratio is greater than 1 \citep{madhusudhan12,moses13}. Therefore, the presence of water in the transmission spectrum suggests that the atmosphere does not have a high C/O ratio. Additionally, the amplitude of the measured water features match solar abundance models with no cloud cover, in contrast to observations of HD~189733b, HD~209458b and XO-1b \citep{gibson12b,pont13,deming13}. However, with our low resolution, it is hard to tell whether the water features may be slightly muted at a similar level to XO-1b, since 1~$\sigma$ shifts in only two of our bins would show a smaller feature consistent with low cloud cover. Water features muted at the level observed in HD~209458b are likely ruled out.

The probable lack of a TiO detection in transmission agrees with previous secondary eclipse observations of an atmosphere without a strong temperature inversion \citep{anderson11b,madhusudhan12}. A lack of TiO features, if not caused by a high C/O ratio, could mean that TiO is trapped in lower regions of the atmosphere that are opaque to transmission spectroscopy, as described by \citet{spiegel09}, \citet{showman09} and \citet{parmentier12}. \citet{spiegel09} find that in the atmosphere of OGLE-TR-56, a planet only $\sim 100$~K cooler than WASP-19b, there could be a TiO cold-trap on the planet's day side. Even without a day side cold-trap, vertical mixing may still not be enough to lift gaseous TiO into the observable upper atmosphere. Condensation on the night side is also possible if TiO particles reach sizes over a few microns, or if TiO forms condensates with other species \citep{parmentier12}. Transmission spectroscopy measures the spectrum at the planetary terminator, and it is currently unclear how day-side and night side conditions affect the terminator region. Since the strength of vertical mixing is coupled with the horizontal flow, the terminator region experiences time-variations in the abundances of atmospheric species, as the upward flow changes \citep{parmentier12}. 

An alternative explanation for a lack of TiO features could be that stellar UV radiation breaks down the molecule as proposed by \citet{knutson10}. WASP-19 is an active star, with $\log (R'_{\mathrm{HK}})$ = -4.660 compared to -4.970 for the inactive HD~209458, which hosts a planet with a stratosphere. WASP-19b is also very close to the star at only 0.0166~AU, and so is very highly irradiated.



It is also possible that a less than solar abundance of gaseous TiO exists in the upper atmosphere, but is concealed by high-altitude hazes, clouds or dust. Since the WFC3 data have not been corrected for unocculted star spots, we would expect the measured WFC3 radii to be high compared to predictions from the G750L data. However, when compared to the clear atmosphere model, the G750L measured radii are too high when the model is fitted to the WFC3 level, which could indicate an extra absorber or scatterer higher than the predicted continuum of a clear atmosphere. However, if water is observed in the infrared, such an obscuring layer would have to be optically thin in the near-infrared. Also, at the high temperatures of the atmosphere of WASP-19b, many dominant condensates are not able to form \citep{fortney05}. Without being able to use the G430L data to construct a transmission spectrum, we cannot distinguish between Rayleigh scattering from small-particle hazes or clouds or dust composed of larger grains, which may also show features blueward of 4500 \AA, such as absorption from HS \citep{zahnle09}. Other features specifically just in the G750L wavelength regime could also dilute TiO features. For example, the cross-sections of VO would cause absorption that will hide the TiO bands if the abundance of VO is high compared to TiO in the upper atmosphere (e.g. both TiO and observable quantities of VO are consistent with the observed transmission spectrum of HD~209458b although not confirmed \citep{desert08}). VO hiding TiO signatures would require a significant departure of abundances from equilibrium in the upper atmosphere, since VO is already included in the equilibrium chemistry models at the solar abundance ($\sim 1/20$ times the abundance of TiO).



\section{Conclusions}
\label{sec_conclusions}

We measured the optical transmission spectrum of WASP-19b with HST STIS G750L from 5300-10300~\AA\ during one planetary transit and with HST STIS G430L from 2900-5700~\AA\ during two transits. The STIS results are the first from a large HST survey of 8 hot Jupiters. We combined these with archive WFC3 data of one transit of WASP-19b from 1.1-1.7~$\umu$m.

The WFC3 data show evidence of water in the atmosphere of WASP-19b at the 4~$\sigma$ confidence level and in agreement with solar-abundance models. Although we cannot rule out low levels of muting, features muted to the extent of those seen in HD~189733b and HD~209458b are ruled out. The presence of non-muted water features suggests a non-cloudy atmosphere or only low-level clouds and also indicates that the planet does not have a high C/O ratio. Followup observations with WFC3 G141 can confirm the water detection as well as determine more accurately the strength of the feature, as the newly available spatial scanning mode can be used to enable higher resolution spectra to be obtained. 
Observations using the WFC3 G102, with wavelength coverage of 8000-11500~\AA\ can also be used to place the optical and near infrared existing transmission spectra on the same scale. 

The STIS G750L data show no evidence for TiO features in either the broadband transmission spectral data (2.8-2.9 sigma confidence) or two TiO specifically tailored bands (2 sigma for the red-edge and cross-correlation indices). There could be several possible reasons for a lack of observable TiO features, including no TiO being present in the upper atmosphere due to rainout or low vertical mixing, 
possible breakdown of TiO by the intense activity of the host star, or an obscuring haze, cloud or dust layer of unknown composition. A high C/O ratio is unlikely if water is observed in the transmission spectrum at the amplitude seen here. A lack of TiO features in the transmission spectrum is not unexpected for a planet which likely has no strong stratospheric thermal inversion.

We were not able to determine the planetary spectrum in the G430L wavelength range because of having to fit for occulted starspots, radii and systematics simultaneously with very few exposures and with no spot-free exposures between 2$^{\mathrm{nd}}$ and 3$^{\mathrm{rd}}$ contact. If the stellar spot temperature is known, then the spot amplitude parameter can be much better constrained for each wavelength. One possible method of inferring the mean stellar spot temperature would be to obtain data for the 3000-5000~\AA\ range from the ground, where complete light curves should increase the chance of having at least some in-transit exposures not occulting star spots. Having non-spotted exposures during the transit between 2$^{\mathrm{nd}}$ and 3$^{\mathrm{rd}}$ contact should help break the degeneracy between occulted spot amplitudes in each band and measured planetary radii. Simultaneous ground and space based observations from 3000-5000~\AA\ could also provide an even better constraint since there would be no difference in unocculted star spot level between the two observations and systematic trends could be better characterised. We would then have to assume that the occulted spots represent the mean temperature of unocculted spots.

We find that WFC3 performs very well for transit observations, with the majority of systematic trends repeating very closely from orbit to orbit. The new spatial scanning mode will likely improve the data obtained with WFC3, by allowing longer exposures and hence an improved duty cycle. It will also spread the light over more pixels reducing the effects of persistence, charge trapping and the contribution of specific pixels to the spectrum. We found that nonlinearity in pixels was the only data-quality issue that affected the spectrum by greater than 1~$\sigma$ and even then, only in the flagged pixels. It will also be easier to avoid saturation with spatial scanning mode.

\section*{Acknowledgments}

This work is based on observations with the NASA/ESA Hubble Space Telescope. This work makes use of data from the Cerro Tololo Inter-American Observatory, National Optical Astronomy Observatory, which are operated by the Association of Universities for Research in Astronomy, under contract with the National Science Foundation. We thank the anonymous referee for a thorough and thoughtful report. C.M. Huitson acknowledges support from STFC. This research has made use of NASA's Astrophysics Data System, and components of the IDL astronomy library. The authors acknowledge support by NASA through grant HST-GO-12473. D. K. Sing and N. Nikolov acknowledge support from STFC consolidated Grant ST/J001627/1.

\bibliographystyle{mn2e} 
\bibliography{wasp19_hst}

\end{document}